\documentclass[fleqn,usenatbib]{mnras}
\usepackage[T1]{fontenc}
\usepackage{ae,aecompl}
\usepackage{hyperref}

\usepackage{graphicx}	% Including figure files
\usepackage{amsmath}	% Advanced maths commands
\usepackage{amssymb}	% Extra maths symbols
\usepackage{multirow}
\usepackage[usenames,dvipsnames,svgnames,table]{xcolor}
\usepackage{verbatim}

\newcommand{\msun}{{\rm M_\odot}}

\newcommand{\hMpc}{\, h^{-1}{\rm{Mpc}} }
\newcommand{\magn}{\, {\rm mag} }

\newcommand{\hsmsun}{\,h_{70}^{-2} {\rm M_\odot}}

\newcommand{\hsMpc}{\, h_{70}^{-1}{\rm{Mpc}} }

\newcommand{\lcdm}{{\rm \Lambda CDM}}
\newcommand{\am}{\, {\rm arcmin}}

\newcommand*{\mean}[1]{\overline{#1}}
\newcommand*{\E}[1]{\times 10^{#1}}

\title[Trough and Ridge Lensing with KiDS]{Studying galaxy troughs and ridges using Weak Gravitational Lensing with the Kilo-Degree Survey}
\author[M. M. Brouwer et al.]{Margot M. Brouwer$^{1,2}$\thanks{E-mail:brouwer@strw.leidenuniv.nl},
	 Vasiliy Demchenko$^{3}$,
	 Joachim Harnois-D{\'e}raps$^{3}$, \and
	 Maciej Bilicki$^{1,4}$,
	 Catherine Heymans$^{3}$,
	 Henk Hoekstra$^{1}$,
	 Konrad Kuijken$^{1}$, \and
	 Mehmet Alpaslan$^{5}$,
	 Sarah Brough$^{6}$,
	 Yan-Chuan Cai$^{3}$, \and
	 Marcus V. Costa-Duarte$^{7}$,
	 Andrej Dvornik$^{1}$,
	 Thomas Erben$^{8}$, \and
	 Hendrik Hildebrandt$^{8}$,
	 Benne W. Holwerda$^{9}$,
	 Peter Schneider$^{8}$, \and
	 Crist\'obal Sif\'on$^{10}$,
	 Edo van Uitert$^{11}$
	\\
	\\
	$^{1}$Leiden Observatory, Leiden University, Niels Bohrweg 2, 2333 CA Leiden, The Netherlands.\\
	$^{2}$Kapteyn Astronomical Institute, University of Groningen,
	PO Box 800, NL-9700 AV Groningen, the Netherlands.\\
	$^{3}$SUPA, Institute for Astronomy, University of Edinburgh, Royal Observatory, Blackford Hill, Edinburgh, EH9 3HJ, UK.\\
	$^{4}$National Centre for Nuclear Research, Astrophysics Division, P.O. Box	447, PL-90-950 Lodz, Poland. \\
	$^{5}$Center for Cosmology and Particle Physics, New York University, 726 Broadway, New York, NY 10003, USA. \\
	$^{6}$School of Physics, University of New South Wales, NSW 2052, Australia.\\
	$^{7}$Institute of Astronomy, Geophysics and Atmospheric Sciences, University of S{\~a}o Paulo, 05508-090 S{\~a}o Paulo, Brazil. \\
	$^{8}$Argelander-Institut f{\"u}r Astronomie, Auf dem H{\"u}gel 71, D-53121 Bonn, Germany.\\
	$^{9}$Department of Physics and Astronomy, University of Louisville, Louisville, KY 40292, USA. \\
	$^{10}$Department of Astrophysical Sciences, Peyton Hall, Princeton University, Princeton, NJ 08544, USA. \\
	$^{11}$Department of Physics and Astronomy, University College London, Gower Street, London WC1E 6BT, UK. \\
}

\date{Accepted 2018 September 16. Received 2018 September 13; in original form 2018 May 1}

\pubyear{2018}

\begin{document}
\label{firstpage}
\pagerange{\pageref{firstpage}--\pageref{lastpage}}
\maketitle

\begin{abstract}
We study projected underdensities in the cosmic galaxy density field known as `troughs', and their overdense counterparts, which we call `ridges'. We identify these regions using a bright sample of foreground galaxies from the photometric Kilo-Degree Survey (KiDS), specifically selected to mimic the spectroscopic Galaxy And Mass Assembly survey (GAMA). Using background galaxies from KiDS, we measure the weak gravitational lensing profiles of the troughs/ridges. We quantify the amplitude of their lensing strength $A$ as a function of galaxy density percentile rank $P$ and galaxy overdensity $\delta$, and find that the skewness in the galaxy density distribution is reflected in the total mass distribution measured by weak lensing. We interpret our results using the mock galaxy catalogue from the Marenostrum Institut de Ci{\`e}ncies de l'Espai (MICE) simulation, and find a good agreement with our observations. Using signal-to-noise weights derived from the Scinet LIghtCone Simulations (SLICS) mock catalogue we optimally stack the lensing signal of KiDS troughs with an angular radius $\theta_{\rm A} = \{5,10,15,20\}\,\am$, resulting in $\{16.8,14.9,10.13,7.55\} \, \sigma$ detections. Finally, we select troughs using a volume-limited sample of galaxies, split into two redshift bins between $0.1<z<0.3$. For troughs/ridges with transverse comoving radius $R_{\rm A}=1.9\hsMpc$, we find no significant difference in the comoving Excess Surface Density as a function of $P$ and $\delta$ between the low- and high-redshift sample. Using the MICE and SLICS mocks we predict that trough and ridge evolution could be detected with gravitational lensing using deeper and wider lensing surveys, such as those from the Large Synoptic Survey Telescope and Euclid.
\end{abstract}


%\newpage
\clearpage

%%%%%%%%%%%%%%%%%%%%%%%%%%%%%%%%%%%%%%%%%%%%%%%%%%

%%%%%%%%%%%%%%%%% BODY OF PAPER %%%%%%%%%%%%%%%%%%

\section{Introduction}
\label{sec:introduction}

Over the past two decades large-scale galaxy redshift surveys, such as the 2dF Galaxy Redshift Survey \cite[2dFGRS,][]{colless2001} and the Sloan Digital Sky Survey \cite[SDSS,][]{abazajian2009}, have provided an ever more accurate picture of the distribution of galaxies in the Universe. They show that galaxies form an intricate `cosmic web' of clusters and filaments, separated by largely empty voids. This distribution is also observed in large-scale hydrodynamical simulations based on the concordance $\lcdm$ cosmology, such as the Illustris \cite[]{vogelsberger2014} and EAGLE \cite[]{schaye2015} projects. These simulations show the gravitational collapse of dark matter (DM) into a web-like structure, establishing the `skeleton' for baryonic matter, which falls into the DM's potential well. Within this framework, the growth factor of voids with redshift can be used to constrain the energy density and equation of state parameter of dark energy (DE) \cite[]{lavaux2010,demchenko2016}, which causes the Universe's accelerated expansion. The low density in voids also makes them clean probes of global cosmological parameters, as their interior is less affected by baryonic physics than denser regions \cite[]{bos2012}. In addition to testing the standard model of cosmology, voids can also be used to detect signatures of modified gravity models, which aim to provide an alternative explanation for the accelerating expansion of the Universe \cite[for reviews, see][]{jain2010,clifton2012}. Because these theories should converge to standard general relativity inside the Solar System, most implement a screening mechanism that suppresses their `$5^{\rm th}$ force' in high-density regions. Simulations based on modified gravity show that low-density regions, like voids, are excellent probes for testing these theories \cite[]{li2012, clampitt2013, cai2015, lam2015,zivick2015, falck2017}.

Studying, detecting, or even defining voids, however, is not a simple matter. There exist numerous void finding algorithms, each one operating with a different void definition \cite[for a comparison study, see e.g.][]{colberg2008}. Moreover, applying the algorithm of choice to detect voids in observational data requires accurate redshift measurements for every individual galaxy. Such accuracy is only available through spectroscopic surveys, which are far more costly than their photometric counterparts. Using the highly complete spectroscopic Galaxy And Mass Assembly survey, \cite{alpaslan2014} discovered that voids found in other surveys still contain a large number of galaxies, which implies that void sizes strongly depend on a survey's galaxy number density and sensitivity limits. Finally, the true DM structure of voids can be different than that of the galaxies that trace them, an effect known as `galaxy bias' \cite[]{benson2000,tinker2010}. Currently, the only way to study the total mass distribution of voids is through gravitational lensing, a statistical method that measures the gravitational deflection (or shear $\gamma$) of the light of background galaxies (sources) by foreground mass distributions (lenses). The first detection of the lensing signal from cosmic voids was presented by \cite{melchior2014}, who stacked the gravitational shear around $901$ voids detected in SDSS. The depth of their void lensing signal corresponded to the prediction from the analytical model by \cite{krause2013}, who concluded that lensing measurements of medium-sized voids with sufficient precision (i.e. with a signal-to-noise ratio $S/N\gtrsim10$) will only be possible with Stage IV surveys such as the Euclid mission \cite[]{laureijs2011} and the Large Synoptic Survey Telescope \cite[LSST,][]{lsst2012}. One of the reasons this signal is so difficult to measure is that lensing measures the average density contrast along the entire line-of-sight (LOS). If a dense cluster is located in the same LOS as the void, it can contaminate the lensing signal. Another challenge of studying voids using stacked gravitational lensing signals is that this method only measures the average shear as a function of the transverse separation from the void centre \cite[]{hamaus2014,nadathur2015}. This means that the detailed void shape information will not be captured, and that stacking voids that are not radially symmetric can even diminish the lensing signal. Moreover, the centre and the radius of these non-spherical voids are difficult to define, and choosing the wrong value reduces the lensing signal even further \cite[for an analysis of these effects, see e.g.][]{cautun2016}.

To circumvent the aforementioned problems, \cite{gruen2016} (hereafter G16) devised a definition for projected voids named `troughs'. These are very simply defined as the most underdense circular regions on the sky, in terms of galaxy number density. Being circular in shape, troughs evade the problem of the centre definition, and are perfectly suited for measuring their stacked shear as a function of transverse separation. Because they are defined as \emph{projected} circular regions of low galaxy density, they have the 3D shapes of long conical frusta\footnote{Frusta, the plural form of frustum: the part of a solid, such as a cone or pyramid, between two (usually parallel) cutting planes.} protruding into the sky. Since this definition only includes regions of low average density over the entire LOS, it automatically excludes LOS's where the total mass of overdensities exceeds that of the underdense regions. Moreover, defining underdensities in projected space alleviates the need for spectroscopic redshifts. Even when projected underdensities are defined in a number of redshift slices, as was done by e.g. \cite{clampitt2015} and \cite{sanchez2017}, photometric redshifts are sufficiently accurate as long as the slices are thicker than the redshift uncertainties.

In summary, troughs have the disadvantage of losing all detailed shape information in projected and in redshift space, but have the advantage that they are simple to define and are specifically designed to provide straightforward and high-$S/N$ weak lensing measurements. This allows for significant lensing measurements of underdensities with currently available surveys. In particular, G16 used the Dark Energy Survey \cite[DES,][]{flaugher2015} Science Verification Data to measure the gravitational lensing signal of projected cosmic underdensities with a significance above $10\sigma$. To achieve this, they counted the number of redMaGiC \cite[]{rozo2016} Luminous Red Galaxies (LRGs) in a large number of circular apertures on the sky. Defining troughs as the $20\%$ lowest density circles, they found a set of $\sim110\,000$ troughs of which they measured the combined shear signal. In their more recent paper, \cite{gruen2017} generalized the concept of troughs to `density split statistics' by splitting the circular apertures into 5 samples of increasing redMaGiC galaxy number density, each sample containing $20\%$ of the circles. They measured the galaxy counts and stacked lensing signals of these 5 samples using both DES First Year \cite[]{drlica2017} and SDSS DR8 data, in order to study the probability distribution function (PDF) of large-scale matter density fluctuations.

The ways in which this new probe can be used for cosmology are still under examination. G16 found the trough shear measurements in their work to be in agreement with a theoretical model based on the assumption that galaxies are biased tracers in a Gaussian mass density distribution. Although the lensing profile of their smallest troughs was marginally sensitive to galaxy bias, the trough-galaxy angular correlation function allowed for much stronger constraints. Using density split statistics in combination with the improved lognormal-based density model from \cite{friedrich2017}, \cite{gruen2017} were able to constrain the total matter density $\Omega_{\rm m}$, the power spectrum amplitude $\sigma_8$, the galaxy bias, galaxy stochasticity and the skewness of the matter density PDF.
	
Another very promising venue for trough lensing is to test models of modified gravity. Using ray-tracing simulations \cite{higuchi2016} found that, while 3D voids could not distinguish between $f(R)$ and $\lcdm$ even in future ($\sim1000 \deg^2$) lensing surveys, the lensing profiles from troughs showed a clear deviation. A recent comparison from \cite{cautun2017} focusing on future surveys (Euclid and LSST) also found that the shear profiles of projected (2D) underdensities will be able to constrain chameleon $f(R)$ gravity with confidence levels of up to $\sim30$ times higher than those of 3D void profiles. \cite{barreira2017} found that another type of modified gravity, the normal branch of the Dvali-Gabadadze-Porrati (nDGP) model, would strengthen the lensing signal of both projected under- and overdensities compared to $\lcdm$. In conclusion, the potential of projected underdensities for cosmology compels the weak lensing community to observationally explore these new probes.

Following up on the work by G16, our goal is to measure and study the gravitational lensing profiles of circular projected underdensities (troughs) and overdensities (which we henceforth call `ridges') using the spectroscopic Galaxy And Mass Assembly survey \cite[GAMA,][]{driver2011} and the photometric Kilo-Degree Survey \cite[KiDS,][]{dejong2017}. By comparing the results from both surveys, we aim to find: 1) whether an analysis of troughs performed using a highly complete spectroscopic survey can be accurately reproduced using only photometric measurements, and 2) which of these surveys is best suited for our trough analysis. Once this is established we study troughs and ridges as a function of their galaxy number density, in order to find the relation between galaxy number density and the total mass density measured by lensing (known as `galaxy bias'). Based on this relation, we aim to find the optimal method of stacking the trough/ridge lensing signals, in order to obtain the highest possible detection significance.

We apply the same trough/ridge selection and lensing methods to two sets of mock observations. The first is the Marenostrum Institut de Ci{\`e}ncies de l'Espai (MICE) Galaxy and Halo Light-cone catalog \cite[]{carretero2015,hoffmann2015} based on the MICE Grand Challenge lightcone simulation \cite[][MICE-GC hereafter]{fosalba2015a,fosalba2015b,crocce2015}. This catalogue is well-suited for comparison to our observations, since the cosmological parameters used to construct the MICE-GC simulations are very similar to those measured in the KiDS-450 cosmic shear analysis \cite[]{hildebrandt2017}. The other set of galaxy lensing mocks is based on the Scinet LIghtCone Simulations (SLICS hereafter), introduced in \cite{harnois2018}. Owing to its large ensemble of independent realisations, this simulation can be used to estimate accurately the covariance matrix and error bars of current and future lensing observations. The goal of this exercise is to find whether these simulations accurately reproduce our trough/ridge lensing measurements, and what possible discrepancies can teach us about cosmology (e.g. information on galaxy bias and cosmological parameter values). In addition, we use the covariance estimates from SLICS to test the accuracy of our analytical covariance method \cite[as described in][]{viola2015} used to find the errors and covariance of our lensing measurements.

G16 also studied the lensing signals of troughs/ridges as a function of redshift, by splitting the LRG sample that defined them into two redshift samples. However, they did not account for possible differences between the galaxy samples or trough/ridge geometry at different redshifts, nor did they correct for the variation in distance between the lenses and the background sources that measured the shear signal. As a result, they did not find any signs of \emph{physical} redshift evolution of troughs/ridges. By correcting the selection method and lensing signal measurement for all known differences between the two redshift samples, we explore the physical evolution of troughs and ridges. Our final goal is to discover whether troughs and ridges can be used as a tool to probe large-scale structure evolution over cosmic time.

Our paper is structured as follows: In Sect. \ref{sec:data} we introduce the KiDS and GAMA data which we use to define the troughs/ridges and measure their lensing profiles, and the MICE-GC and SLICS mock data used to interpret our observations. Section \ref{sec:analysis} describes the classification of troughs/ridges and explains the gravitational lensing method in detail. In Sect. \ref{sec:results} we show the resulting trough lensing profiles as a function of galaxy density and size, and define our optimal trough stacking method. Our study of troughs/ridges as a function of redshift is described in Sect. \ref{sec:redshift}. We end with the discussion and conclusion in Sect. \ref{sec:discon}.

Throughout this work we adopt the cosmological parameters used in creating the MICE-GC simulations ($\Omega_{\rm m}=0.25$, $\sigma_8=0.8$, $\Omega_{\rm \Lambda}=0.75$, and $H_0 = 70 \, {\rm km \, s^{-1} Mpc^{-1}}$) when handling the MICE mock catalogue and the KiDS and GAMA data. Only when handling the SLICS mock catalogue, which is based on a different cosmology, we use: $\Omega_{\rm m} = 0.2905$, $\sigma_8=0.826$, $\Omega_{\rm \Lambda} = 0.7095$, and \mbox{$H_0 = 68.98 \, {\rm km \, s^{-1} Mpc^{-1}}$}. Throughout the paper we use the reduced Hubble constant $h_{70} \equiv \ H_0 / (70 \, {\rm km \, s^{-1} Mpc^{-1}})$.

% With its very uniform and compact Point Spread Function (PSF), the KiDS galaxies are accurate tracers of the gravitational lensing effect induced by foreground mass distributions. 

\section{Data}
\label{sec:data}

We use two samples of foreground galaxies to define the locations of troughs and ridges: one from the spectroscopic GAMA survey and one from the photometric KiDS survey. Comparing the results obtained from these two samples allows us to test the strength and reliability of trough studies using only photometric data. Table \ref{tab:samples} in Sect. \ref{sec:redshift_selection} shows a summary of the galaxy selections used to define the troughs/ridges. Their gravitational lensing signal is measured using a sample of KiDS background galaxies. The combination of the KiDS and GAMA datasets and the lensing measurement method, which is used for the observations described in this work, closely resembles earlier KiDS-GAMA galaxy-galaxy lensing papers. For more information we recommend reading Sect. 3 of \cite{viola2015}, which discusses the galaxy-galaxy lensing technique in detail, and \cite{dvornik2017} which makes use of exactly the same KiDS and GAMA data releases as this work. In order to compare our observational results to predictions from simulations, the same process of selecting troughs and measuring their lensing profiles is performed using the MICE-GC and SLICS mock galaxy catalogues. In this section we introduce the KiDS, GAMA, MICE and SLICS galaxy catalogues, including their role in the trough selection and lensing measurement.

\subsection{KiDS source galaxies}
\label{sec:kids}

In order to derive the mass distribution of troughs, we measure their gravitational lensing effect on the images of background galaxies. Observations of these source galaxies are taken from KiDS, a photometric lensing survey in the $u$, $g$, $r$ and $i$ bands, performed using the OmegaCAM instrument \cite[]{kuijken2011} mounted on the VLT Survey Telescope \cite[]{capaccioli2011}. For this work we use the photometric redshift, magnitude, and ellipticity measurements from the third data release \cite[KiDS-DR3,][]{dejong2017}, which were also used for the KiDS-450 cosmic shear analysis \cite[]{hildebrandt2017}. These measurements span $449.7 \deg^2$ on the sky, and completely cover the $180\deg^2$ equatorial GAMA area (see Sect. \ref{sec:gama} below).

The galaxy ellipticity measurements are based on the $r$-band observations, which have superior atmospheric seeing constraints (a maximum of $0.8$ arcsec) compared to the other bands \cite[][]{dejong2017}. The galaxies are located with the {\scshape SExtractor} detection algorithm \cite[]{bertin1996} from the co-added $r$-band images produced by the {\scshape Theli} pipeline \cite[]{erben2013}. The ellipticity of each galaxy is measured using the self-calibrating \emph{lens}fit pipeline \cite[]{miller2007,miller2013,fenechconti2017}.

Galaxies in areas surrounding bright stars or image defects (such as read-out spikes, diffraction spikes, cosmic rays, satellite tracks, reflection haloes and ghosts) are removed. After removing masked and overlapping areas from all survey tiles, the effective survey area is $360.3 \deg^2$ ($\sim80\%$ of the original area) \cite[]{hildebrandt2017}. This means that, even though the total area of KiDS-450 is $2.5$ times larger than that of the GAMA survey, the effective KiDS/GAMA area ratio is $360.3/180\approx2$.

\begin{figure*}
	\includegraphics[width=1.0\textwidth]{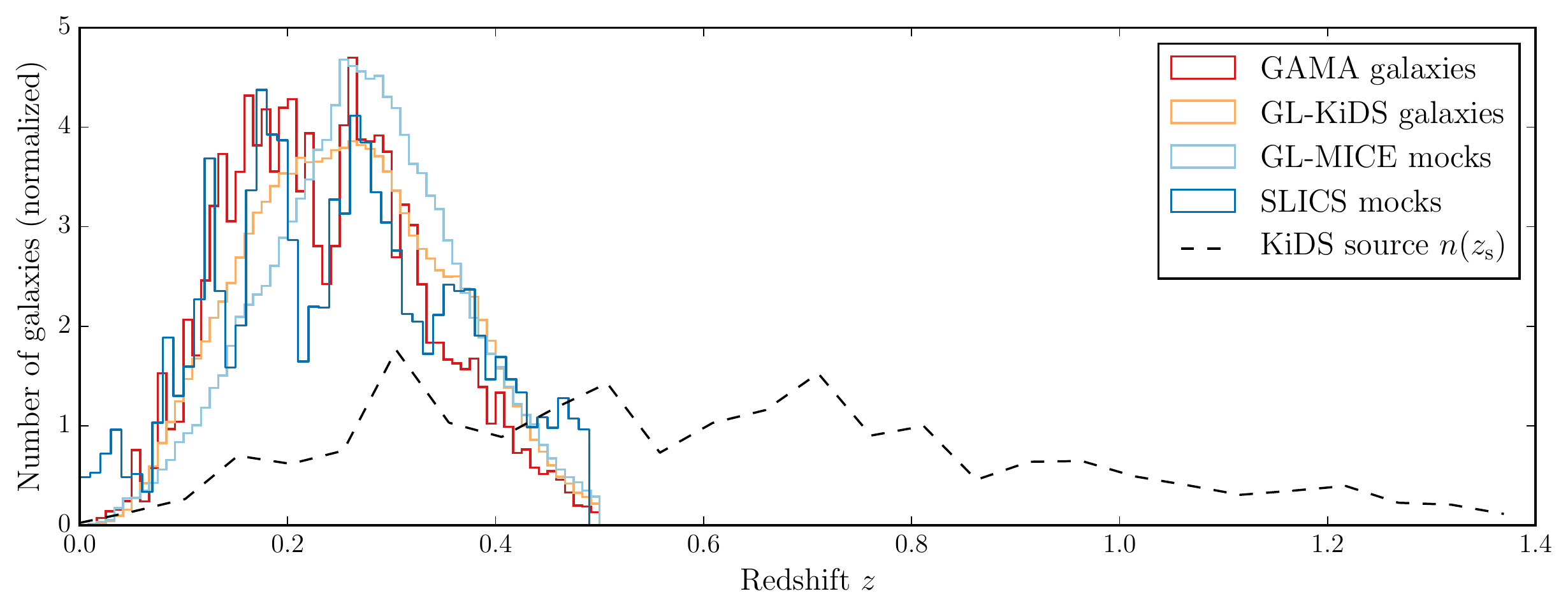}
	\caption{The normalized redshift histograms of the GAMA (red), GL-KiDS (orange), GL-MICE (light blue) and SLICS (dark blue) galaxy samples used to define the troughs/ridges, and the redshift distribution $n(z_{\rm s})$ of the KiDS sources (dashed line) used to measure the trough lensing signals. The histograms show that all foreground samples have similar redshift distributions. Although the average redshifts of the GL-KiDS and GL-MICE samples are slighly higher than those of the GAMA and SLICS samples, this does not significantly affect the lensing signals. The best-fit redshifts of the sources are limited to $0.1<z_{\rm B}<0.9$, but the full $n(z_{\rm s})$ stretches between $0.0 < z < 3.5$.}
	\label{fig:redshift_hist}
\end{figure*}

The photometric redshifts of the sources are estimated from co-added $ugri$ images, which were reduced using the Astro-WISE pipeline \cite[]{mcfarland2013}. From the galaxy colours measured by the Gaussian Aperture and PSF pipeline \cite[GAaP,][]{kuijken2008,kuijken2015}, the total redshift probability distribution $n(z_{\rm s})$ of the full source population is calculated using the direct calibration (DIR) method described in \cite{hildebrandt2017}. We use this full $n(z_{\rm s})$ for our lensing measurements (as described in Sect. \ref{sec:esd_measurements}), in order to circumvent the bias inherent in individual photometric source redshift estimates. In this analysis we do not include any systematic uncertainty on the calibration correction to the shear measurements or the redshift distributions, as these are both expected to be small. Following \cite{hildebrandt2017} we use the best-fit photometric redshift $z_{\rm B}$ \cite[]{benitez2000,hildebrandt2012} of each galaxy to limit the redshift range to $0.1<z_{\rm B}<0.9$. The final $n(z_{\rm s})$ of the KiDS sources is shown in Fig. \ref{fig:redshift_hist}. This distribution shows that the $n(z_{\rm s})$ extends beyond these $z_{\rm B}$ limits, due to the uncertainty on the individual photometric source redshifts (where the full distribution lies between $0 < z_{\rm s} < 3.5$, as shown in Fig. A5 of \citealp{dvornik2017}).

\subsection{GAMA foreground galaxies}
\label{sec:gama}

One of the galaxy samples we use to define the troughs is obtained using the spectroscopic GAMA survey \cite[]{driver2011}, which was performed with the AAOmega spectrograph mounted on the Anglo-Australian Telescope. The galaxy locations were selected from the Sloan Digital Sky Survey \cite[SDSS,][]{abazajian2009}. For this study we use the three equatorial regions (G09, G12 and G15) from the GAMA II data release \cite[]{liske2015}, which span a total area of $180\deg^2$ on the sky, since these areas completely overlap with the KiDS survey. GAMA has a redshift completeness of $98.5\%$ down to Petrosian $r$-band magnitude $m_{\rm r}=19.8 \magn$, resulting in a catalogue containing $180\,960$ galaxies with redshift quality $n_{\rm Q}\geq2$. As recommended, we only use the galaxies with redshift quality $n_{\rm Q}\geq3$, which amounts to $99.74\%$ of the full catalogue. In order to indicate regions where the survey is less complete, GAMA provides a `mask' which contains the redshift completeness of galaxies on a $0.001 \deg$ Cartesian grid. We use this mask to account for incomplete regions during the trough classification.

To mimic the galaxy sample corresponding to resolved haloes in the mock catalogues (see Sect. \ref{sec:mice_mocks} and \ref{sec:slics_mocks}), we only use galaxies with absolute $r$-band magnitude $M_{\rm r}<-19.67 \magn$. The GAMA rest-frame $M_{\rm r}$ is determined by fitting \cite{bruzual2003} stellar population synthesis models to the $ugrizZYJHK$ spectral energy distribution of SDSS and VIKING observations \cite[]{abazajian2009,edge2013}, and corrected for flux falling outside the automatically selected aperture \cite[]{taylor2011}. Together, the $n_{\rm Q}$ and $M_{\rm r}$ cuts result in a sample of $159\,519$ galaxies ($88.15 \%$ of the full catalogue), with a redshift range between $0<z_{\rm G}<0.5$ and a mean redshift of $\mean{z_{\rm G}} = 0.24$. The total redshift distribution of the GAMA sample is shown in Fig. \ref{fig:redshift_hist}. The average number density of this sample (including masks) is $\mean{n_{\rm g}} = 0.25 \am^{-2}$. The projected number density of this sample of GAMA galaxies, together with their completeness mask, is used to define the troughs as detailed in Sect. \ref{sec:trough_selection}.

\subsection{KiDS foreground selection}
\label{sec:gamalike_kids}

Since the currently available area of the KiDS survey is $2.5$ times larger than that of the GAMA survey (and will become even larger in the near future) it can be rewarding to perform both the trough selection and lensing measurement using the KiDS galaxies alone, employing the full $454\deg^2$ area of the current KiDS-450 dataset. To be able to compare the KiDS troughs to those obtained using GAMA, we select a sample of `GAMA-like' (GL) KiDS galaxies that resembles the GAMA sample as closely as possible. Because GAMA is a magnitude-limited survey ($m_{\rm r, Petro}<19.8 \magn$), we need to apply the same magnitude cut to the (much deeper) KiDS survey. Since there are no Petrosian $r$-band magnitudes available for the KiDS galaxies, we use the KiDS magnitudes that have the most similar $m_{\rm r}$-distribution: the extinction-corrected and zero-point homogenised isophotal $r$-band magnitudes $m_{\rm r,iso}$ \cite[]{dejong2017}. These magnitude values, however, are systematically higher than the Petrosian magnitudes from GAMA. We therefore match the KiDS and GAMA galaxies using their sky coordinates, and select the magnitude cut based on the completeness of this match. Using $m_{\rm r, iso}<20.2 \magn$, the completeness of the match is $99.2\%$. Although this is slightly higher than that of the real GAMA sample, this small difference does not significantly affect our results which are primarily based on the relative number density (compared to other areas or the mean density).

In addition, we wish to cut the KiDS galaxies at the maximum redshift of GAMA: $z_{\rm G} < 0.5$. Contrary to the KiDS source redshifts used for the lensing measurement, where we can use the redshift probability distribution of the full population (see Sect. \ref{sec:lensing}), the application of this cut and the use of KiDS galaxies as lenses both require individual galaxy redshifts. These photometric redshifts $z_{\rm ANN}$ are determined using the machine learning method ANNz2 \cite[]{sadeh2016} as described in Sect. 4.3 of \cite{dejong2017}. Following \cite{bilicki2017} the photo-z's are trained exclusively on spectroscopic redshifts from the equatorial GAMA fields.\footnote{\cite{bilicki2017} use a slightly different apparent magnitude cut to select the GL-KiDS galaxy sample: $m_{\rm r, auto}<20.3 \magn$. However, since this is an a-posteriori cut it does not influence the determination of the photo-z values.} This is the first work that uses the KiDS photometric redshifts measured through machine learning to estimate the distances of the lenses. Compared to the spectroscopic GAMA redshifts $z_{\rm G}$, the mean error on the ANNz2 photometric redshifts is $(z_{\rm ANN} - z_{\rm G})/(1 + z_{\rm G}) = −3.3\E{-4}$, with a standard deviation of $0.036$ (much smaller than the width of the redshift selections used in this work; see Sect. \ref{sec:redshift_selection}). Finally, to mimic the galaxy sample corresponding to resolved haloes in the mock catalogues (see Sect. \ref{sec:mice_mocks} and \ref{sec:slics_mocks}), we apply the absolute $r$-band magnitude cut $M_{\rm r}<-19.67 \magn$. These absolute magnitudes: $M_{\rm r} = m_{\rm r, iso} - D_{\rm M} + K_{\rm cor}$, are determined using distance moduli $D_{\rm M}$ based on the $z_{\rm ANN}$ redshifts. The K-corrections $K_{\rm cor}$ are calculated from the isophotal $g$- and $i$-band magnitudes of the KiDS galaxies, using the empirical relation in Table 4 of \cite{beare2014}.

To remove stars from our galaxy sample, we use a star/galaxy separation method based on the object's morphology \mbox{\cite[described in Sect. 4.5 of][]{dejong2015}}. We also mask galaxies that have been affected by readout and diffraction spikes, by saturation cores and primary haloes of bright stars, or by bad pixels in any band ($u$, $g$, $r$ or $i$). We do not remove galaxies affected by secondary and tertiary stellar haloes because these do not heavily affect bright galaxies.\footnote{Our masking choice corresponds to MASK values 1, 2, 4, 8 and 64 as described in Sect. 4.4 (Table 4) of \cite{dejong2015}.} In addition, we remove galaxies that have an unreliable magnitude measurement in any band, as recommended in App. 3.2 of \cite{dejong2017}. Using this selection we obtain a sample of $309\,021$ KiDS galaxies that resemble the GAMA and MICE-GC galaxy populations. This is $\sim2$ times the number of selected GAMA galaxies, which is a consequence of the completeness of GAMA compared to KiDS (where the latter has a relatively large area that is covered by the aforementioned masks; see also Sect. \ref{sec:kids}). The average galaxy number density of the final GL-KiDS sample (including masks) is $\mean{n_{\rm g}} = 0.33 \am^{-2}$, and the average redshift $\mean{z_{\rm ANN}}=0.26$. This is $7.9\%$ higher than the average redshift of the GAMA sample, due to the slightly higher magnitude cut. However, by calculating the values of the lensing efficiency ($\Sigma_{\rm crit}^{-1}$) using the average redshifts of both lens and source samples, we estimate that the effect of this difference on the lensing signal is not significant ($\sim1\%$). The total redshift distribution $n(z_{\rm ANN})$ is shown in Fig. \ref{fig:redshift_hist}.

Based on the aforementioned image defects, the KiDS survey provides an automatic mask that flags affected pixels. We use these pixel maps to account for the masked areas in the trough selection (see Sect. \ref{sec:trough_selection}). For simplicity we only use the $r$-band pixel mask, which has a less than $1\%$ difference with the pixel mask based on all bands. We use this map to account for incomplete regions during the trough classification procedure (see Sect. \ref{sec:trough_selection}). In order to save computational time, we create a map that provides the survey completeness on a $0.04 \deg$ Cartesian grid, by calculating the ratio of `good pixels' in the square area surrounding each grid point. The grid spacing of the resulting mask ($2.4 \am$) is the same as that used for the trough selection, and is chosen such that it is at least two times smaller than the aperture radius of the smallest troughs ($\theta_{\rm A} = 5 \am$).

\subsection{MICE mock galaxies}
\label{sec:mice_mocks}

We wish to apply the same trough detection and analysis to simulated data, in order to compare and interpret our observational results. The MICE-GC $N$-body simulation presented by \cite{fosalba2015b} contains $\sim 7$$\E{10}$ DM particles in a $(3072 \hsMpc)^3$ comoving volume, allowing the construction of an all-sky lightcone with a maximum redshift of $z=1.4$. From this lightcone \cite{crocce2015} built a halo and galaxy catalogue using a Halo Occupation Distribution (HOD) and Halo Abundance Matching (HAM) technique, resulting in an average galaxy bias of $b_{\rm MICE}\sim0.9$ at scales above $2\hMpc$ (see Fig. 4 of \citealp{carretero2015}, bottom left panel). Its large volume and fine spatial resolution make MICE-GC mocks ideally suited for accurate modelling of both large-scale (linear) and small-scale (non-linear) clustering and structure growth. The mock galaxy clustering as a function of luminosity has been constructed to reproduce observations from SDSS \cite[]{zehavi2011} at lower redshifts ($z<0.25$), and has been validated against the COSMOS catalogue \cite[]{ilbert2009} at higher redshifts ($0.45 < z < 1.1$). The MICE-GC simulation resolves DM halos down to a mass of $6\E{11} \hsmsun$ (corresponding to 20 particles), which host galaxies with an absolute magnitude $<-18.9$. Since this absolute magnitude includes a cosmology correction such that: $M_{\rm r,MICE} = M_{\rm r} - 5\log_{10}(h)$, where $h=0.7$ is their reduced Hubble constant, we apply an $M_{\rm r}<-18.9-0.77=-19.67 \magn$ cut to the GAMA and GL-KiDS samples in order to resemble the mock galaxy population.

From the MICE-GC catalogue\footnote{The MICE-GC catalogue is publicly available through CosmoHub (\url{http://cosmohub.pic.es}).} we obtain the sky coordinates, redshifts, comoving distances, absolute magnitudes and SDSS apparent magnitudes of the mock galaxies. In order to create a GL-MICE sample, we limit the mock galaxy redshifts to $z < 0.5$. When considering the choice of magnitude cut, we find that the distribution of the SDSS magnitudes in the MICE catalog is very similar to that of the isophotal KiDS magnitudes. We therefore limit the MICE galaxies to $m_{\rm r} < 20.2 \magn$, and find that indeed the galaxy number density of the GL-MICE sample, $\mean{n_{\rm g}} = 0.3 \am^{-2}$, is almost equal to that of the GL-KiDS sample (which is also visible in Fig. \ref{fig:trough_hist} of Sect. \ref{sec:trough_selection}). In addition (as can be seen in Fig. \ref{fig:redshift_hist}) the redshift distribution of the GL-MICE sample resembles that of the GL-KiDS galaxies, with an average redshift $\mean{z_{\rm MICE}}=0.27$. As with GL-KiDS this average redshift is slightly higher than that of the GAMA sample. Again calculating the lensing efficiency ($\Sigma_{\rm crit}^{-1}$) for the average redshifts of both samples, we estimate that the effect on the lensing signal is less than $3\%$. Like the GAMA and GL-KiDS galaxies, this sample of MICE foreground galaxies is used to define troughs following the classification method described in Sect. \ref{sec:trough_selection}.

Each galaxy in the lightcone also carries the lensing shear values $\gamma_1$ and $\gamma_2$ (with respect to the Cartesian coordinate system) which were calculated from the all-sky weak lensing maps constructed by \cite{fosalba2015a}, following the `onion shell' method presented in \cite{fosalba2008}. In this approach the DM lightcone is decomposed and projected into concentric spherical shells around the observer, each with a redshift thickness of ${\rm d}z \approx 0.003 (1+z)$. These 2D DM density maps are multiplied by the appropriate lensing weights and combined in order to derive the corresponding convergence and shear maps. The results agree with the more computationally expensive `ray-tracing' technique within the Born approximation. We use these shear values (in the same way we used the ellipticities observed by KiDS) to obtain mock lensing profiles around troughs, following the weak lensing method described in Sect. \ref{sec:lensing}. To this end we create a MICE background source sample with $0.1 < z < 0.9$ and $m_{\rm r} > 20 \magn$. This apparent magnitude cut is equal the one applied to the KiDS background sources by \cite{hildebrandt2017}, and the redshift cut is analogous to their limit on the best-fit photometric redshift $z_{\rm B}$ (although uncertainties in these KiDS redshifts are not accounted for in this selection). Also, in order to resemble the KiDS source redshift distribution more closely, we choose to apply an absolute magnitude cut of $M_{\rm r} > -19.3 \magn$ on the mock galaxies. Note that any cut on the mock galaxy sample does not affect the shear values (which do not depend on any mock galaxy property) but only the redshift distribution of the sources, which is used in Sect. \ref{sec:esd_measurements} to calculate the excess surface density profiles.

Because all quantities in the mock catalogue are exactly known, we do not need to take into account measurement errors in the calculation of the mock lensing signals. However, simulations are affected by sample variance: the fact that there exist differences between astrophysical measurements from different parts of the sky. To accurately measure the variance of mock shear profiles, one needs a large ensemble of mock realisations (such as those of the SLICS, see Sect. \ref{sec:slics_mocks}) in order to compute a covariance matrix. The MICE simulations, however, consist of one large realisation with an area of $90^\circ \times 90^\circ$. In order to obtain a rough estimate of the mentioned uncertainties, we split the MICE-GC public lightcone area into 16 patches of $20^\circ \times 20^\circ = 400 \deg^2$ (approximately the same size as the used KiDS area). Comparing the results obtained from the full lightcone area with those of the 16 sub-samples provides an estimate of the sample variance within the MICE mocks.

\subsection{SLICS mock galaxies}
\label{sec:slics_mocks}

We conduct our measurement on a second set of simulated data, which were created by \cite{harnois2018} based on the Scinet LIghtCone Simulations \cite[]{harnois2015}. The SLICS consist of a large ensemble of $N$-body runs, each starting from a different random noise realisation. These realisations can be used to make quantitative estimates of the covariance matrices and error bars of the trough lensing signals (as described in Sect. \ref{sec:covariance}), which can be compared to those from our observations and used to predict the success of future measurements. All realisations have a fixed cosmology: $\Omega_{\rm m} = 0.2905$, $\Omega_\Lambda = 0.7095$, $\sigma_8=0.826$, $n_s = 0.969$, \mbox{$H_0 = 68.98 \, {\rm km \, s^{-1} Mpc^{-1}}$} and $\Omega_{\rm b} = 0.0473$. The SLICS followed the non-linear evolution of $1536^3$ particles of mass $2.88\E{9} \, \msun$ in a box size of $(505 \, {\rm Mpc})^3$, writing mass sheets and halos on-the-fly at 18 different redshifts up to $z = 3.0$.
The matter power spectrum has been shown to agree within $5\%$ with the Cosmic Emulator \cite[]{heitmann2014} up to $k = 2.0 \, {\rm Mpc^{-1}}$, while haloes with a mass greater than $2.88 \times 10^{11} \msun$ are resolved with at least 100 particles. Haloes of this mass host galaxies with a mean absolute magnitude $M_{\rm r}\sim-20$, close to the absolute magnitude limit of MICE ($M_{\rm r}<-19.67$) which we use throughout this work.

The SLICS are then ray-traced onto 100 ${\rm deg}^2$ lightcones in the multiple thin lens approximation to extract shear maps and halo catalogues. The lightcones are first populated with source galaxies placed at random angular coordinates and reproducing the KiDS-450 number density and $n(z)$ \citep[measured using the DIR method in][]{hildebrandt2017}. For each galaxy, the $\gamma_{1}$ and $\gamma_{2}$ shear components are interpolated from the enclosing shear planes at the galaxy position.
The halo catalogues are then populated with galaxies following a HOD prescription from \citet{smith2017}, in which the parameters are slightly modified to enhance the agreement in clustering with the GAMA data. A cut in apparent $r$-band magnitude ($m_r < 19.8$) and in redshift ($z<0.5$) is applied to the catalogues, after which the apparent and absolute magnitudes, the number density ($\mean{n_{\rm g}} = 0.244 \am^{-2}$) and the redshift distribution (as seen in Fig. \ref{fig:redshift_hist}) of the GL-SLICS mocks closely match the GAMA data.
The match in projected clustering $w(\theta)$ is better than $20\%$ over the angular scales $0.1 < \theta < 40 \am$, with the mocks being overall more clustered. The value of the galaxy bias ($b_{\rm SLICS} = 1.2$) is slightly higher than that of MICE ($b_{\rm MICE} \sim 0.9$). However, we have checked that the effect of this difference in galaxy bias on the amplitudes of the trough/ridge shear profiles is at most $5\%$, such that it does not affect our conclusions.

\section{Data analysis}
\label{sec:analysis}

The two most important aspects of the data analysis are the classification of the troughs, and the subsequent measurement of their gravitational lensing profiles. For the galaxies used in the trough classification we compare the spectroscopic GAMA sample to the GL-KiDS sample, which has photometric redshifts (see Sect. \ref{sec:gamalike_kids}). For the measurement of the gravitational lensing effect around these troughs, we use the shapes of the KiDS background galaxies. In this section we discuss the trough classification and lensing measurement methods in detail.

\subsection{Trough \& ridge classification}
\label{sec:trough_selection}

Our approach to trough detection is mainly inspired by the method devised by G16. This effectively comprises measuring the projected number density of galaxies within circular apertures on the sky, and ranking the apertures by galaxy density. We first define a finely spaced Cartesian grid of positions on the sky. Around each sky position $\vec{x}$, we count the number of galaxies within a circular aperture of chosen radius $\theta_{\rm A}$. We perform this method for apertures with different angular radii: $\theta_{\rm A} = \{5, 10, 15, 20\} \, \am$, which allows us to study cosmic structure at different scales. To make sure that no information is lost through under-sampling we choose a grid spacing of $0.04 \deg$ ($=2.4 \am$), which is smaller than $\theta_{\rm A}/2$ even for the smallest aperture size.

The projected galaxy number density $n_{\rm g}(\vec{x}, \theta_{\rm A})$ of each aperture is defined as the galaxy count within angular separation $\theta_{\rm A}$ of the sky position $\vec{x}$, divided by the effective area of the corresponding circle on the sky, determined using the appropriate (KiDS or GAMA) mask. Each mask provides the survey area completeness on a finely spaced grid, which we average to a $0.04 \deg$ Cartesian grid to save computational time. Following G16 we exclude those circles that are less than $80\%$ complete from our sample. We also tested a trough selection procedure that excludes circles with less than $60\%$, $70\%$ and $90\%$ completeness, and found that the specific choice of completeness threshold does not significantly affect the trough shear profiles.

\begin{figure}
	\includegraphics[width=1.0\columnwidth]{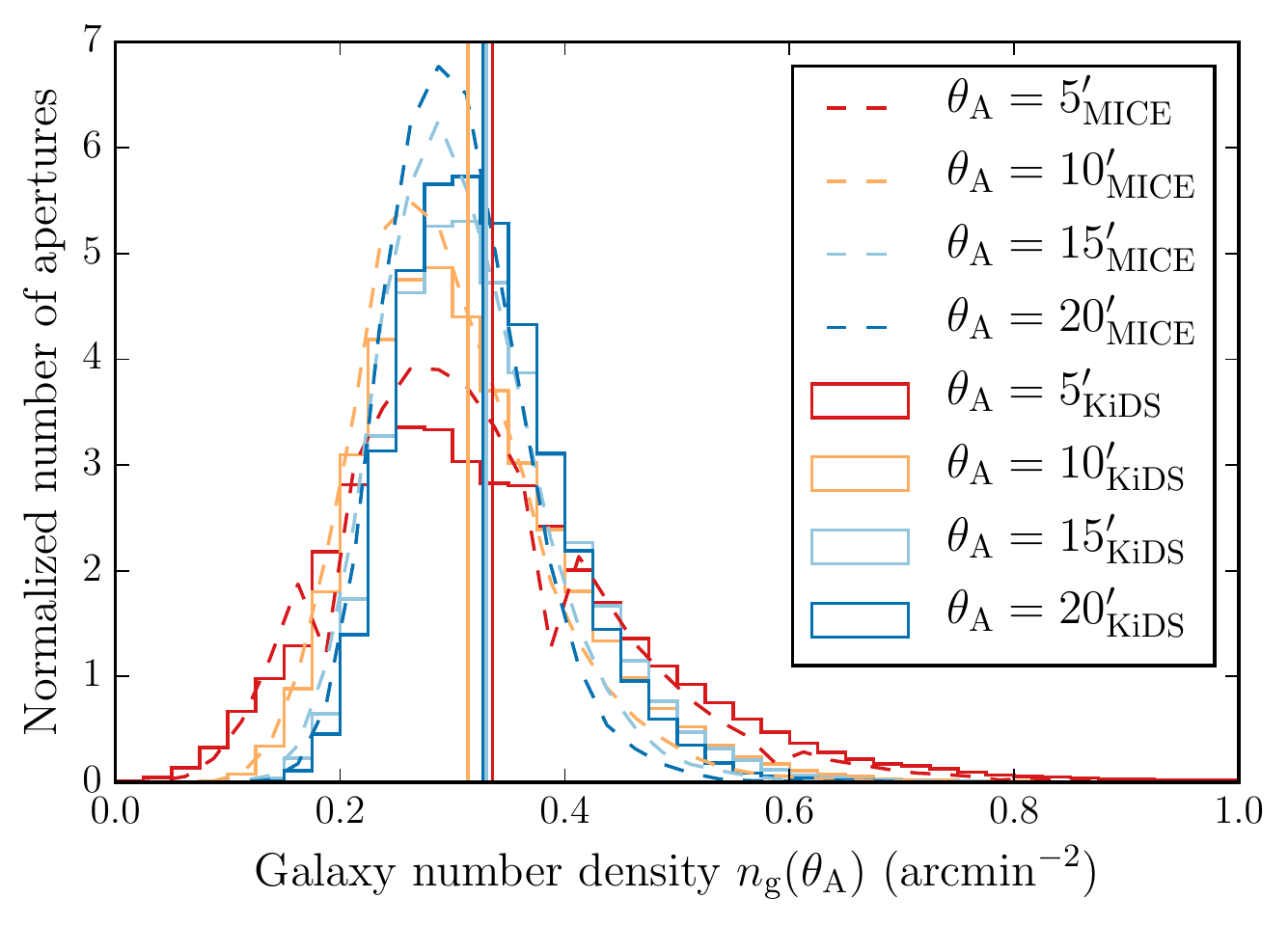}
	\caption{This histogram shows the distribution of the normalized number density $n_{\rm g}$ of the GL-KiDS (solid steps) and MICE (dashed lines) galaxies used to define the troughs, inside all used apertures (those with an effective area $>80\%$). The colors designate apertures of different radius $\theta_{\rm A}$, and the solid vertical lines indicate the mean of each distribution. As expected, the density distribution of circles with a smaller area is more asymmetric, and has a larger dispersion from the mean density $\mean{n_{\rm g}}(\theta_{\rm A})$. The `troughs' are defined as all underdense apertures (i.e. $n_{\rm g}<\mean{n_{\rm g}}(\theta_{\rm A})$), while all overdense apertures (i.e. $n_{\rm g}>\mean{n_{\rm g}}(\theta_{\rm A})$) are called `ridges'.}
	\label{fig:trough_hist}
\end{figure}

\begin{figure*}
	\includegraphics[width=0.9\textwidth]{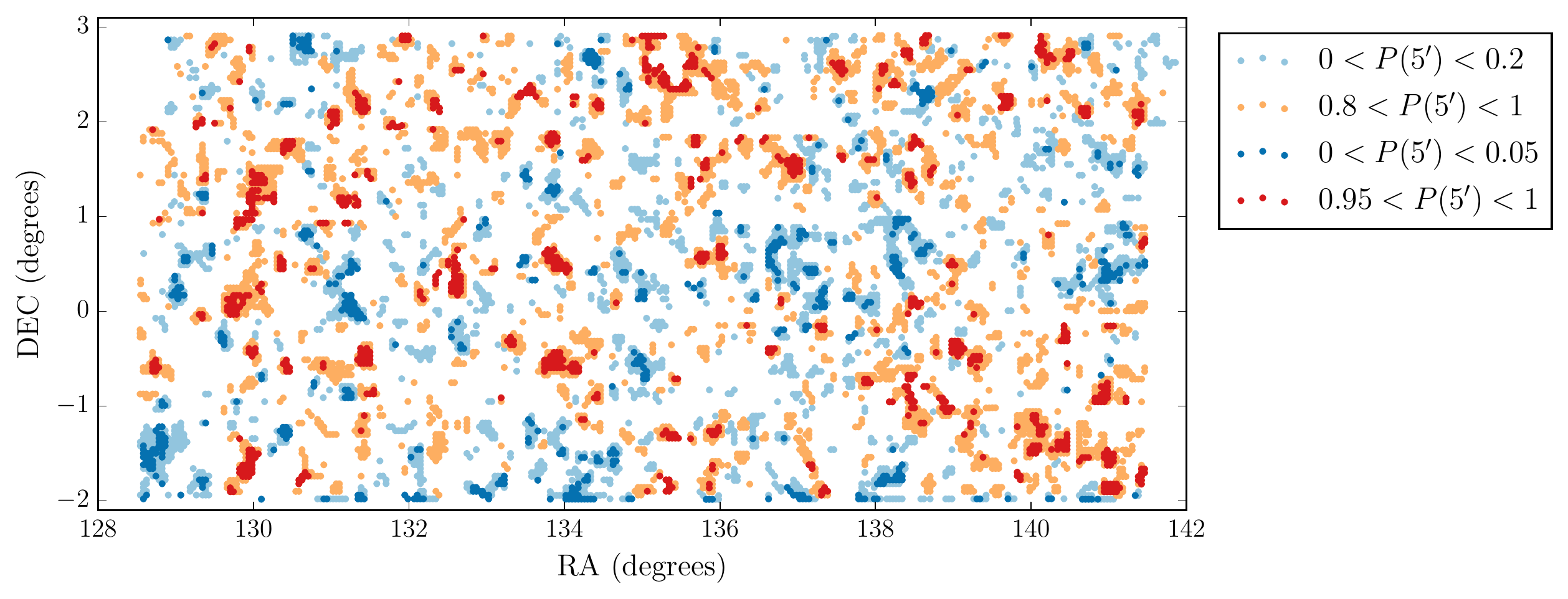}
	\caption{This sky map of the G09 equatorial field shows the spatial distribution of different trough and ridge samples with aperture radius $\theta_{\rm A} = 5 \am$, defined using the GL-KiDS galaxies. The coloured dots represent the centers of troughs ($P<0.2$, light blue) and ridges ($P>0.8$, orange) selected using the fiducial G16 definition, as well as a set of lower-density troughs ($P<0.05$, dark blue) and higher-density ridges ($P>0.95$, red). These `deeper' troughs (and `higher' ridges) tend to reside at the centers of `shallower' ones, resulting in a more clustered distribution.}
	\label{fig:trough_map}
\end{figure*}

The histogram in Fig. \ref{fig:trough_hist} shows the normalized GL-KiDS and MICE galaxy number density distributions (represented by solid steps and dashed lines respectively) for apertures with different radii $\theta_{\rm A}$. The density roughly follows a log-normal distribution, as was originally modeled by \cite{coles1991}. The skewness of the distribution is larger for circles with a smaller area, which is expected since larger apertures measure the average density over a larger area, diluting the influence of individual \mbox{(under-)}density peaks. The smaller apertures are therefore more sensitive to small-scale non-Gaussianities, while the density distribution of the larger apertures tends more towards a Gaussian shape. This is visible in both the observational KiDS and MICE mock data (we verify that this skewness is also observed in the density distribution of troughs selected using GAMA galaxies).

Following G16 we determine, for each of these circles, the percentile rank $P(\vec{x}, \theta_{\rm A})$: the fraction of equally sized apertures that have a lower galaxy density than the circle considered. Ranking the apertures by galaxy density in this way means that low-density circles have a low value of $P$ (down to $P=0$), while high-density circles have a high $P$-value (up to $P=1$). A circle containing the median density has the value $P=0.5$. In the fiducial definition of G16, all apertures in the lowest quintile ($20\%$) of galaxy density (i.e. $P(\vec{x}, \theta_{\rm A})<0.2$) are called troughs, while apertures in the highest quintile (i.e. $P(\vec{x}, \theta_{\rm A})>0.8$) are considered overdensities (which we call `ridges'). A map of the G09 KiDS field showing the spatial distribution of troughs/ridges as defined by G16 (which we henceforth call the `fiducial' troughs/ridges) is shown in Fig. \ref{fig:trough_map}. In addition, we show the distribution of a set of `deeper' (i.e. lower-density) troughs ($P(\vec{x}, \theta_{\rm A})<0.05$) and `higher' (i.e. higher-density) ridges ($P(\vec{x}, \theta_{\rm A})>0.95$). Each coloured dot represents the centre of a $\theta_{\rm A}=5\am$ aperture. The map clearly shows that deeper troughs (and higher ridges) tend to reside at the centers of `shallower' ones, and are hence more strongly clustered. This clustering is accounted for in our error propagation through the calculation of the analytical covariance matrix (see Sect. \ref{sec:covariance}).

By arbitrarily narrowing/expanding the density percentile limit one can define deeper/shallower trough samples (which include fewer/more apertures). However, whether a region is underdense or overdense is not directly determined by its $P$-value, but by its galaxy number density $n_{\rm g}$ with respect to the mean galaxy number density $\mean{n_{\rm g}}$ of the survey. We will therefore define the terms `trough' and `ridge' based on the apertures' galaxy overdensity:
\begin{equation}
\delta(\vec{x},\theta_{\rm A}) = \frac{n_{\rm g}(\vec{x},\theta_{\rm A}) - \mean{n_{\rm g}} }{ \mean{n_{\rm g}} } \, .
\label{eq:delta}
\end{equation}
In our classification, all underdense apertures (i.e. $\delta(\vec{x}, \theta_{\rm A}) < 0$) are called troughs, while all overdense apertures (i.e. $\delta(\vec{x}, \theta_{\rm A}) > 0$) are called ridges. This definition does not \emph{a priori} exclude any apertures from our combined sample of troughs and ridges, allowing us to take advantage of all available data. We will further specify sub-samples of troughs and ridges, selected as a function of both $P$ and $\delta$, where necessary throughout the work.

\subsection{Lensing measurement}
\label{sec:lensing}

In order to measure the projected mass density of the selected troughs and ridges, we use weak gravitational lensing \cite[see][for a general introduction]{bartelmann2001,schneider2006}. This method measures systematic tidal distortions of the light from many background galaxies (sources) by foreground mass distributions (lenses). This gravitational deflection causes a distortion in the observed shapes of the source images of $\sim 1\%$, which can only be measured statistically. This is done by averaging, from many background sources, the projected ellipticity component $\epsilon_{\rm t}$ tangential to the direction towards the centre of the lens, which is an estimator of the `tangential shear' $\gamma_{\rm t}$. This quantity is averaged within circular annuli around the center of the lens, to create a shear profile $\gamma_{\rm t}(\theta)$ as a function of the separation angle $\theta$ to the lens centre. For each annulus, $\gamma_{\rm t}(\theta)$ is a measure of the density contrast of the foreground mass distribution. In order to obtain a reasonable signal-to-noise ratio ($S/N$), the shear measurement around many lenses is `stacked' to create the average shear profile of a specified lens sample. In this work, the centres of the lenses are the grid points that define our circular troughs and ridges (as defined in Sect. \ref{sec:trough_selection}).

The background sources used to measure the lensing effect are the KiDS galaxies described in Sect. \ref{sec:kids}. Following \cite{hildebrandt2017}, we only use sources with a best-fit photometric redshift $0.1<z_{\rm B}<0.9$. For troughs defined at a specific redshift we only select sources situated beyond the troughs, including a redshift buffer of $\Delta z=0.2$ (see Sect. \ref{sec:esd_measurements}). This cut is not applied when troughs are selected over the full redshift range. This can allow sources that reside at similar redshifts as the lenses to be used in the measurement, which would result in a contamination of the lensing signal by sources that are not lensed (`boost factor') and/or by sources that are intrinsically aligned with the troughs. However, even without a redshift cut $80\%$ of the KiDS source galaxies have a best-fit photometric redshift $z_{\rm B}$ above the mean redshift ($\mean{z_{\rm G}} = 0.24$) of our GAMA sample. Also, the intrinsic alignment effect has proven to be very small and difficult to detect, and primarily plays a role in very high-density regions on small ($\lesssim 1 \hsMpc$) scales. On the large scales probed by the troughs, the contamination of the lensing signal from intrinsic alignment is expected to be at most a few percent \cite[]{heymans2006,blazek2012}. Regarding the boost factor, this effect is also reproduced in the results obtained from the mock catalogues to which we compare our observations.

The ellipticities of the source galaxies are measured using the self-calibrating \emph{lens}fit pipeline \cite[]{miller2007,miller2013,fenechconti2017}. For each galaxy this model fitting method also produces the \emph{lens}fit weight $w$, which is a measure of the precision of the shear estimate it provides. We incorporate the \emph{lens}fit weight of each source into the average tangential shear in each angular bin as follows:
\begin{equation}
\mean{\gamma} = \frac{1}{1+\mu} \frac{\sum_{ls} w_{s} \, \epsilon_{{\rm t},ls} }{ \sum_{ls}{w_{s}} }  \, .
\label{eq:shear_measured}
\end{equation}
Here the sum goes over each lens $l$ in the lens sample (e.g. all apertures with a specified size and galaxy number density) and over each source $s$ inside the considered bin in angular separation from the centre of the lens. The factor $1+\mu$ is used to correct for `multiplicative bias'. Based on extensive image simulations \cite{fenechconti2017} showed that, on average, shears are biased at the $1-2\%$ level, and how this can be corrected using a multiplicative bias correction $m$ for every ellipticity measurement. Following \cite{dvornik2017}, the value of $\mu$ is calculated from the $m$-corrections in $8$ redshift bins (with a width of $0.1$) between $0.1 < z_{\rm B} < 0.9$. The average correction in each bin is defined as follows:
\begin{equation}
\mu=\frac{\sum_{s} w_{s} m_{s}}{\sum_{ls} w_{s}} \, .
\label{eq:biascorr}
\end{equation}
The required correction is small ($\mu\approx0.014$) independent of angular separation, and reduces the residual multiplicative bias to $\lesssim1\%$. The errors on our shear measurement are estimated by the square-root of the diagonal of the analytical covariance matrix (see Sect. \ref{sec:covariance}). The analytical covariance is based on the contribution of each individual source to the lensing signal, and takes into account the covariance between sources that contribute to the shear profile of multiple lenses. Its calculation is described in Sect. 3.4 of \cite{viola2015}.

\begin{figure}
	\includegraphics[width=1.0\columnwidth]{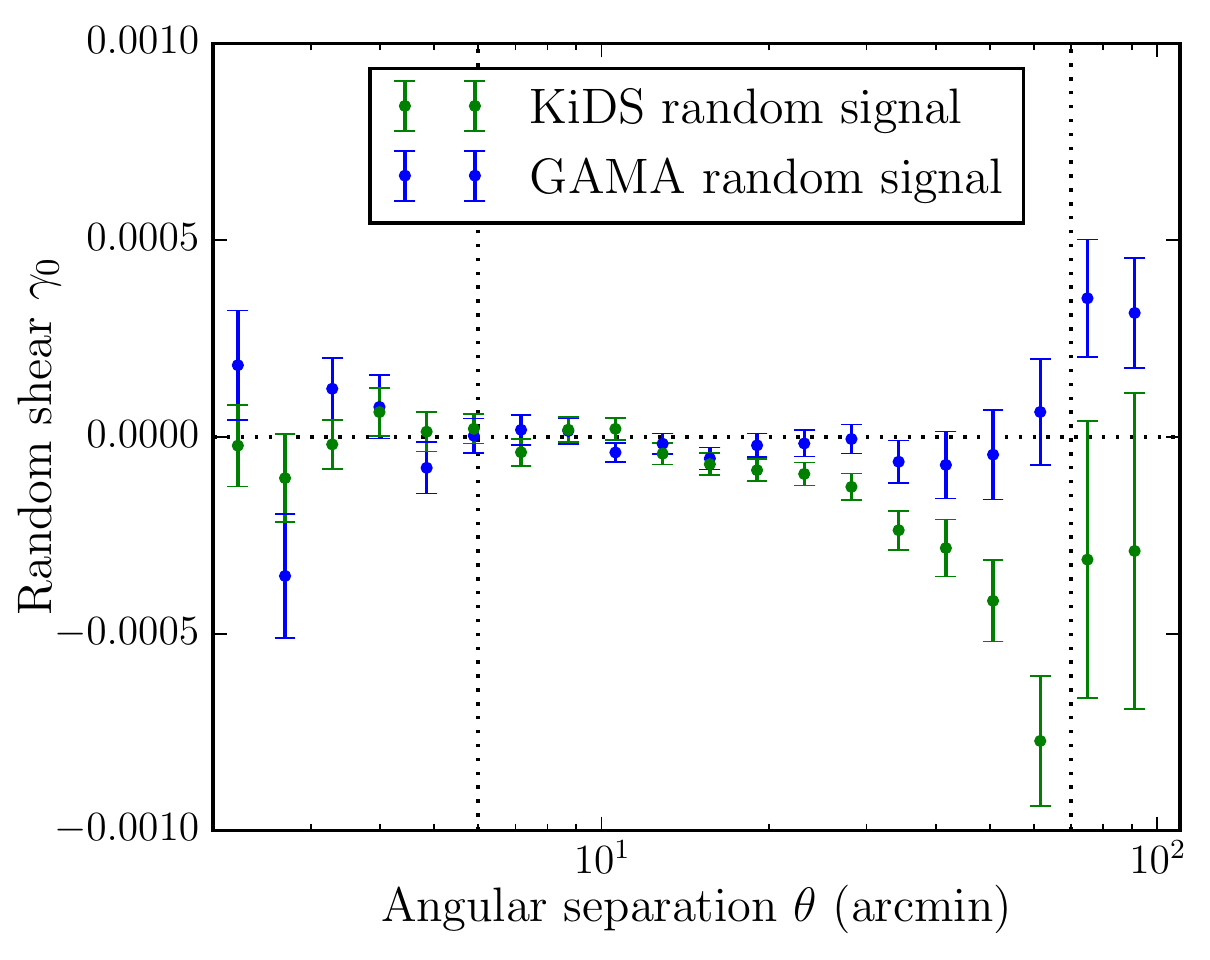}
	\caption{The random shear profile $\gamma_0$ (including $1\sigma$ analytical covariance errors) as a function of angular separation $\theta$, which results from stacking \emph{all} $\theta_{\rm A}=5 \am$ apertures with an area $>80\%$ complete. Using the GAMA area and mask, the systematic effects are consistent with zero up to $\theta = 70 \am$, while the KiDS random signal already starts to deviate from zero at $\theta \approx 20 \am$ as a result of the patchy survey coverage of KiDS outside the GAMA overlap. Only the range within the dotted vertical lines is used to study the trough lensing profiles in this work.}
	\label{fig:kids_vs_gama_randoms}
\end{figure}

In addition to measuring the lensing profile around troughs and ridges, we stack the shear around \emph{all} grid points ($262\,507$ in the case of KiDS, $112\,500$ in the case of GAMA). In accordance with the real trough measurements, the apertures with an effective area less than $80\%$ of the total circle area are removed (see Sect. \ref{sec:trough_selection}). This `random' tangential shear signal, that we henceforth denote as $\gamma_0$, does not contain a coherent shear profile, but only systematic effects resulting from the imperfect correction of any low-level PSF anisotropy in combination with the survey edges and masks. Subtracting $\gamma_0$ from our shear profiles will both remove these systematic effects and reduce the noise in the measured signals \cite[]{gruen2017,singh2017}. The random signals for KiDS and GAMA are shown in Fig. \ref{fig:kids_vs_gama_randoms}. When using the GAMA survey area and mask, $\gamma_0$ is consistent with zero (within $1\sigma$ error bars) up to $\theta = 70 \am$, where it rises to $\gamma_0 \sim 3\E{-3}$ for all values of $\theta_{\rm A}$, while the KiDS random signal already starts to deviate from zero at $\theta \approx 20 \am$. This difference does not significantly depend on the choice of area completeness threshold, and also occurs when we apply no completeness mask at all. However, when we perform the $\gamma_0$ measurement using the KiDS mask on the GAMA area only, the systematic effect is significantly reduced. This shows that the difference between the random signals is primarily caused by the patchy surface coverage of the KiDS-450 dataset beyond the GAMA area \cite[see e.g. Fig. 1 of][]{hildebrandt2017}. The same effect can be seen in Fig. 15 of \cite{uitert2016b}, who conclude that it originates from the boundaries of the survey tiles.

To correct for this effect at larger scales, we subtract the appropriate $\gamma_0$ from all lensing measurements in this work. Based on the radius where the random signal becomes significant ($\theta\sim70 \am$), and on our grid spacing of $0.04 \deg = 2.4 \am$ (see Sect. \ref{sec:trough_selection}), we compute our lensing profiles within the angular separation: $2 < \theta < 100 \am$. We split this range into 20 logarithmically spaced bins.

\subsection{Covariance}
\label{sec:covariance}

\begin{figure*}
	\includegraphics[width=0.9\textwidth]{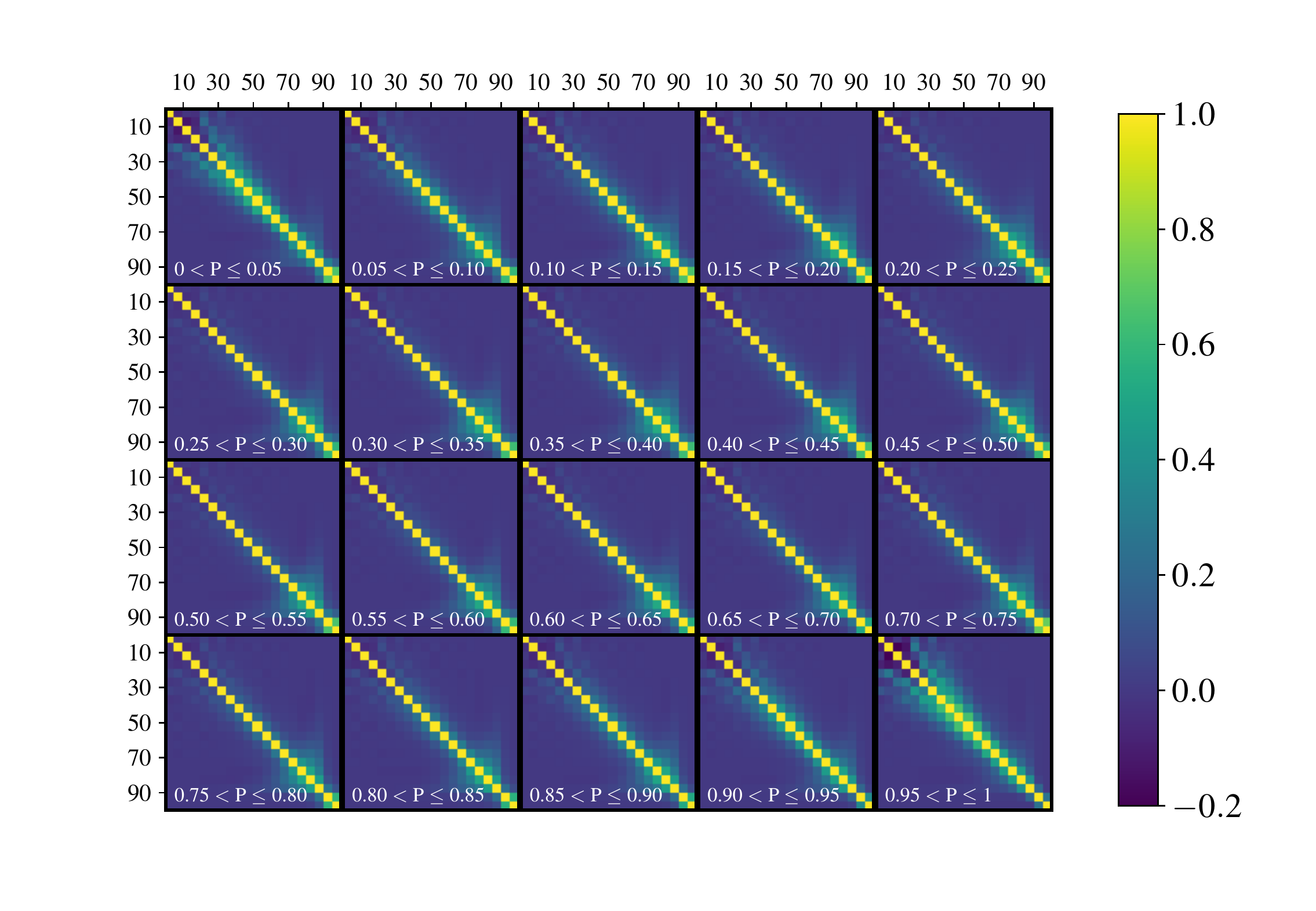}
	\includegraphics[width=0.9\textwidth]{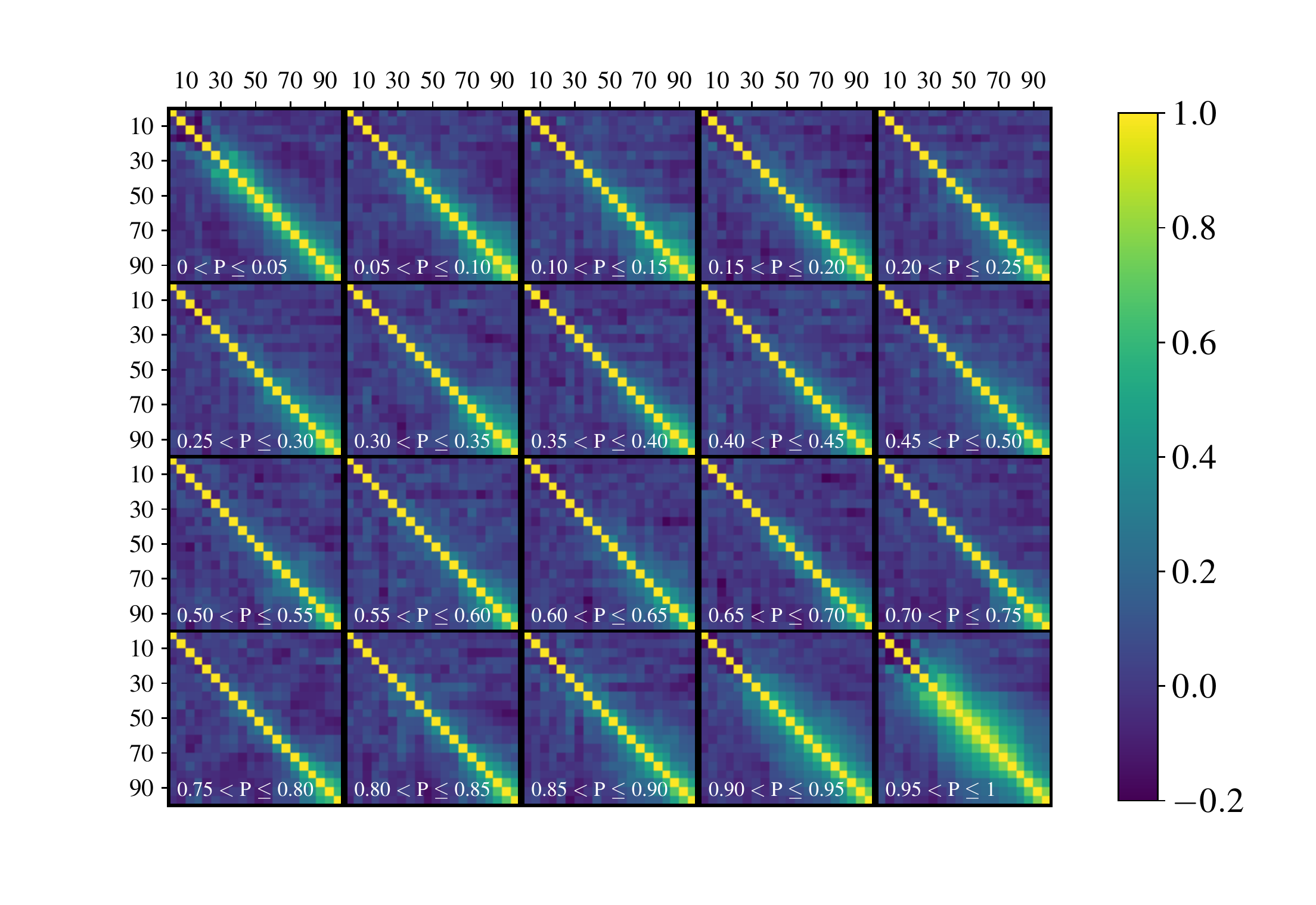}	
	\caption{The two panels show the analytical GL-KiDS (top) and SLICS GAMA HOD (bottom) correlation matrices, resulting from apertures with an angular radius $\theta_{\rm A} = 5 \am$. The correlation matrices are computed for 20 bins of increasing galaxy density percentile rank $P(\vec{x}, \theta_{\rm A}=5\am)$, corresponding to the shear profiles shown in Fig. \ref{fig:amplitude_fits}. The increased correlation at large radii is caused by the overlap between sources (in the case of both KiDS and SLICS) and by sample variance (in the case of SLICS). The increased correlation at extreme $P$-values is caused by the spatial clustering of low- and high-density regions.}
	\label{fig:cov_matrices}
\end{figure*}

\begin{figure}
	\includegraphics[width=1.0\columnwidth]{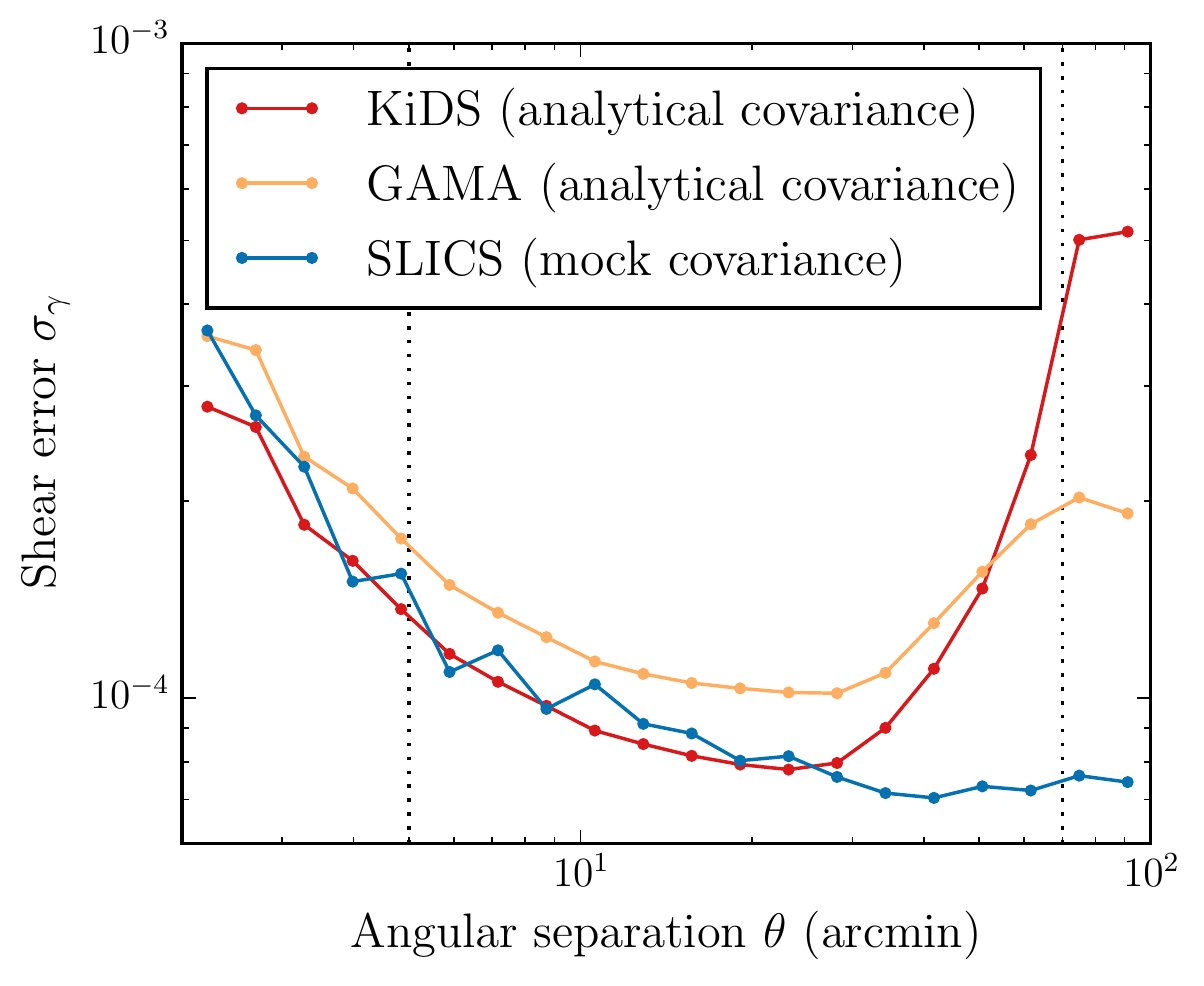}
	\caption{The error values $\sigma_\gamma(\theta)$ (as a function of angular separation $\theta$) on the shear profile of the fiducial G16 troughs ($P<0.2$) with a radius of $\theta_{\rm A}=5\am$. The KiDS and GAMA errors are estimated using the diagonal of the analytical covariance matrix, while the mock errors are estimated from the covariance matrix calculated using $349$ SLICS mock realisations. The GAMA errors are higher than those of KiDS, as expected from its smaller survey area. The KiDS errors are in reasonable agreement with the SLICS mock errors up to $\theta=30\am$, where they rise steeply as a result of the patchiness of the survey.}
	\label{fig:cov_errors}
\end{figure}

For all shear and ESD measurements created using the KiDS and GAMA data, we compute the analytical covariance matrix as described in Sect. 3.4 of \cite{viola2015}. This covariance matrix is based on the contribution of each individual source to the stacked lensing signal, and takes into account the correlation between sources that contribute to the shear profile of multiple lenses. The errors on our shear profiles are estimated by the square-root of the diagonal of this analytical covariance matrix. However, these error bars could underestimate the uncertainties at larger scales, where sample variance starts to play a significant role \cite[]{viola2015}. Here we compare the analytical covariance calculated using our KiDS data to those based on the large ensemble of mock realisations from the SLICS mocks, in order to find whether the analytical covariance is sufficient for our analysis.

Utilising the SLICS HOD mock catalogues described in Sect. \ref{sec:slics_mocks} we compute the covariance matrix using the following equation:
\begin{equation}
C^{ij} = \frac{1}{N-1}\sum_{n=1}^{N} ({\gamma^i_{{\rm t},n}}-\mean{\gamma_{\rm t}}^i)({\gamma^j_{{\rm t},n}}-\mean{\gamma_{\rm t}}^j) \, ,
\label{eq:cov}
\end{equation}
where $N$ is the number of mock realisations, ${\gamma_{\rm t}}^i$ is the tangential shear signal in the $i$-th angular bin of the $n$-th mock realisation, and $\bar{\gamma_{\rm t}}^i$ is the tangential shear average of the $i$-th bin from all used realisations. The covariance is then multiplied by the area factor:
\begin{equation}
f_{\rm area} = \frac{100}{360.3} \, ,
\label{eq:err_fact}
\end{equation}
in order to account for the difference in area between the SLICS mocks and the KiDS data. The errors on the shear are then calculated using the square root of the diagonal of this scaled covariance matrix. Since we calculate the mock covariance from multiple realisations and use the total modeled ellipticities of the galaxies to calculate the tangential shear signal, the mock covariance accounts for shape noise, shot noise, and sample variance. Fig. \ref{fig:cov_matrices} shows the correlation matrices, $r_{\rm corr}$, for the mock and analytical covariances, respectively, where the correlation matrix is calculated using:
\begin{equation}
r^{ij}_{\rm corr} = \frac{C^{ij}}{\sqrt{C^{ii}C^{jj}}} \, .
\label{eq:corr_mat}
\end{equation}
We calculate the SLICS shear profiles and covariance matrices using 349 line-of-sight realisations. We found no significant difference in the shear profiles or covariance matrices (Fig. \ref{fig:amplitude_fits} and \ref{fig:cov_matrices} respectively) of $5 \am$ troughs/ridges when we increased the number of realisations to 608, concluding that using 349 realisations is therefore sufficient for all following analyses.

In Fig. \ref{fig:cov_matrices} we show the data-based analytical (top) and mock-based SLICS (bottom) correlation matrices for the shear profiles $\gamma(\theta)$ of apertures with radius $\theta_{\rm A} = 5 \am$, split into 20 bins based on their galaxy density percentile rank $P(\theta_{\rm A})$ (corresponding to the shear profiles shown in Fig. \ref{fig:amplitude_fits} of Sect. \ref{sec:amplitudes}). Comparing the analytical and mock correlation matrices, we notice that those from the SLICS mocks are noisier compared those calculated analytically, due to the limited number of mock realisations in combination with the effects of sample variance. In addition, the correlation at large scales appears to be stronger for the mock results, which is also expected since the mock correlation incorporates the effects of sample variance (which the analytical covariance does not). Nevertheless, the analytically calculated correlation also increases at large scales, due to the increasing overlap of source galaxies with increasing radius. For both data and mocks, the covariance depends significantly on density, increasing at extremely low and high $P$-values. This is expected, since extremely low-density troughs (high-density ridges) tend to cluster at the centres of larger low-density (high-density) regions, as can be seen in Fig. \ref{fig:trough_map}. This clustering of extreme density regions increases the correlation between the lensing signals of the more extreme troughs and ridges.

Most importantly, we assess the agreement between the \emph{diagonals} of the covariance matrices created by both methods, since the square-root of these diagonals defines the errors $\sigma_\gamma$ on the measured shear profiles. Fig. \ref{fig:cov_errors} shows the $\sigma_\gamma(\theta)$ values of KiDS and GAMA-selected fiducial G16 troughs ($P(\vec{x}, \theta_{\rm A})<0.2$), with a radius of $\theta_{\rm A}=5\am$.\footnote{We have performed the error comparison not only for this trough sample, but for all 20 galaxy density percentile bins shown in Fig. \ref{fig:cov_matrices} and for all four aperture sizes used in this work ($\theta_{\rm A}=\{5, 10, 15, 20\} \, \am$), finding similar results.} As expected from its smaller survey area, the small-scale ($\theta<30 \am$) error values from GAMA are a factor $\sim1.3$ higher than those from KiDS. We compare these analytical covariance errors to those calculated from $349$ SLICS mock realisations, adjusted using the area factor in Eq. \ref{eq:err_fact} to resemble the KiDS survey. Up to a separation $\theta=30\am$ (half the size of a $1 \times 1 \deg$ KiDS tile) the KiDS and SLICS error values are in excellent agreement. Due to the patchy KiDS survey coverage beyond the GAMA fields, the KiDS errors increase rapidly at larger angular separations. For the GAMA survey, whose area is more contiguous, this increase in error values is much smaller. For the SLICS mocks, which consist of $10 \times 10 \deg$ patches, it is completely absent. Because this effect dominates the error values at larger scales, we conclude that we do not need to worry about a possible underestimation of the analytical covariance errors at larger scales due to the lack of sample variance. We therefore use the analytical covariance matrix to estimate the errors on the observed trough/ridge profiles throughout this work. However, we do use SLICS mock covariances to devise an optimal trough and ridge weighting scheme (in Sect. \ref{sec:weighting}), and to predict the significance of future trough measurements (in Sect. \ref{sec:higher_redshifts}).

\section{Trough \& ridge shear profiles}
\label{sec:results}

After a general classification of the troughs and ridges, we define more specific samples and measure their lensing profiles. First, we compare the trough shear profiles of the GAMA vs. GL-KiDS selected troughs, to decide on the best trough sample to use in this work. Using these troughs, we measure the shear amplitude of the lensing profiles as a function of their galaxy density percentile rank $P(\vec{x}, \theta_{\rm A})$, for apertures of different sizes $\theta_{\rm A}$. This allows us to study non-linearities in cosmic structure formation, and to define an optimal way to stack the shear signals of troughs and ridges in order to optimize the $S/N$.

\subsection{KiDS vs. GAMA troughs}
\label{sec:kids_vs_gama}

The very complete and pure sample of GAMA galaxies (see Sect. \ref{sec:gama}) allows us to define a clean sample of troughs. However, since the currently available area of the KiDS survey is $2.5$ times larger than that of the GAMA survey, we also use a set-up that uses the GL-KiDS galaxies (see Sect. \ref{sec:gamalike_kids}) to define the troughs. For this initial comparison, we use the fiducial trough/ridge definition of G16: the apertures with the lowest/highest $20\%$ in density (i.e. $P<0.2$ / $P>0.8$). We construct both fiducial trough samples following the same classification method (see Sect. \ref{sec:trough_selection}), using both galaxy catalogues as our trough-defining samples. We use the corresponding completeness mask to remove unreliable troughs (i.e. with an area $<80\%$ complete).

\begin{figure}
	\includegraphics[width=1.\columnwidth]{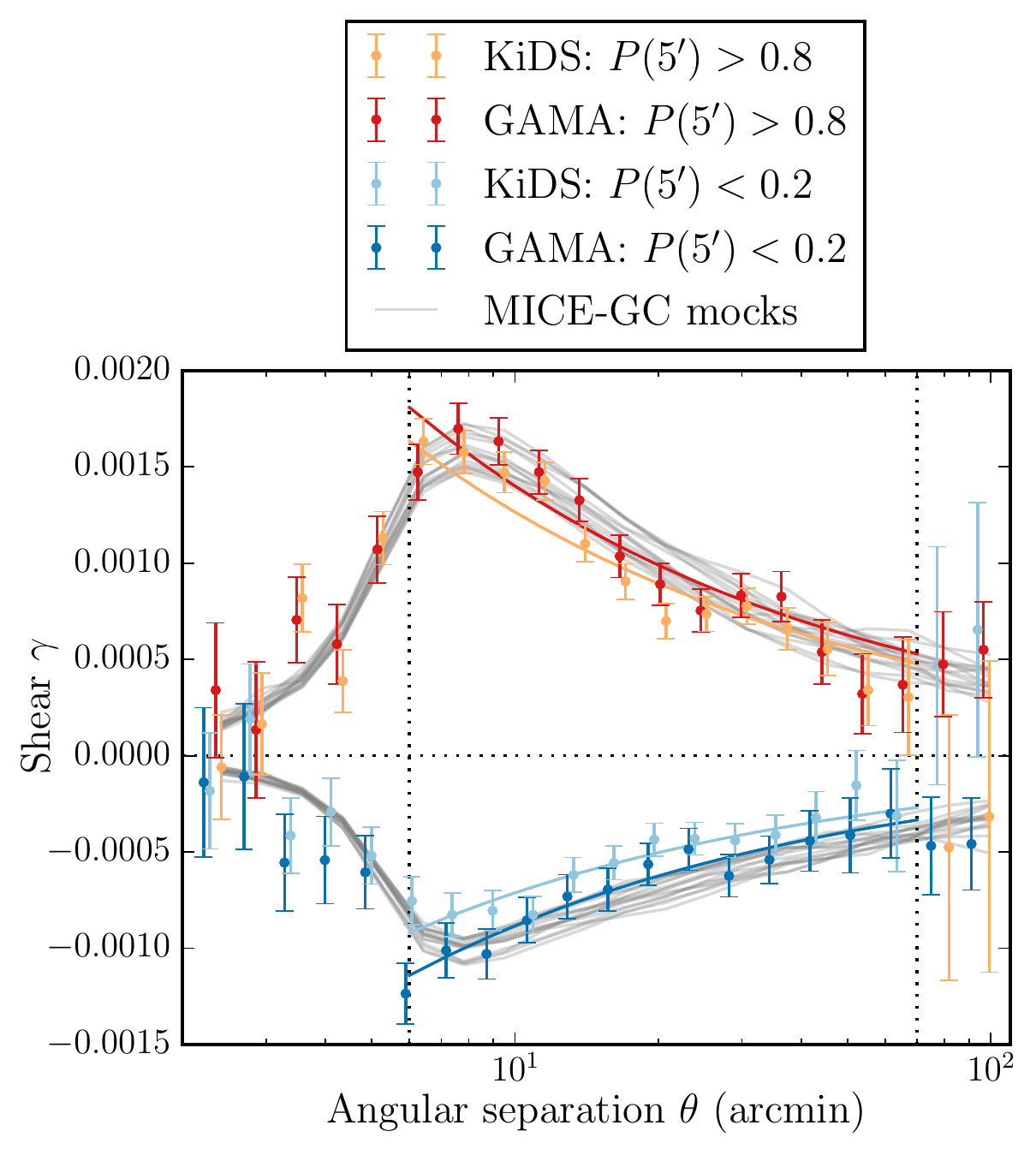}
	\caption{The gravitational shear profile $\gamma_{\rm t}(\theta)$ (with $1\sigma$ errors) of the G16 fiducial troughs and ridges, selected using the GL-KiDS (orange and light blue dots) and GAMA (red and dark blue dots) foreground galaxy sample, including a comparison with the MICE-GC mock troughs/ridges from 16 independent patches (grey lines). All troughs and ridges are selected following the fiducial trough/ridge definition in G16 (i.e. $P < 0.2$ / $P > 0.8$), and have a radius $\theta_{\rm A} = 5 \am$. We fit a simple $A/\sqrt{\theta}$ function (solid coloured lines) within the indicated range (dotted vertical lines) to determine the best-fit amplitude $A$ of the KiDS and GAMA fiducial troughs/ridges.}
	\label{fig:kids_vs_gama}
\end{figure}

\begin{figure*}
	\includegraphics[width=1.0\textwidth]{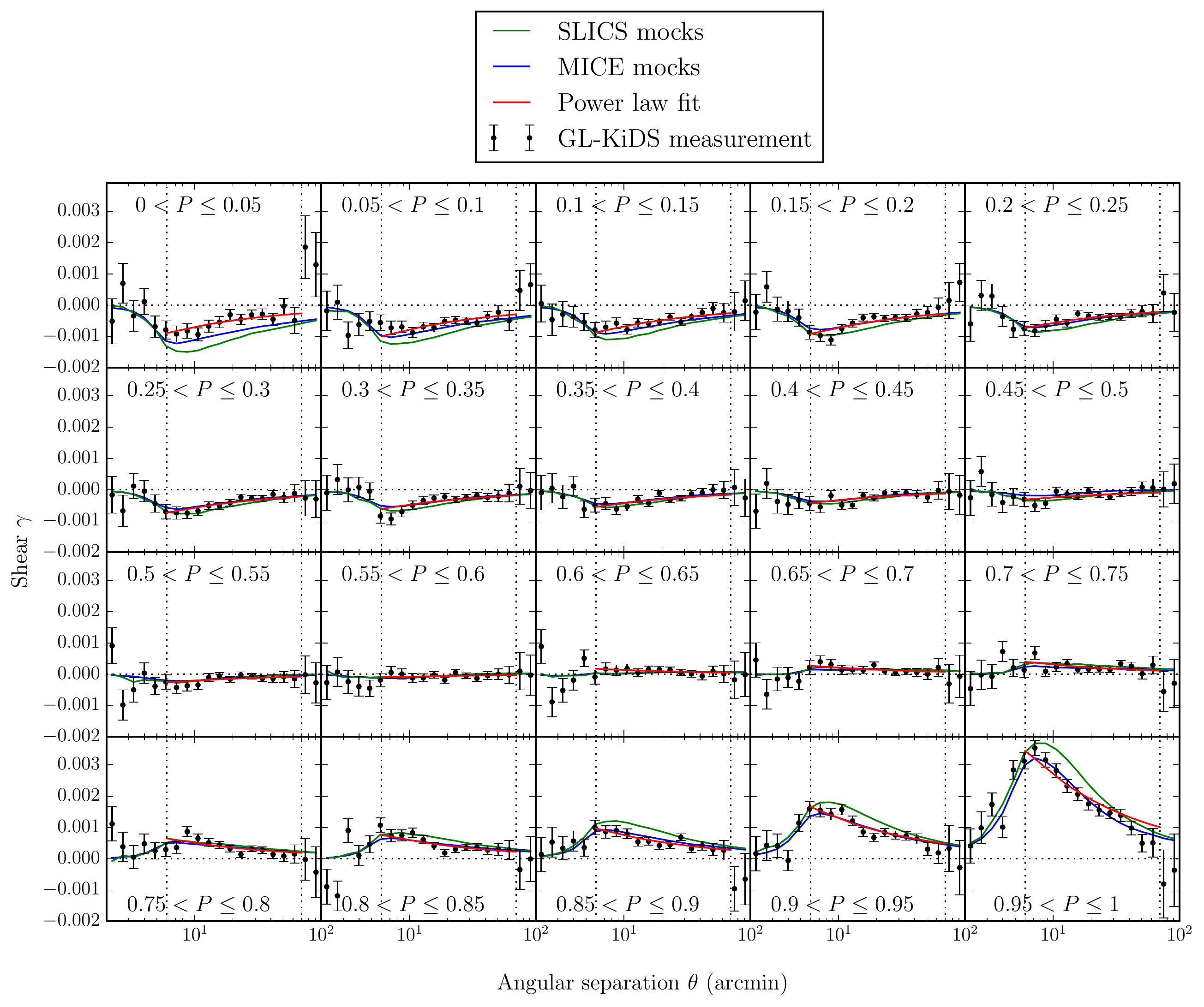}
	\caption{Each panel shows the GL-KiDS (black dots with $1\sigma$ errors), MICE (blue line) and SLICS (green line) shear profiles $\gamma_{\rm t}(\theta)$, resulting from apertures of angular radius $\theta_{\rm A} = 5 \am$. The shear profile of these apertures is stacked in 20 bins of increasing galaxy density percentile rank $P(\vec{x}, \theta_{\rm A}=5\am)$. For underdense apertures (troughs) the amplitude $A$ of the lensing signal becomes negative outside the trough radius, while for overdense apertures (ridges) $A$ becomes positive. A simple power law fit: $A/\sqrt{\theta}$ (red line), within the fitting range (dotted vertical lines) is used to obtain $A$ as a function of $P$.}
	\label{fig:amplitude_fits}
\end{figure*}

The main goal of this exercise is to find whether trough lensing measurements can accurately be reproduced using only the photometric KiDS data, without the help of the spectroscopic GAMA survey. In addition, we wish to find which galaxy survey provides the trough lensing profiles with the highest $S/N$. In Fig. \ref{fig:kids_vs_gama} we show the stacked shear profiles $\gamma_{\rm t}(\theta)$ of G16 fiducial troughs with radius $\theta_{\rm A} = 5 \am$, selected using the GL-KiDS or GAMA galaxies. For comparison we also include the fiducial trough shear profiles obtained using all 16 patches of the MICE mock catalogue, where the vertical spread in the 16 profiles gives an estimate of the sample variance. The absolute values of the amplitudes (which we will henceforth call `absolute amplitudes') of the GAMA-selected fiducial trough/ridge profiles are slightly higher than those of the KiDS-selected troughs. Nevertheless, within the $1\sigma$ analytical covariance errors both profiles agree with the predictions from the MICE-GC simulation. However, when we use the GL-KiDS galaxies to select troughs but restrict the used area to the GAMA equatorial fields, we find that the KiDS trough profiles have the same amplitude as those from GAMA. This suggests that, like the systematic effects measured by the randoms, the shallower trough lensing profile is caused by the patchy survey coverage of the non-equatorial KiDS fields. This reduces the completeness of the circles, which diminishes the accuracy of the density measurements and results in slightly shallower shear profiles.

The dotted vertical lines in Fig. \ref{fig:kids_vs_gama} indicate the angular separation range: $1.2 \, \theta_{\rm A} < \theta < 70 \am$, that we consider in our analysis. Our reasons for selecting this range are: 1) inside $\theta_{\rm A}$ the lensing is not sensitive to the full trough mass (where we leave a $20\%$ buffer outside the trough edge), and 2) the random signal $\gamma_0$ in Fig. \ref{fig:kids_vs_gama_randoms} shows that at $\theta > 70 \am$ our measurement is sensitive to systematic effects (see Sect. \ref{sec:lensing}). Within this range we observe that the fiducial trough and ridge shear signals are well-described by a power law. We can therefore fit a relation $\gamma_{\rm t}(\theta) = A \, \theta^{\alpha}$ within the specified angular range, to obtain the best-fit amplitude $A$ and index $\alpha$ of the lensing signal. Because we are mainly interested in the amplitude, we fix the value of $\alpha$ with the help of the MICE-GC simulations. By fitting the power law (with both $A$ and $\alpha$ as free parameters) to all 16 fiducial MICE lensing signals, we find a mean best-fit index value $\mean{\alpha}$ of $-0.45$ for the fiducial troughs and $-0.55$ for ridges. We therefore choose to fit all trough lensing profiles in this work with the function:
\begin{equation}
	\gamma_{\rm t}(\theta) = A / \sqrt{\theta} \, .
	\label{eq:trough_fit}
\end{equation}
However, we verify that our conclusions do not significantly depend on the specific choice of $\alpha$ by performing the same analysis with $\alpha=-1$, and finding similar results in terms of the amplitude comparison between various trough/ridge profiles. This indicates that, as long as we use one function of $A$ that provides a good fit to all profiles, the comparison between the resulting amplitude values is robust.

From the best-fit amplitudes thus obtained, we wish to find a measure of the signal-to-noise ratio $S/N$ in order to select the best trough measurement. We define $S/N \equiv A /\sigma_{\rm A}$, where $\sigma_{\rm A}$ is the $1\sigma$ error on the best-fit amplitude based on the full analytical covariance matrix of the shear profile. Using this definition we find that the fiducial trough lensing signal is detected at $\lvert S/N \rvert = 12.0$ with the GAMA selection, and $\lvert S/N \rvert = 12.3$ when GL-KiDS is used: evidently the KiDS-450 area advantage compared to GAMA is almost completely offset by the greater patchiness. However, we can conclude from this exercise that the larger KiDS dataset provides trough lensing measurements with a slightly higher $S/N$ than the GAMA dataset. In what follows we will therefore primarily use the full KiDS sample, but we have verified throughout that similar results are obtained using the GAMA galaxies instead.

\subsection{Lensing amplitudes}
\label{sec:amplitudes}

\begin{figure*}
	\includegraphics[width=1.0\columnwidth]{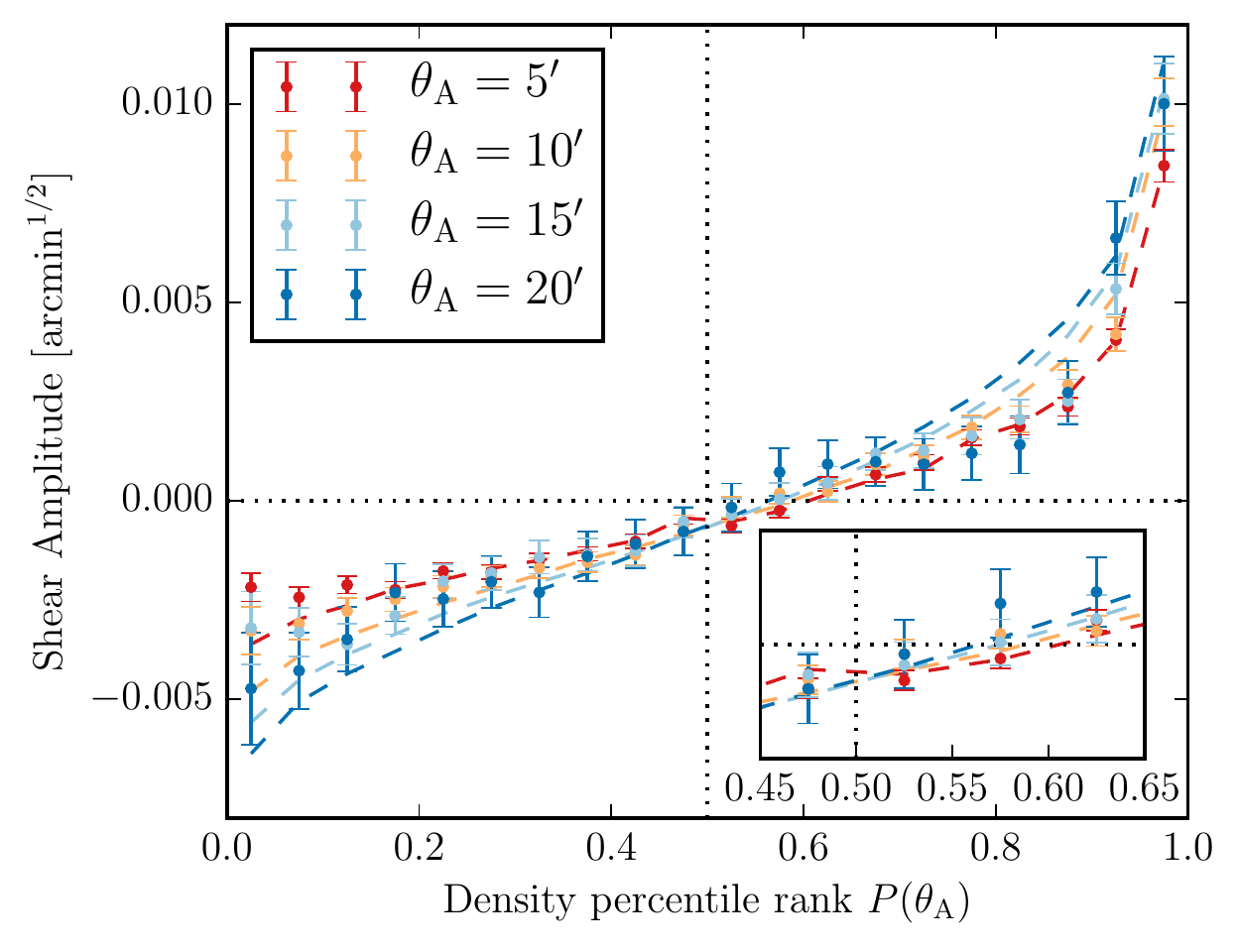}
	\includegraphics[width=1.0\columnwidth]{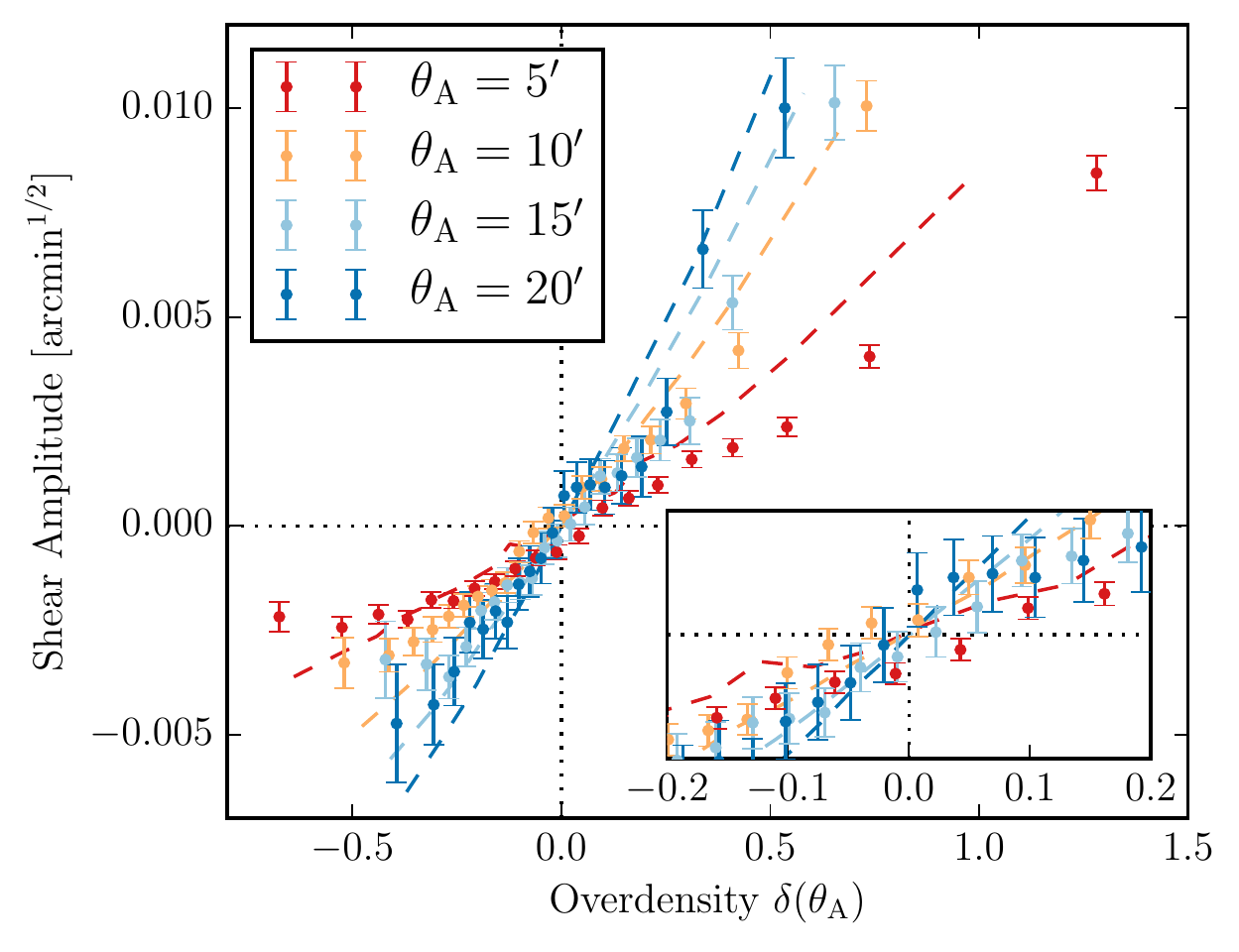}
	\caption{The amplitude $A$ of the KiDS (dots with $1\sigma$ errors) and MICE (dashed lines) shear signals as a function of the galaxy density percentile rank $P$ (left) and galaxy overdensity $\delta$ (right), for apertures of different angular radius $\theta_{\rm A}$. The crossing point between negative and positive $A$ is situated at the mean density ($\delta=0$) as is expected when linear galaxy bias dominates. This crossing point, however, is not situated at the median density ($P=0.5$) but at $P\approx0.55-0.6$, which means that the density distribution is skewed. The smallest apertures also reveal the skewness of the density distribution, since their distribution extends to more extreme values of $P$, $\delta$ and $A$ for the ridges than for the troughs, while larger apertures have a more symmetrical $A(\delta)$ distribution.}
	\label{fig:amplitudes_perc_delta}
\end{figure*}

After this initial test, which uses only the lowest and highest $20\%$ of the troughs/ridges, we wish to study \emph{all} troughs and ridges as a function of their galaxy density percentile rank $P(\theta_{\rm A})$. Our aim is to gain more insight into the relation between the total mass distribution (measured by lensing) and the galaxy number density, generally called `galaxy bias'. Considering apertures of fixed radius $\theta_{\rm A}$ we split them into 20 samples of increasing $P$-value, using a bin width of ${\rm d}P = 0.05$. We measure the shear profile $\gamma_{\rm t}(\theta)$ of each sample (using the method described in Sect. \ref{sec:lensing}). Figure \ref{fig:amplitude_fits} shows the GL-KiDS, MICE and SLICS lensing profiles in the 20 $P$-bins for $\theta_{\rm A} = 5\am$. To each shear measurement we fit Eq. \ref{eq:trough_fit} within the indicated angular range, to measure the shear amplitude $A$. Throughout this work, all amplitude fits take into account the full covariance matrix of each shear profile, in this case shown in Fig. \ref{fig:cov_matrices}. However, we find that the off-diagonal elements only have a minor effect on the amplitude estimates. For the amplitudes of these 20 $P$-bins, using the full covariance matrix versus using only the diagonal errors yields an average difference of only $2.3\%$ ($3.6\%$) for the KiDS (SLICS) amplitudes, where for each percentile bin this difference is much lower than the error estimate on that amplitude.

As expected the apertures with lowest/highest $P$-values correspond to the strongest negative/positive shear signals. The absolute amplitudes of the profiles predicted by the MICE mocks tend to be lower than those from SLICS, where the former predictions are in better agreement with the GL-KiDS measurements. This offset is expected given the different background cosmologies chosen for the SLICS and MICE simulations \cite[]{friedrich2017}, where higher values of $\sigma_8$ and $\Omega_{\rm m}$ give rise to higher absolute amplitudes. Interestingly the cosmological constraints from the cosmic shear analysis with KiDS-450 \cite[]{hildebrandt2017} suggest that the KiDS data prefer a cosmology with lower values of $\Omega_{\rm m}$ and $\sigma_8$. These values are close to those adopted by the MICE simulations ($\sigma_8 = 0.8$, $\Omega_{\rm m} = 0.25$), and in slight tension with the Planck cosmology which is adopted by the SLICS simulation ($\sigma_8 = 0.826$, $\Omega_{\rm m} = 0.29$). Therefore, the tension in cosmology with Planck seen in the KiDS-450 cosmic shear results is also reflected in the trough/ridge measurements in this paper.

It is also apparent that troughs and ridges are not symmetrical, but that the lensing signal is stronger for the highest ridges than for the deepest troughs. This is an indication that the skewness of the galaxy number density distribution (seen in Fig. \ref{fig:trough_hist}) is reflected by the total (baryonic + DM) density distribution. This skewness is also indicated by apertures with $P\sim0.5$. Fig. \ref{fig:amplitudes_perc_delta} (left panel) shows the best-fit $A$ as a function of $P$ for apertures of different radius $\theta_{\rm A}$. For both the KiDS and MICE data the crossing point $A=0$ is not reached at $P=0.5$, but at $P\approx0.55-0.6$. The right panel of Fig. \ref{fig:amplitudes_perc_delta} shows $A$ as a function of the mean galaxy overdensity $\delta(\theta_{\rm A})$ (defined in Eq. \ref{eq:delta}) in each $P$-bin, for both KiDS and MICE troughs/ridges. The $\delta$-value of each bin is taken to be the mean galaxy overdensity $\mean{\delta}(\theta_{\rm A})$ of all apertures in each $P$-bin. For all aperture sizes the $A(\delta)$ relation is approximately linear, with the crossing point between negative and positive $A$ roughly situated at the mean density ($\delta=0$). This is expected when linear galaxy bias dominates, i.e.: there exists a linear relation between the density distributions of galaxies and DM.

The difference between the crossing points ($A = 0$) of the $A(P)$ and $A(\delta)$ relations shows that (like the galaxy number density distribution in Fig. \ref{fig:trough_hist}) the mass distribution measured using lensing is skewed. Note that the crossing point of the $A(P)$-relation occurs at larger $P$ for smaller $\theta_{\rm A}$: the smaller the aperture (i.e. smoothing scale of the density distribution), the larger is the skewness of the distribution. This skewness is caused by the fact that, during cosmic structure formation through clustering, the density of matter is bound to a strict lower limit (a completely empty region) but not to an upper limit. This is also revealed by the fact that, especially for smaller apertures, the positive amplitudes are significantly larger than the negative amplitudes, while larger apertures have more symmetrical $A(\delta)$ relations.

In conclusion, the trough/ridge measurements as a function galaxy number density show that both the galaxy number density and total mass distributions are skewed, and that this skewness increases with decreasing aperture size. These non-linearities can in principle be used as a statistic to constrain cosmological parameters, analogous to performing shear peak statistics \cite[]{liu2015,kacprzak2016, shan2018, martinet2018}. In fact, \cite{gruen2017} and \cite{friedrich2017} used trough and ridge lensing measurements to constrain $\Omega_{\rm m}$ and $\sigma_8$, also finding that the total density field is skewed.

\subsection{Optimal weighting}
\label{sec:weighting}

\begin{figure}
	\includegraphics[width=1.0\columnwidth]{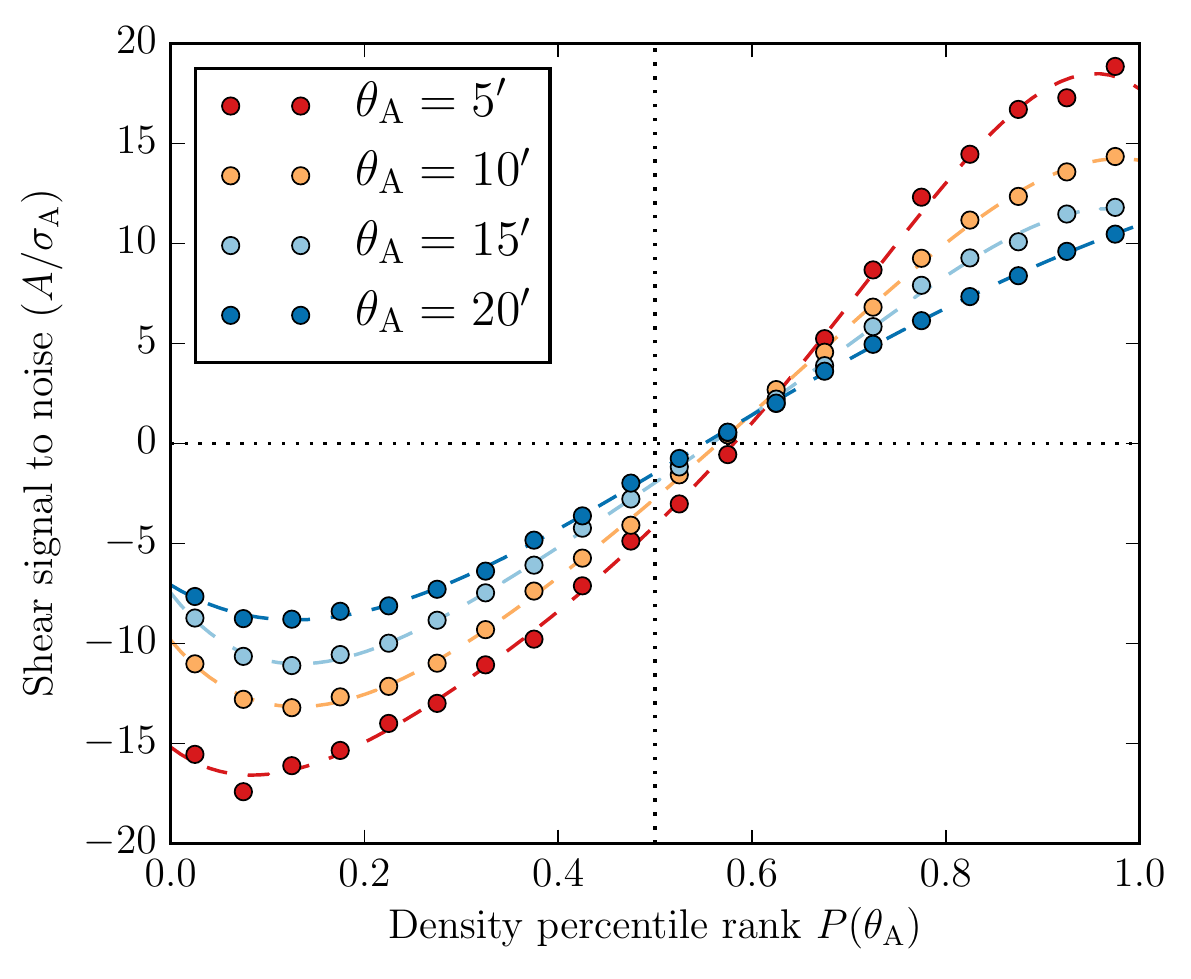}
	\caption{The signal-to-noise ratio, defined as $S/N \equiv A/\sigma_{\rm A}$, of the SLICS mock profiles as a function of the galaxy density percentile rank $P$. To obtain the optimal weight to stack the troughs and ridges, we fit a $5^{\rm th}$-order polynomial (dashed lines) to the measured $S/N$ values (dots). The resulting weight function $w(P)$ allows us to obtain a (positive) stacking weight $w_{\rm P} = |w(P)|$ for each individual lens.}
	\label{fig:weigts_perc}
\end{figure}

Instead of selecting troughs and ridges using a `hard cut' in the percentile rank $P(\vec{x}, \theta_{\rm A})$ of the apertures, one can apply a more sophisticated $S/N$-based weighting scheme to stack the shear profiles of the apertures. In order to obtain the most significant stacked lensing detection, the optimal weighting of each individual trough/ridge contributing to the stacked signal should be based on the $S/N \equiv A/\sigma_{\rm A}$ of that contribution. Our motivation for obtaining the highest possible $S/N$ is to facilitate the most accurate comparison with predictions from simulations (e.g. to constrain cosmological parameters or alternative gravity models, see Sect. \ref{sec:introduction}), as the $S/N$ of these predictions is currently higher than that of trough observations. To prevent \emph{a-posteriori} selection and boosting of random fluctuations in the data we use the SLICS mocks, which provide both the shear signal and corresponding covariance matrices (see Fig. \ref{fig:cov_matrices}, bottom panel), to obtain the appropriate $S/N$ weights. The $S/N$ measurement of the SLICS profiles takes into account the full covariance between the angular separation bins (as is done throughout this work), but not the possible correlation between the 20 galaxy density percentile bins. Future work which also takes this effect into account might enable an even better optimization of the trough and ridge shear signals. The $S/N$ of the SLICS mock profiles as a function of $P$ is shown in Fig. \ref{fig:weigts_perc}. In this relation the peaks at very high and low $P$ are reduced compared to those in the $A(P)$ relation, since very low-density troughs (and very high-density ridges) tend to cluster at the centres of large voids (or large clusters), as seen in Fig. \ref{fig:trough_map}. This increases the covariance between the lensing signals of the very `deep' troughs (or `high' ridges), thereby increasing the error values.

We fit $5^{\rm th}$-order polynomials (the dashed lines in Fig. \ref{fig:weigts_perc}) to the SLICS $A/\sigma_{\rm A}$ values in order to provide a lens weight $w_{\rm P}$ for every individual aperture. We define the weight as the absolute value of this fit, in order to obtain a positive weight for both ridges and troughs. Finally, when we compute the combined lensing profile of all troughs or ridges, we use these weights to scale the contribution of each lens to the combined shear signal. The $w_{\rm P}$-value of each lens $l$ is incorporated into Eq. \ref{eq:shear_measured}, such that it becomes:
\begin{equation}
\mean{\gamma}_{\rm P} = \frac{1}{1+K_{\rm P}} \frac{\sum_{l} \left( w_{{\rm P},l} \sum_{s} w_{s} \epsilon_{\rm t} \right) }{\sum_{l} \left( w_{{\rm P},l} \sum_{s} {w_{s}} \right) } \, .
\end{equation}
In this way we give higher weights to troughs/ridges that provide a higher $S/N$, which thus contribute more heavily to the combined shear signal.
These same weights are also applied to the average multiplicative bias correction from Eq. \ref{eq:biascorr}:
\begin{equation}
K_{\rm P} = \frac{\sum_{l} \left( w_{{\rm P},l} \sum_{s}  w_{s} m_{s} \right) }{ \sum_{l} \left( w_{{\rm P},l} \sum_{s}  w_{s} \right) } \, .
\end{equation}
Likewise, the lens weight is incorporated into the uncertainty through the calculation of the analytical covariance matrix (see Sect. \ref{sec:kids}). 

\begin{figure*}
	\includegraphics[width=1.0\textwidth]{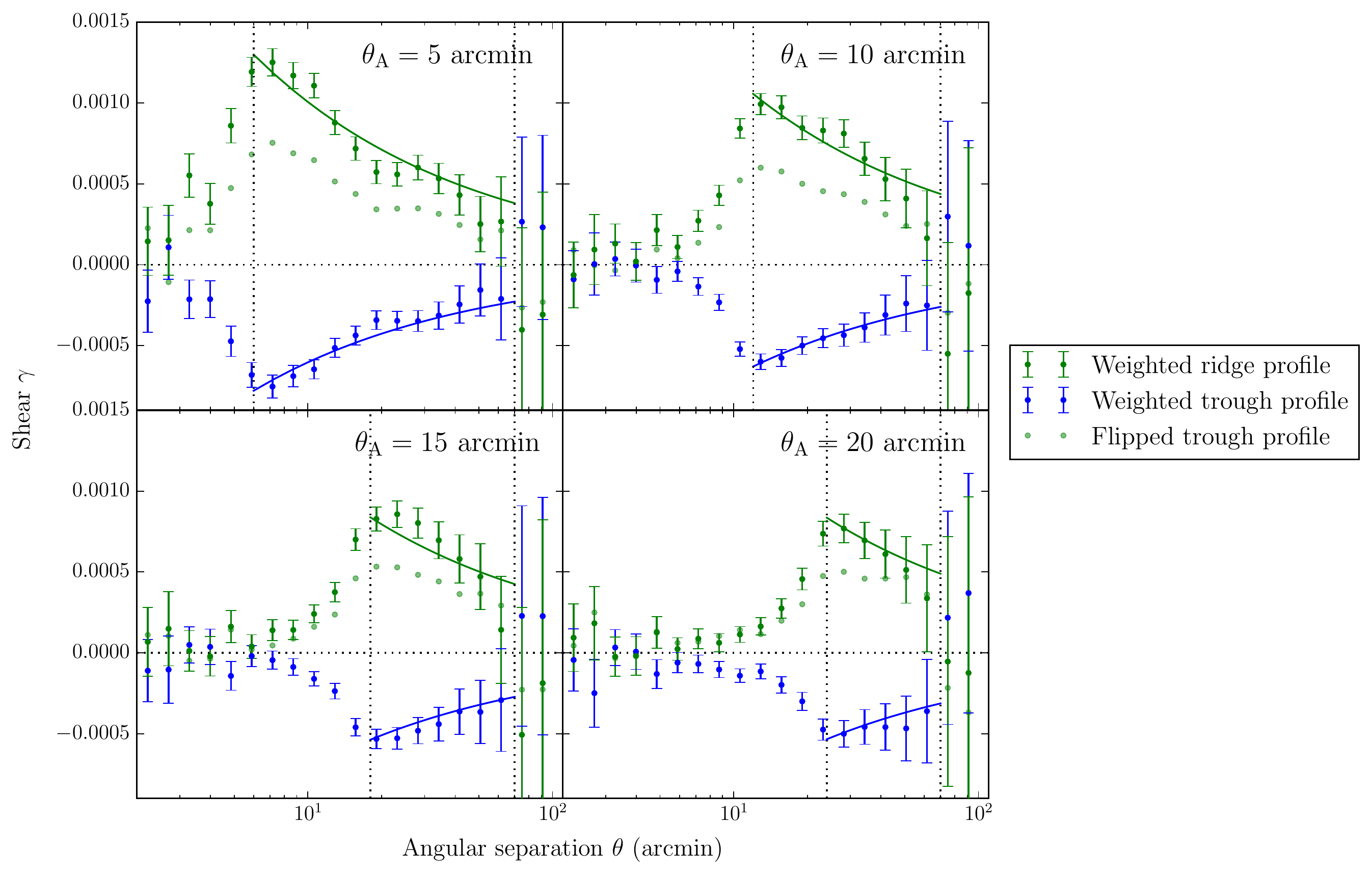}
	\caption{The optimally weighted KiDS trough (blue) and ridge (green) shear profiles $\gamma_{\rm t}(\theta)$ (dots with $1\sigma$ errors), for apertures of increasing radius $\theta_{\rm A}$. The weights of each stack are based on the $S/N$ of the SLICS mock profiles as a function of galaxy density percentile rank $P$ (shown in Fig. \ref{fig:weigts_perc}). The mirror image $-\gamma_{\rm t}(\theta)$ of each trough profile (light green dots) is added to allow for a better visual comparison between troughs and ridges. We fit a simple power law: $A/\sqrt{\theta}$ (solid lines), fitting range (dotted vertical lines) to obtain the amplitude $A$ of the lensing signals. For all aperture sizes, the shear from overdensities (ridges) is stronger than that of underdensities (troughs). This difference, which gives an indication of the skewness of the total (baryonic + DM) density distribution of troughs/ridges, is slightly larger for the smallest apertures.}
	\label{fig:weighted_signal}
\end{figure*}

We combine all troughs (ridges) into a single negative (positive) shear signal using the weighting scheme described above. The optimally stacked GL-KiDS lensing profiles are shown in Fig. \ref{fig:weighted_signal}, for different aperture sizes $\theta_{\rm A}$.
The best-fit $A$ and $\lvert S/N \rvert$ of these troughs and ridges are shown in Table \ref{tab:results}. As a comparison, the table also shows the best-fit parameters for the fiducial G16 definition of troughs/ridges: the lowest/highest $20\%$ in density fraction ($P<0.2$/$P>0.8$). These show that, performing an optimally weighted stack of trough lensing profiles based on accurate mock predictions, we can obtain $S/N$ values that are on average $32\%$ higher than those of the fiducial stacks (while the average $S/N$ of the optimally stacked ridges is $7\%$ higher).

\begin{table*}
	\centering
	\caption{The best-fit amplitude $A$ and absolute signal-to-noise ratio $\lvert S/N \rvert$ of the shear profiles for troughs/ridges following the fiducial G16 definition ($P<0.2$/$P>0.8$), and those optimally stacked based on the SLICS mock $S/N$ as function of galaxy density percentile rank $P$. This demonstrates that, using the same KiDS dataset, the optimally stacked troughs have $S/N$ values that are on average $32\%$ higher than those from the fiducial stacks.}
	\label{tab:results}
	\begin{tabular}{lllllllll}
		\hline
		\multicolumn{1}{l}{ $\theta_{\rm A} \, [\am]$} & \multicolumn{2}{l}{{\bf Fiducial (G16)}: $\lvert S/N \vert$} & \multicolumn{2}{l}{ $A \, [10^{-3} \am^\frac{1}{2}]$ } & \multicolumn{2}{l}{{\bf Optimal stack}: $\lvert S/N \rvert$} & \multicolumn{2}{l}{ $A \, [10^{-3} \am^\frac{1}{2}]$ } \\ \hline
		
		\multicolumn{1}{l}{} & \multicolumn{1}{l}{$\,\,\,$ Troughs} & \multicolumn{1}{l}{Ridges} & \multicolumn{1}{l}{Troughs} & \multicolumn{1}{l}{Ridges} & \multicolumn{1}{l}{$\,\,\,$ Troughs} & \multicolumn{1}{l}{Ridges} & \multicolumn{1}{l}{Troughs} & \multicolumn{1}{l}{Ridges} \\ 
		
		\multicolumn{1}{l}{$5$} & \multicolumn{1}{l}{$\,\,\,\,\,\, 12.3$} & \multicolumn{1}{l}{$20.9$} & \multicolumn{1}{l}{$-2.25\pm0.18$} & \multicolumn{1}{l}{$4.00\pm0.19$} & \multicolumn{1}{l}{$\,\,\,\,\,\, 16.8$} & \multicolumn{1}{l}{$21.7$}  & \multicolumn{1}{l}{$-1.91\pm0.11$} & \multicolumn{1}{l}{$3.18\pm0.15$} \\ 
		
		\multicolumn{1}{l}{$10$} & \multicolumn{1}{l}{$\,\,\,\,\,\, 10.7$} & \multicolumn{1}{l}{$16.8$} & \multicolumn{1}{l}{$-2.81\pm0.26$} & \multicolumn{1}{l}{$4.59\pm0.27$} & \multicolumn{1}{l}{$\,\,\,\,\,\, 14.9$} & \multicolumn{1}{l}{$17.9$} & \multicolumn{1}{l}{$-2.18\pm0.15$} & \multicolumn{1}{l}{$3.66\pm0.20$} \\ 
		
		\multicolumn{1}{l}{$15$} & \multicolumn{1}{l}{$\,\,\,\,\,\, 8.41$} & \multicolumn{1}{l}{$11.5$} & \multicolumn{1}{l}{$-3.24\pm0.39$} & \multicolumn{1}{l}{$4.58\pm0.40$} & \multicolumn{1}{l}{$\,\,\,\,\,\, 10.1$} & \multicolumn{1}{l}{$12.3$}  & \multicolumn{1}{l}{$-2.62\pm0.23$} & \multicolumn{1}{l}{$3.55\pm0.29$} \\ 	
		
 		\multicolumn{1}{l}{$20$} & \multicolumn{1}{l}{$\,\,\,\,\,\, 5.73$} & \multicolumn{1}{l}{$9.01$} & \multicolumn{1}{l}{$-3.34\pm0.58$} & \multicolumn{1}{l}{$5.12\pm0.57$}& \multicolumn{1}{l}{$\,\,\,\,\,\, 7.55$} & \multicolumn{1}{l}{$9.94$}  & \multicolumn{1}{l}{$-3.18\pm0.35$} & \multicolumn{1}{l}{$4.09\pm0.41$} \\ \hline
	\end{tabular}
\end{table*}

To allow for easier visual comparison between the shape of trough and ridge profiles, we include the trough lensing signal with its sign flipped (i.e. $-\gamma_{\rm t}(\theta)$) in Fig. \ref{fig:weighted_signal}. We find that, for all aperture sizes, the shear resulting from ridges is stronger than that from troughs, which again indicates skewness in the total density distribution. Like G16, we observe that the fractional amplitude difference between troughs and ridges slightly decreases with aperture size. This can be explained by the fact that non-linearities affect the density field more strongly at smaller scales, as we derived earlier from Fig. \ref{fig:amplitudes_perc_delta}.

\section{Redshift evolution}
\label{sec:redshift}

So far we have studied troughs which extend across the entire redshift range of the GAMA galaxies ($0<z<0.5$). We can, however, define troughs that cover only a part of this range, and attempt to study the evolution of troughs and ridges over cosmic time. In this section we define the foreground galaxy and trough samples as a function of redshift and discuss the resulting lensing measurements. For the GAMA galaxies this selection is based on their spectroscopic redshifts, while for the GL-KiDS sample we use the photometric ANNz2 redshifts determined through machine learning (see Sect. \ref{sec:gamalike_kids} and \citealp{bilicki2017}).

\subsection{Redshift-dependent selection}
\label{sec:redshift_selection}

\begin{figure}
	\includegraphics[width=1.0\columnwidth]{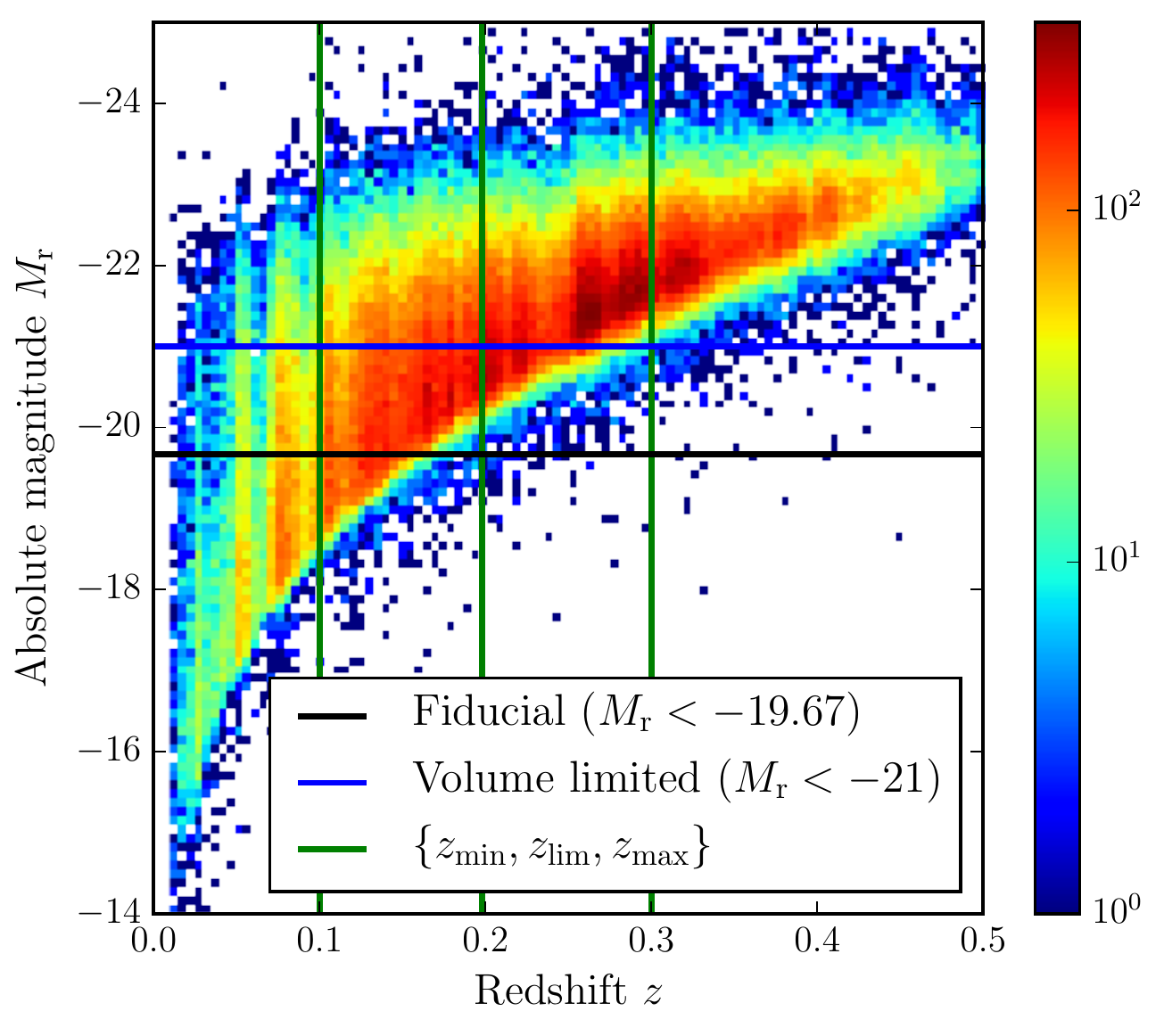}
	\caption{The distribution of GAMA galaxies as a function of redshift $z$ ($x$-axis) and absolute $r$-band magnitude $M_{\rm r}$ ($y$-axis). The color-scale indicates the number of galaxies in each pixel. The black line indicates the minimum $M_{\rm r}$ of the fiducial galaxy sample, while the blue line indicates the volume-limited sample, split into a high- and low-redshift sample by the green lines.}
	\label{fig:redshift_limit}
\end{figure}

To study the redshift evolution of troughs we create two foreground galaxy samples, a low- and a high-redshift sample, which are used to select the low- and high-redshift troughs. These two galaxy samples need to be physically similar to ensure that the troughs detected at different redshifts can be compared in a meaningful way. One requirement is that the two samples should consist of similar galaxy populations, since different kinds of galaxies might be subject to a varying amount of clustering. Another condition is that the galaxy samples should be complete in both redshift slices. In order to meet these two requirements, we define a volume-limited sample of galaxies by applying a cut in redshift: $0.1 < z < 0.3$, and in absolute $r$-band magnitude: $M_{\rm r} < -21 \magn$. Figure \ref{fig:redshift_limit} shows the distribution of GAMA galaxies as a function of redshift $z$ and absolute $r$-band magnitude $M_{\rm r}$, with coloured lines indicating the fiducial and volume-limited galaxy samples.

When defining troughs as a function of redshift, we also need to take into account their spatial shape. We choose to match the radial lengths and transverse radii of the troughs at different redshifts, such that their shapes describe (as much as possible) the same length scales. In addition, as the frusta are defined to have the same length and radius, their volumes are by construction also very similar. In the case of this work, the trough volumes of consecutive redshift bins are always equal within $\sim5\%$. In combination with the volume limited galaxy sample, this also ensures a similar galaxy count in each trough, leading to equal levels of shot noise at each redshift.

\begin{figure}
	\begin{center}
		\includegraphics[width=0.8\columnwidth]{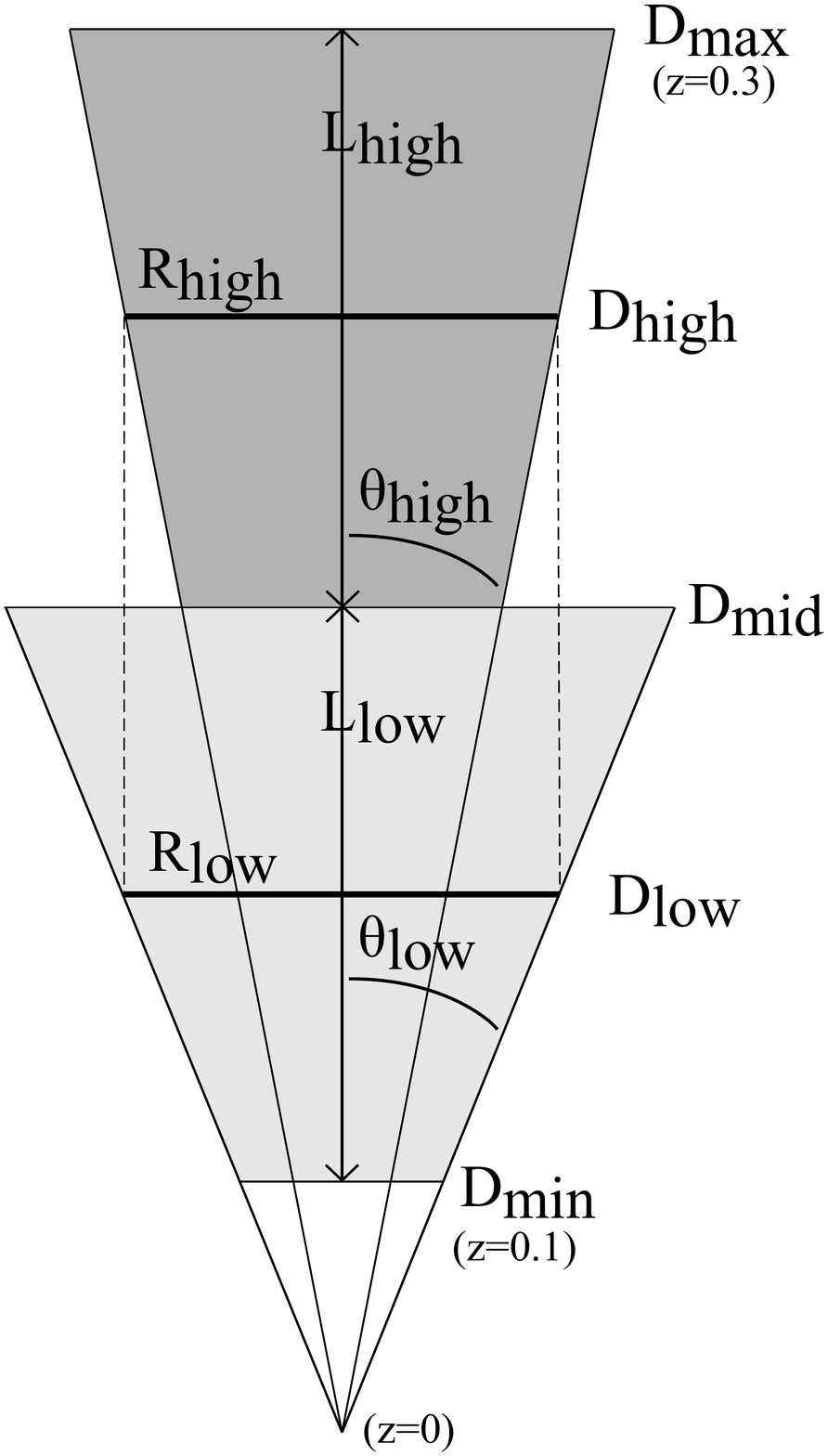}
		\caption{A visualization of the trough selection as a function of redshift. The two conical frusta used to define the low-redshift troughs (light grey) and the high-redshift troughs (dark grey) are separated at the comoving distance limit $D_{\rm mid}$. In order to select similar troughs at different redshifts, $D_{\rm mid}$ is chosen such that both volumes have the same comoving length: $L_{\rm low}=L_{\rm high}$. Moreover, the opening angles $\theta_{\rm low}$ and $\theta_{\rm high}$ of the cones are chosen such that the transverse comoving radius $R_{\rm low}$ ($R_{\rm high}$) at the mean comoving distance $D_{\rm low}$ ($D_{\rm high}$) of the low-/high-redshift galaxies are the same.}
		\label{fig:redshift_limit_sketch}
	\end{center}
\end{figure}

A visualization of the trough geometry is given by Fig. \ref{fig:redshift_limit_sketch}, which shows a cross section of the volumes that define the low- and high-redshift troughs. Inside these two conical frusta, the projected number density of the low-/high-redshift galaxy samples is measured in order to define the low-/high-redshift troughs. We split the redshift range at $z_{\rm mid}$, which corresponds to a comoving distance limit $D_{\rm mid}$. This limit is chosen in such a way that the comoving radial lenghts ($L_{\rm low}$ and $L_{\rm high}$ in Fig. \ref{fig:redshift_limit_sketch}) of the two volumes are equal, i.e.:
\begin{equation}
	D_{\rm mid} - D_{\rm min} = D_{\rm max} - D_{\rm mid} \, .
\end{equation}
For our chosen redshift range: $0.1 < z < 0.3$, and the corresponding comoving distances (see Table \ref{tab:samples}) we find that $z_{\rm mid} = 0.198$, very close to the `half-way' redshift of $0.2$. Of course $z_{\rm mid}$ depends on our chosen values for the cosmological parameters, but this effect would only cause a $\sim1\%$ difference in distance at these low redshifts (for reasonable values of the cosmological parameters).

In addition to having equal radial lengths, the cones need to have the same transverse radius. Selecting troughs to have equal \emph{physical} radii would cause a decrease in the galaxy density in troughs at lower redshifts (i.e. later cosmic times), due to the expansion of the Universe. Therefore, we select low- and high-redshift troughs that have the same \emph{comoving} radius, by choosing their opening angles $\theta_{\rm low}$ and $\theta_{\rm high}$ so that:
	\begin{equation}
	\theta_{\rm low} \, D_{\rm low} = \theta_{\rm high} \, D_{\rm high} \, .
	\end{equation}
Here $D_{\rm low}$ ($D_{\rm high}$) is defined as the mean comoving distance of the GAMA galaxies in the low- (high-)redshift sample.\footnote{We use the spectroscopic GAMA redshifts for this calculation to avoid any possible effects of photo-z scatter, but in principle the whole selection could be done using only KiDS photo-z's.} We find the mean distances: $D_{\rm low} = 653.5 \hsMpc$ and $D_{\rm high} = 1037 \hsMpc$. Choosing low-redshift radius $\theta_{\rm low} = 10 \am$, we find the corresponding high-redshift radius $\theta_{\rm high} = 6.3 \am$. This relatively small opening angle provides a high-$S/N$ shear signal, while still avoiding unreliable (i.e. noisy) density estimates resulting from the low number of galaxies inside smaller apertures (because $\theta_{\rm high}$ is larger than our smallest aperture, $\theta_{\rm A} = 5\am$, which has proved adequate in our results and those of G16). This choice corresponds to a transverse comoving size $R_{\rm A} = 1.9 \hsMpc$ of the troughs/ridges. The information on the low- and high-redshift galaxy samples is summarized in Table \ref{tab:samples}.

\begin{table*}
	\centering
	\caption{The names and sizes of the different trough definitions used in this work, including information on the galaxy samples used to select these troughs/ridges: the redshift range, the comoving distance range, and the absolute magnitude limits.}
	\label{tab:samples}
	\begin{tabular}{lllll}
		\hline
		Troughs/Galaxies & Trough radius $\theta_{\rm A}$ & Redshift range & Distance $[\hsMpc]$ & $M_{\rm r}$-limit $[\magn]$ \\ 
		\hline
		Fiducial & $5,10,15,20\am$ & $0<z<0.5$ & $0<D_{\rm c}<1922.5$ & $<-19.67$  \\
		Low-redshift & $10\am \, (1.9\hsMpc)$ & $0.1<z<0.198$ & $420.0<D_{\rm c}<813.9$ & $<-21.0$  \\ 
		High-redshift & $6.3\am \, (1.9\hsMpc)$ & $0.198<z<0.3$ & $813.9<D_{\rm c}<1207.7$ & $<-21.0$ \\ \hline
	\end{tabular}
\end{table*}

\subsection{Excess surface density measurements}
\label{sec:esd_measurements}

For lenses at a given redshift $z_{\rm l}$, the measured shear depends on the distance between the lens, the source and the observer. In order to take this effect into account, we convert the shear profile $\gamma_{\rm t}(\theta)$ to the physical excess surface density (ESD) profile $\Delta\Sigma(R_{\rm p})$ as a function of the transverse physical separation $R_{\rm p}$. The ESD is defined as the surface mass density $\Sigma (R_{\rm p})$, subtracted from the mean surface density $\mean{\Sigma}(<R_{\rm p})$ within that radius:
\begin{equation}
\Delta \Sigma (R_{\rm p}) = \mean{\Sigma}(<R_{\rm p}) - \Sigma (R_{\rm p}) = \Sigma_{\rm crit} \gamma_{\rm t}(R_{\rm p}) \, .
\label{eq:deltasigma}
\end{equation}
The conversion factor between the shear and the physical ESD is the critical surface density 
$\Sigma_{\rm crit}$.\footnote{We note that within the literature different conventions are used to define $\Sigma_{\rm crit}$. In this work we use the `proper' critical surface mass density, in contrast to a co-moving convention, refering the reader to Appendix C of \cite{dvornik2018} for a full discussion.} It depends on the angular diameter distance from the observer to the lens $D(z_{\rm l})$, to the source $D(z_{\rm s})$, and between the lens and the source $D(z_{\rm l}, z_{\rm s})$, as follows:
	\begin{equation}
	\Sigma_{\rm crit,ls}^{-1} = \frac{4\pi G}{c^2} D(z_{\rm l}) \int_{z_{\rm l}}^{\infty} \frac{D(z_{\rm l}, z_{\rm s})}{D(z_{\rm s})} n(z_{\rm s}) \, {\rm d}z_{\rm s} \, .
	\label{eq:sigmacrit1}
	\end{equation}
Here $c$ denotes the speed of light and $G$ the gravitational constant. As the lens redshifts $z_{\rm l}$ of the low-/high-redshift troughs, we use the mean redshift of the low-/high-redshift galaxy sample which is used to define the troughs. Based on their best-fit photometric redshifts $z_{\rm B}$ we limit the sample of sources whose shear contributes to the lensing signal to those situated behind the lens, including a redshift buffer $\Delta z=0.2$, such that: $z_{\rm B} > z_{\rm l} + \Delta z$. This same photometric redshift limit is applied to a galaxy catalogue that also includes spectroscopic redshift information, and has been weighted to reproduce the galaxy colour-distribution of KiDS \citep[]{hildebrandt2017}. The spectroscopic source redshifts $n_{\rm s}$ remaining after this $z_{\rm B}$ cut determine the source redshift distribution $n(z_{\rm s})$ at each lens redshift. We calculate $\Sigma_{\rm crit}$ by integrating over the part of $n(z_{\rm s})$ situated behind the lens, following the method described in Sect. 4.2 of \cite{dvornik2018}.

Since lenses with a higher lensing efficiency ($=\Sigma_{\rm crit}^{-1}$) produce a stronger shear, we give them more weight in the combined ESD measurement. We incorporate $\Sigma_{\rm crit}$ into the total weight:
	\begin{equation}
	W_{\rm{ls}} = w_{\rm s} \left( \Sigma_{\rm crit,ls}^{-1} \right)^2 \, ,
	\label{eq:weights}
	\end{equation}
which is used to calculate our combined ESD measurement as follows:
	\begin{equation}
	\Delta\Sigma = \frac{1}{1+\mu} \frac{\sum_{ls} W_{ls} \, \epsilon_{{\rm t},ls} \, \Sigma_{{\rm crit},ls} }{ \sum_{ls}{W_{ls}} }  \, .
	\label{eq:ESDmeasured}
	\end{equation}
The correction for the multiplicative bias is weighted by the same total weight.

The angular separation range $2 < \theta < 100 \am$, used to measure the shear profiles in Sect. \ref{sec:results}, corresponds to a transverse physical separation of $0.44 < R_{\rm p} < 22.24 \hsMpc$ at the mean angular diameter distance of the fiducial GAMA sample (see Table \ref{tab:samples}). We therefore measure the ESD profiles of the low-/high-redshift troughs for $10$ logarithmically spaced bins within $0.5 < R_{\rm p} < 20 \hsMpc$. The reason we use only half the number of angular bins, is that splitting the tracer galaxies as function of redshift results in trough profiles with a lower $S/N$. Although it is customary to use physical distances to measure the ESD profile around galaxies and other bound structures, the trough lensing measurements need to take the expansion of the Universe into account. We therefore translate our physical $\Delta\Sigma(R_{\rm p})$ profiles into the comoving surface density $\Delta\Sigma^{\rm c}(R)$ as a function of comoving radius $R$, by dividing each measured $\Delta\Sigma$ by $(1+z_{\rm l})^2$, and multiplying each $R_{\rm p}$ with $(1+z_{\rm l})$.

\subsection{Results}
\label{sec:redshift_results}

\begin{figure*}
	\includegraphics[width=1.0\columnwidth]{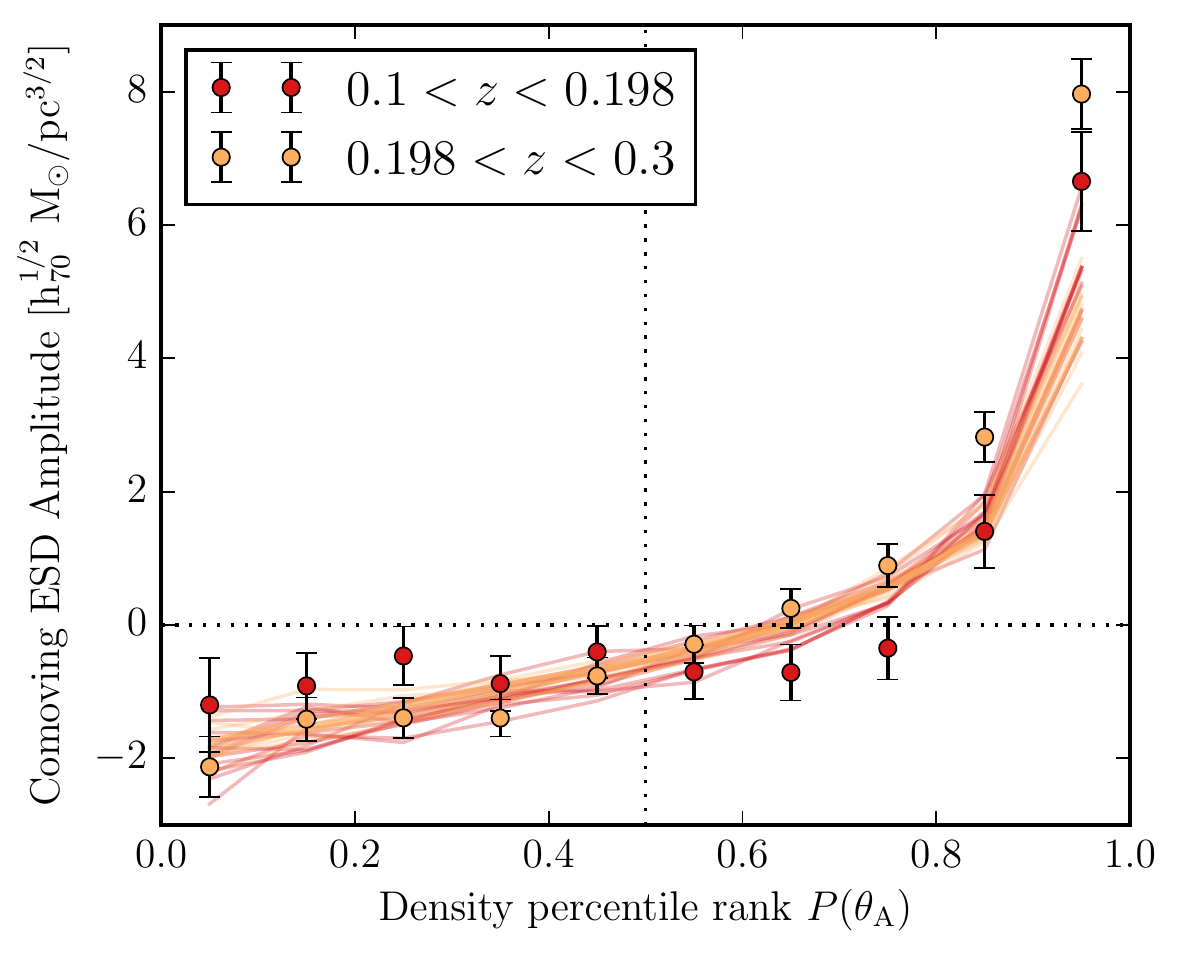}
	\includegraphics[width=1.0\columnwidth]{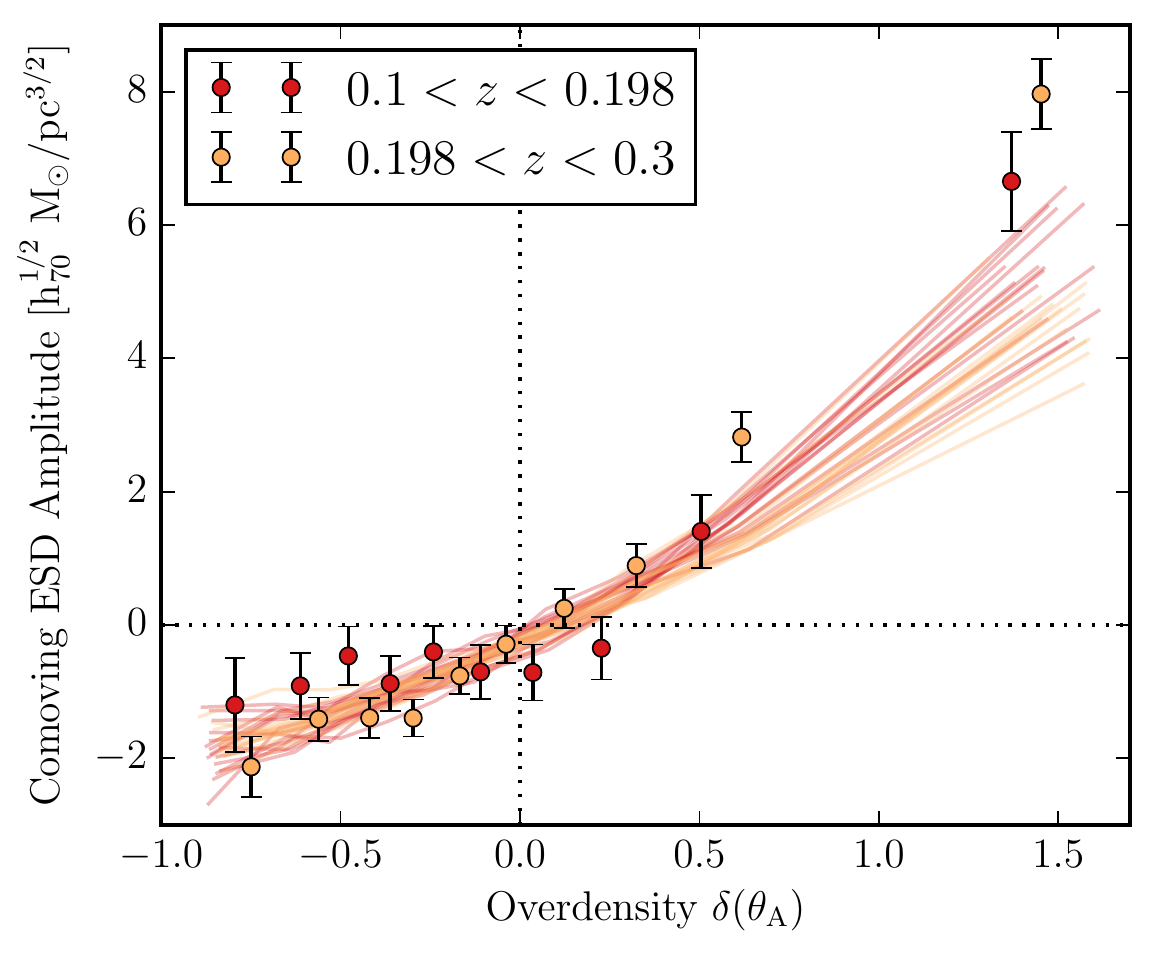}
	\caption{The amplitude $A'$ of the comoving ESD profiles as a function of the galaxy density percentile rank $P$ (left) and galaxy overdensity $\delta$ (right), for troughs and ridges with comoving radius $R_{\rm A} = 1.9 \hsMpc$, selected at two different redshifts. The observed amplitudes from KiDS (dots with $1\sigma$ errors) are in reasonable agreement with those from 16 independent patches of the MICE mocks (solid lines). For the ridges ($\delta>0$) in the MICE mocks, the amplitude is slightly higher at low redshifts. This effect, however, is not found in the observations, where we find no significant physical difference between the observed amplitudes at low and high redshifts.}
	\label{fig:amplitudes_perc_delta_redshifts}
\end{figure*}

We measure the comoving ESD profiles of the troughs/ridges selected at different redshifts, and apply the same method of amplitude fitting as discussed in Sect. \ref{sec:kids_vs_gama} and \ref{sec:amplitudes} to both the KiDS and MICE data.  Similar to Eq. \ref{eq:trough_fit}, we define a fitting function for the comoving ESD profiles:
	\begin{equation}
	\Delta\Sigma^{\rm c}(R) = A' / \sqrt{R} \, ,
	\label{eq:ESD_fit}
	\end{equation}
where $A'$ is now the comoving ESD amplitude. The comoving transverse fitting range is $1.2 \, R_{\rm A} < R < 20 \hsMpc$, where the maximum is based on the transverse comoving separation corresponding to $\theta < 70 \am$ (see Sect. \ref{sec:kids_vs_gama}) at the mean distance of the fiducial GAMA sample. In the left panel of Fig. \ref{fig:amplitudes_perc_delta_redshifts} we again show the best-fit amplitude $A'$ (including $1\sigma$ error bars derived from the full analytical covariance matrix) as a function of $P$, this time for the low- and high-redshift troughs/ridges. For both the high- and low-redshift sample the shape of the $A'(P)$ relation resembles that of the fiducial sample: rising gradually from negative $A'$ at low $P$, crossing the turn-over to positive $A'$ at $P\approx0.6$, and peaking at $P=1$. The observed relation is in reasonable agreement with the prediction from 16 independent patches of the MICE mocks. We show the same $A'$ as a function of the galaxy overdensity $\delta$ in the right panel of Fig. \ref{fig:amplitudes_perc_delta_redshifts}. As for the fiducial troughs, the $A'(\delta)$-relation of both trough samples is approximately linear, and crosses to positive $A'$ at the mean density ($\delta\approx0$) for both GL-KiDS and MICE.

Based on these amplitudes, we aim to assess whether there is a significant difference between the measurements of the low- and high-redshift troughs/ridges. This difference is best visible in the $A'(\delta)$-relation (right panel of Fig. \ref{fig:amplitudes_perc_delta_redshifts}), where we see that the amplitudes of the low-redshift ridges ($\delta>0$) in the MICE mocks are slightly higher than those of the high-redshift ridges. This is expected, since the clustering of mass increases the height of ridges (and the depth of troughs) at later cosmic times.
The difference between the mock redshift samples, however, is not significant compared to the large sample variance, indicated by the wide spread in the amplitudes from the 16 MICE patches. Moreover, the trend is not reflected in the amplitudes measured using KiDS, where in fact we see a hint of the opposite effect. We verify that this is in agreement with the results based on GAMA galaxies. This effect is likely not physical, and within the error bars the data is consistent with a null-measurement. Based on this result, we conclude that we find no significant difference between the observed trough and ridge amplitudes at different redshifts, and that more accurate data at higher redshifts will be required to observe trough/ridge evolution.

\subsection{Predictions for higher redshifts}
\label{sec:higher_redshifts}

The physical interpretation of the MICE mock results in Fig. \ref{fig:amplitudes_perc_delta_redshifts} would be that the total density of ridges increases with cosmic time. This is expected, since overdensities in the cosmic structure cluster over cosmic time, forming higher ridges. Since this mass is accreted from more underdense regions, these are expected to form deeper troughs. As we showed in Sect. \ref{sec:esd_measurements}, current data are unable to resolve this effect over the redshift range $0.1 < z < 0.3$. In order to obtain a more solid interpretation of our results, we study the predictions from both the MICE-GC and SLICS mocks at higher redshifts. Our goal is to predict whether the redshift evolution of troughs would be measurable using future high-redshift lensing surveys such as Euclid \cite[]{laureijs2011} and LSST \cite[]{lsst2012}. In particular, the 349 realisations of the SLICS simulation allow us to estimate the uncertainties on the redshift-dependent trough/ridge amplitudes obtained using such a survey.

To define our mock galaxy sample we use the same absolute magnitude limit: $M_{\rm r} < -21 \magn$, but abandon the cut in apparent magnitude such that the sample is complete at every redshift. Using these MICE and SLICS samples we perform the same redshift-dependent trough selection as described in Sect. \ref{sec:redshift_selection}. But instead of splitting galaxies into two redshift bins between $0.1 < z < 0.3$, we split the SLICS galaxies into four bins between $0.1 < z < 0.5$ and the MICE galaxies into five bins between $0.1 < z < 0.6$. These redshift slices of equal comoving length have the following redshift limits: $z_{\rm mid} = \{0.1, 0.192, 0.289, 0.391, 0.5\}$ for SLICS and $\{0.1, 0.191, 0.286, 0.385, 0.489, 0.6\}$ for MICE. As in Sect. \ref{sec:redshift_selection} we wish to select the opening angles $\theta_{\rm A}$ corresponding to these redshifts, such that the comoving radii of the apertures are the same and none of the angles is smaller than $5 \am$. The chosen opening angles for the SLICS mocks, $\theta_{\rm A} = \{ 15.0, 9.554, 7.283, 5.770 \}$, correspond to the same transverse comoving separation $R_{\rm A} = 2.775 \hsMpc$ at the mean GAMA galaxy distance in each redshift bin (calculated using the SLICS cosmological parameters, see Sect. \ref{sec:slics_mocks}). For MICE, which extends to slightly higher redshifts, we choose larger opening angles: $\theta_{\rm A} = \{ 20.0, 12.85, 9.45, 7.44, 6.14 \}$, which all correspond to comoving separation $R_{\rm A} = 3.712 \hsMpc$ at the respective mean MICE galaxy distances.

\begin{figure*}
	\includegraphics[width=1.0\columnwidth]{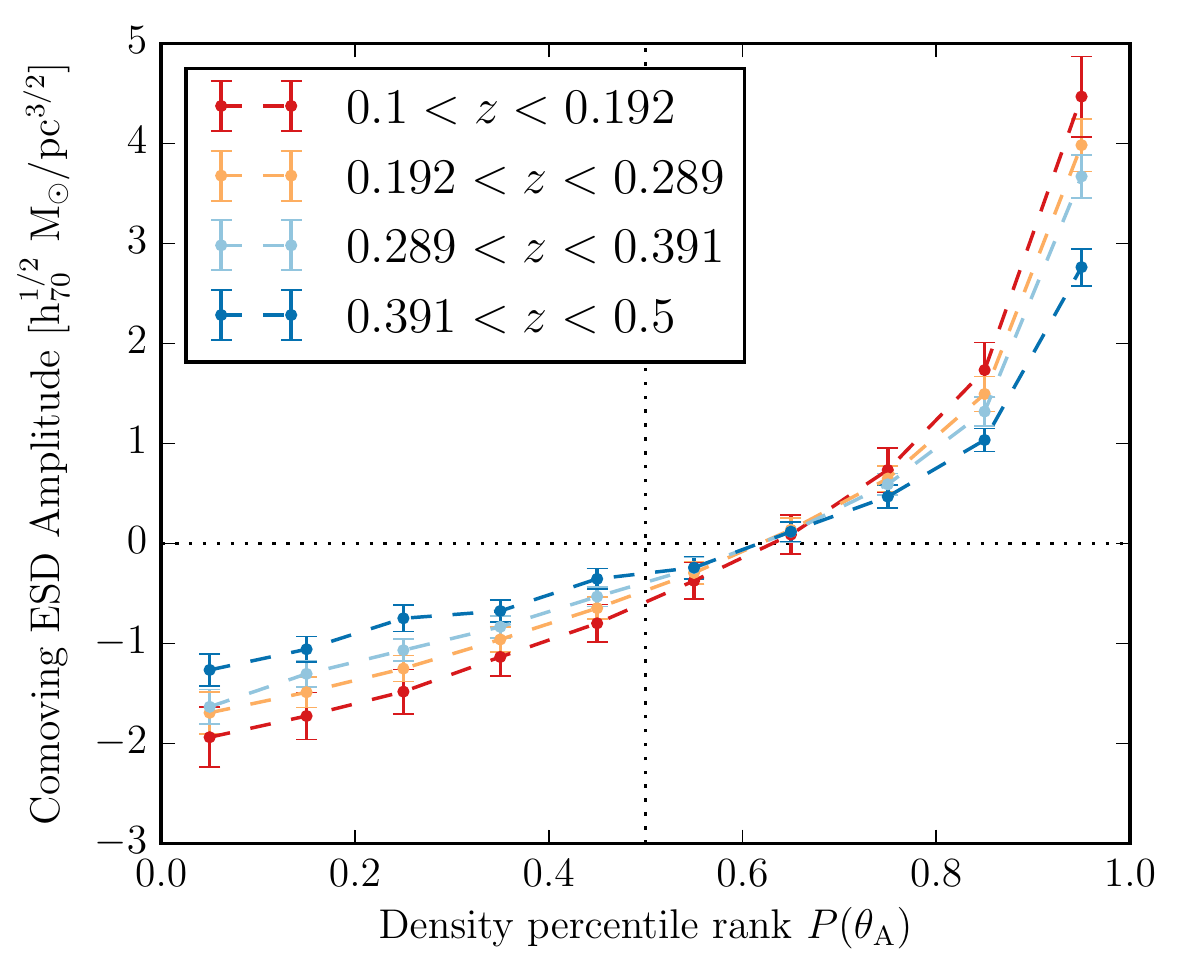}
	\includegraphics[width=1.0\columnwidth]{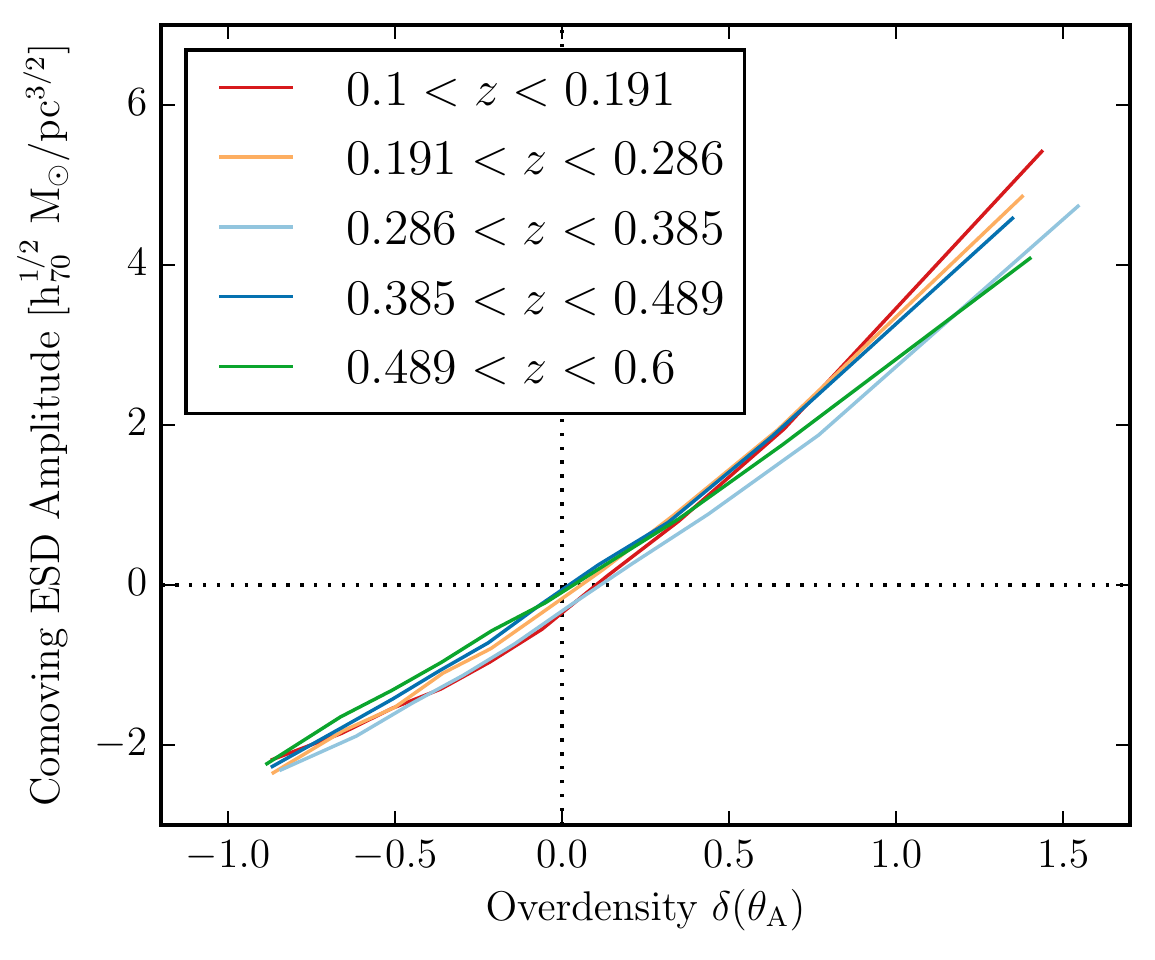}
	\caption{The amplitude $A'$ of the comoving ESD profiles from the SLICS mocks (including $1\sigma$ error estimates for a Euclid-like survey) as a function of the galaxy density percentile rank $P$ (left), and from the MICE mocks as a function of galaxy overdensity $\delta$ (right). The troughs and ridges are selected in different redshift bins. For both troughs and ridges the redshift evolution, that was hinted at by the results at low redshifts, appears to be a continuing trend as the covered range of $A'$ steadily decreases with redshift. This is expected from the clustering of mass with cosmic time, which causes massive ridges to accrete mass from the low-density troughs. Based on these mock results, we predict that future surveys like Euclid and LSST should be able to observe the evolution of troughs and ridges with cosmic time.}
	\label{fig:amplitudes_perc_delta_miceZ}
\end{figure*}

We perform the same measurement of the comoving ESD profiles in the different redshift bins, and fit Eq. \ref{eq:ESD_fit} to the results. In the left panel of Fig. \ref{fig:amplitudes_perc_delta_miceZ} we show the best-fit comoving amplitude $A'$ as a function of $P$ for the SLICS troughs/ridges in five redshift bins. The (tiny) error bars are estimated using the SLICS covariance matrix, this time multiplied by the area factor $f_{\rm Euclid} = \frac{100}{15\,000}$ in order to emulate the $15 \,000 \deg^2$ area that the Euclid satellite aims to observe. It is clear that the difference that was barely visible in Fig. \ref{fig:amplitudes_perc_delta_redshifts} has become a significant trend: as the redshift increases to $z=0.5$, the absolute amplitudes decrease. In order to predict the significance of such a future observation, we calculate the $\chi^2$ between the amplitude differences and a null result. Using the covariance estimate for Euclid, this calculation gives $\chi^2 \gtrsim 73$ for the difference between each of the consecutive redshift bins. Since the $A'(P)$ measurements consist of $N=10$ data-points (corresponding to a Cumulative Distribution Function with $10-1=9$ degrees of freedom) this $\chi^2$ corresponds to a standard deviation $\gtrsim 7 \sigma$. In conclusion, this study of the SLICS mocks suggests that next-generation high-redshift surveys, such as Euclid and LSST, should be able to constrain trough/ridge evolution with a very high significance.

As an additional comparison we show the best-fit amplitude $A'$ as a function of galaxy overdensity $\delta$, this time for the MICE troughs/ridges in six redshift bins, in the right panel of Fig. \ref{fig:amplitudes_perc_delta_miceZ}. The evolution of MICE mock amplitudes with redshift is less pronounced than in the SLICS mocks. This can be explained by the different cosmologies used by the two simulations (as discussed in Sect. \ref{sec:amplitudes}), where the higher values of $\Omega_{\rm m}$ and $\sigma_8$ in the SLICS simulations result in stronger structure evolution. Nevertheless, the amplitudes of the ridges clearly decrease with redshift. This effect is even slightly visible for the troughs where, except for the third redshift bin ($0.286<z<0.385$), the absolute amplitude continues to decrease with $z$. But while the comoving ESD amplitude range spanned by the troughs/ridges increases with cosmic time, the span of the galaxy overdensity remains constant, possibly signifying non-linear galaxy bias.

\section{Discussion and conclusion}
\label{sec:discon}

We used the Kilo-Degree Survey (KiDS) to perform a weak gravitational lensing study of troughs: circular projected underdensities in the cosmic galaxy density field, following up on the work by \citet[][G16]{gruen2016} who used the Dark Energy Survey (DES). We defined the troughs using two different foreground galaxy samples: $159\,519$ galaxies from the equatorial fields of the Galaxy and Mass Assembly survey (GAMA), and a sample of $309\,021$ `GAMA-like' (GL) KiDS galaxies that was limited to photometric redshift $z_{\rm ANN} < 0.5$ and apparent magnitude $m_{\rm r} < 20.2 \magn$ in order to mimic the GAMA selection. Both galaxy samples were limited to an absolute magnitude $M_{\rm r} < -19.67 \magn$ in order to mimic the mock galaxy sample from the MICE Grand Challenge (MICE-GC) lightcone simulation, which was used to interpret our results. Following the fiducial trough definition of G16 (apertures with a galaxy density percentile rank $P(\theta_{\rm A}) < 0.2$), we detected a gravitational lensing signal with an absolute signal-to-noise ratio ($S/N$) of $\lvert S/N \rvert = 12.3$ for the KiDS foreground sample and $12.0$ for GAMA. Since the currently available KiDS area already provided a more significant trough lensing detection than the GAMA survey, we mainly used the GL-KiDS galaxies for this work (although we confirmed all our results using GAMA). As the KiDS survey progresses in the coming years, the available area will become larger and less irregular. The coming KiDS data release, which aims to make a contiguous area of $\sim900 \deg^2$ available for lensing studies, will likely reduce the systematic lensing effects found at large scales and increase the detection significance of the trough signal (by a factor of at most $\sim\sqrt{900/180} = 2.24$ compared to GAMA).

In addition to stacking only the most underdense/overdense $20\%$ of the apertures, we studied troughs and ridges (overdensities) as a function of their galaxy number density $n_{\rm g}$. By fitting the simple function $\gamma_{\rm t}(\theta) = A/\sqrt{\theta}$ to the lensing signal in bins of increasing $n_{\rm g}$, we obtained the amplitude $A$ of troughs and ridges as a function of galaxy density percentile rank $P$ and galaxy overdensity $\delta$. We discovered that the crossing point between negative and positive $A$ was situated at $P\approx0.6$ (and not at the median density $P=0.5$), while $A(\delta)$ did generally pass through the origin (the mean density $\delta=0$). This indicated that the non-linearities in the density field caused by structure formation, which were shown by the skewed distribution of $n_{\rm g}$ (see Fig. \ref{fig:trough_hist}), were reflected in the total (baryonic + dark matter) density distribution measured by gravitational lensing. As expected, these non-linearities were more prominent on smaller scales, i.e. for smaller trough radii. This conclusion is supported by mock trough profiles obtained from the MICE-GC lightcone simulation, which showed exactly the same trend.

The mock catalogue based on the Scinet LIghtCone Simulations (SLICS) was used to estimate $S/N$ of the trough/ridge lensing signals as a function of $P$, which we applied as a weight to optimally stack the shear profiles. On average, the optimally stacked signals had a $32\%$ higher $S/N$ compared to those of the fiducial trough definition (see Table \ref{tab:results}). Inspecting the optimally stacked trough and ridge profiles showed that the shear profiles of ridges are much stronger than those of troughs, especially for the smallest trough radii. This finding, which is in agreement with the results from G16, again revealed the skewness of the total mass density distribution.

In addition, a comparison of both mocks with the KiDS observations showed a higher lensing strength for the SLICS troughs/ridges compared to the KiDS and MICE results. In combination with the increased values of the matter density $\Omega_{\rm m}$ and power spectrum amplitude $\sigma_8$ of SLICS compared to KiDS and MICE, this indicates that trough measurements are sensitive to these cosmological parameters. This confirms the potential of troughs/ridges as a possible probe for measuring $\Omega_{\rm m}$ and $\sigma_8$, as was demonstrated by \cite{gruen2017}.

Finally, we attempted to observe physical evolution of the density field by performing the trough selection in two redshift bins. We created a volume-limited sample of foreground galaxies ($z<0.3$ and $M_{\rm r} < -21 \magn$), and split it into a low- ($0.1 < z < 0.198$) and high- ($0.198 < z < 0.3$) redshift sample of equal comoving length. By adjusting the opening angle $\theta_{\rm high}$ of the high-redshift apertures, we ensured that the transverse comoving radii of the troughs were identical at both redshifts: $R_{\rm A} = 1.9 \hsMpc$. The measured comoving excess surface density (ESD) profiles of the troughs/ridges did not reveal a significant physical evolution of the comoving trough/ridge amplitudes $A'$ as a function of $P$ and $\delta$. Applying the same method to 16 independent patches of the MICE-GC mock catalogue provided a reasonable agreement with the observation, although the decrease in the lensing amplitude of ridges with redshift that was seen in the mocks could not be distinguished with our data. This increase in ridge height with cosmic time is expected from the effects of clustering.

This raised the question whether this trend would continue at higher redshifts, and whether the effects of clustering could also be observed in troughs. We therefore used the SLICS and MICE mock catalogues to gain more insight into our finding, by extending our measurement to four redshift bins between $0.1 < z < 0.5$ for SLICS, and to five redshift bins between $0.1 < z < 0.6$ for MICE. The comoving ESD amplitude of the mock ridges continued to decrease with redshift, indicating that the increasing ridge height with cosmic time is an actual trend. In the mock measurements at high redshifts, we could even distinguish the corresponding deepening of troughs with cosmic time. We used $349$ realisations of the SLICS simulations to estimate the uncertainties on these measurements when performed with future surveys. Based on the SLICS simulations we predicted that large upcoming surveys like Euclid and LSST should be able to observationally constrain the redshift evolution of troughs and ridges with very high significance ($\gtrsim 7 \sigma$ between every consecutive redshift bin), thereby potentially providing a simple, practical way to trace the growth of large scale structure.

\section*{Acknowledgements}

We would like to thank the referee, Daniel Gruen, for constructive comments that have helped to improve this work.

V. Demchenko acknowledges the Higgs Centre Nimmo Scholarship and the Edinburgh Global Research Scholarship. J. Harnois-D{\'e}raps is supported by the European Commission under a Marie-Sk{\l}odowska-Curie European Fellowship (EU project 656869). M. Bilicki is supported by the Netherlands Organization for Scientific Research, NWO, through grant number 614.001.451. C. Heymans acknowledges support from the European Research Council under grant number 647112. H. Hoekstra acknowledges support from Vici grant 639.043.512, financed by the Netherlands Organization for Scientific Research. K. Kuijken acknowledges support by the Alexander von Humboldt Foundation. H. Hildebrandt is supported by an Emmy Noether grant (No. Hi 1495/2-1) of the Deutsche Forschungsgemeinschaft. P. Schneider is supported by the Deutsche Forschungsgemeinschaft in the framework of the TR33 `The Dark Universe'. E. van Uitert acknowledges support from an STFC Ernest Rutherford Research Grant, grant reference ST/L00285X/1.

Computations for the $N$-body simulations were performed in part on the Orcinus supercomputer at the WestGrid HPC consortium (\url{www.westgrid.ca}), in part on the GPC supercomputer at the SciNet HPC Consortium. SciNet is funded by: the Canada Foundation for Innovation under the auspices of Compute Canada; the Government of Ontario; Ontario Research Fund - Research Excellence; and the University of Toronto.

This research is based on data products from observations made with ESO Telescopes at the La Silla Paranal Observatory under programme IDs 177.A-3016, 177.A-3017 and 177.A-3018, and on data products produced by Target OmegaCEN, INAF-OACN, INAF-OAPD and the KiDS production team,  on behalf of the KiDS consortium. OmegaCEN and the KiDS production team acknowledge support by NOVA and NWO-M grants. Members of INAF-OAPD and INAF-OACN also acknowledge the support from the Department of Physics \& Astronomy of the University of Padova, and of the Department of Physics of Univ. Federico II (Naples).

GAMA is a joint European-Australasian project based around a spectroscopic campaign using the Anglo-Australian Telescope. The GAMA input catalogue is based on data taken from the Sloan Digital Sky Survey and the UKIRT Infrared Deep Sky Survey. Complementary imaging of the GAMA regions is being obtained by a number of independent survey programs including GALEX MIS, VST KiDS, VISTA VIKING, WISE, Herschel-ATLAS, GMRT and ASKAP providing UV to radio coverage. GAMA is funded by the STFC (UK), the ARC (Australia), the AAO, and the participating institutions. The GAMA website is \url{www.gama-survey.org}.

This work has made use of CosmoHub \cite[]{carretero2017}. CosmoHub has been developed by the Port d'Informaci{\'o}n Cient{\'i}fica (PIC), maintained through a collaboration of the Institut de F{\'i}sica d'Altes Energies (IFAE) and the Centro de Investigaciones Energ{\'e}ticas, Medioambientales y Tecnol{\'o}gicas (CIEMAT), and was partially funded by the ``Plan Estatal de Investigaci{\'o}n Cient{\'i}fica y T{\'e}cnica y de Innovaci{\'o}n'' program of the Spanish government.

This work has made use of {\scshape python} (\url{www.python.org}), including the packages {\scshape numpy} (\url{www.numpy.org}) and {\scshape scipy} (\url{www.scipy.org}). Plots have been produced with {\scshape matplotlib} \cite[]{hunter2007matplotlib}. The mock shear profiles from MICE are computed using {\scshape TreeCorr} (\url{https://pypi.python.org/pypi/TreeCorr}).

\emph{Author contributions:} All authors contributed to the development and writing of this paper. The authorship list is given in three groups: the lead authors (M. Brouwer, V. Demchenko, J. Harnois-D{\'e}raps), followed by two alphabetical groups. The first alphabetical group includes those who are key contributors to both the scientific analysis and the data products. The second group covers those who have either made a significant contribution to the data products, or to the scientific analysis.

%%%%%%%%%%%%%%%%%%%%%%%%%%%%%%%%%%%%%%%%%%%%%%%%%%



\bibliographystyle{mnras}
\bibliography{biblio}

\newcommand{\SortNoop}[1]{}
\begin{thebibliography}{}
\makeatletter
\relax
\def\mn@urlcharsother{\let\do\@makeother \do\$\do\&\do\#\do\^\do\_\do\%\do\~}
\def\mn@doi{\begingroup\mn@urlcharsother \@ifnextchar [ {\mn@doi@}
  {\mn@doi@[]}}
\def\mn@doi@[#1]#2{\def\@tempa{#1}\ifx\@tempa\@empty \href
  {http://dx.doi.org/#2} {doi:#2}\else \href {http://dx.doi.org/#2} {#1}\fi
  \endgroup}
\def\mn@eprint#1#2{\mn@eprint@#1:#2::\@nil}
\def\mn@eprint@arXiv#1{\href {http://arxiv.org/abs/#1} {{\tt arXiv:#1}}}
\def\mn@eprint@dblp#1{\href {http://dblp.uni-trier.de/rec/bibtex/#1.xml}
  {dblp:#1}}
\def\mn@eprint@#1:#2:#3:#4\@nil{\def\@tempa {#1}\def\@tempb {#2}\def\@tempc
  {#3}\ifx \@tempc \@empty \let \@tempc \@tempb \let \@tempb \@tempa \fi \ifx
  \@tempb \@empty \def\@tempb {arXiv}\fi \@ifundefined
  {mn@eprint@\@tempb}{\@tempb:\@tempc}{\expandafter \expandafter \csname
  mn@eprint@\@tempb\endcsname \expandafter{\@tempc}}}

\bibitem[\protect\citeauthoryear{{Abazajian} et~al.,}{{Abazajian}
  et~al.}{2009}]{abazajian2009}
{Abazajian} K.~N.,  et~al., 2009, \mn@doi [\apjs]
  {10.1088/0067-0049/182/2/543}, \href
  {http://adsabs.harvard.edu/abs/2009ApJS..182..543A} {182, 543}

\bibitem[\protect\citeauthoryear{{Alpaslan} et~al.,}{{Alpaslan}
  et~al.}{2014}]{alpaslan2014}
{Alpaslan} M.,  et~al., 2014, \mn@doi [\mnras] {10.1093/mnrasl/slu019}, \href
  {http://adsabs.harvard.edu/abs/2014MNRAS.440L.106A} {440, L106}

\bibitem[\protect\citeauthoryear{{Barreira}, {Bose}, {Li}  \&
  {Llinares}}{{Barreira} et~al.}{2017}]{barreira2017}
{Barreira} A.,  {Bose} S.,  {Li} B.,   {Llinares} C.,  2017, \mn@doi [\jcap]
  {10.1088/1475-7516/2017/02/031}, \href
  {http://adsabs.harvard.edu/abs/2017JCAP...02..031B} {2, 031}

\bibitem[\protect\citeauthoryear{{Bartelmann} \& {Schneider}}{{Bartelmann} \&
  {Schneider}}{2001}]{bartelmann2001}
{Bartelmann} M.,  {Schneider} P.,  2001, \mn@doi [\physrep]
  {10.1016/S0370-1573(00)00082-X}, \href
  {http://adsabs.harvard.edu/abs/2001PhR...340..291B} {340, 291}

\bibitem[\protect\citeauthoryear{{Beare}, {Brown}  \& {Pimbblet}}{{Beare}
  et~al.}{2014}]{beare2014}
{Beare} R.,  {Brown} M.~J.~I.,   {Pimbblet} K.,  2014, \mn@doi [\apj]
  {10.1088/0004-637X/797/2/104}, \href
  {http://adsabs.harvard.edu/abs/2014ApJ...797..104B} {797, 104}

\bibitem[\protect\citeauthoryear{{Ben{\'{\i}}tez}}{{Ben{\'{\i}}tez}}{2000}]{benitez2000}
{Ben{\'{\i}}tez} N.,  2000, \mn@doi [\apj] {10.1086/308947}, \href
  {http://adsabs.harvard.edu/abs/2000ApJ...536..571B} {536, 571}

\bibitem[\protect\citeauthoryear{{Benson}, {Cole}, {Frenk}, {Baugh}  \&
  {Lacey}}{{Benson} et~al.}{2000}]{benson2000}
{Benson} A.~J.,  {Cole} S.,  {Frenk} C.~S.,  {Baugh} C.~M.,   {Lacey} C.~G.,
  2000, \mn@doi [\mnras] {10.1046/j.1365-8711.2000.03101.x}, \href
  {http://adsabs.harvard.edu/abs/2000MNRAS.311..793B} {311, 793}

\bibitem[\protect\citeauthoryear{{Bertin} \& {Arnouts}}{{Bertin} \&
  {Arnouts}}{1996}]{bertin1996}
{Bertin} E.,  {Arnouts} S.,  1996, \mn@doi [\aaps] {10.1051/aas:1996164}, \href
  {http://adsabs.harvard.edu/abs/1996A%26AS..117..393B} {117, 393}

\bibitem[\protect\citeauthoryear{{Bilicki} et~al.,}{{Bilicki}
  et~al.}{2017}]{bilicki2017}
{Bilicki} M.,  et~al., 2017, preprint, \href
  {http://adsabs.harvard.edu/abs/2017arXiv170904205B} {} (\mn@eprint {arXiv}
  {1709.04205})

\bibitem[\protect\citeauthoryear{{Blazek}, {Mandelbaum}, {Seljak}  \&
  {Nakajima}}{{Blazek} et~al.}{2012}]{blazek2012}
{Blazek} J.,  {Mandelbaum} R.,  {Seljak} U.,   {Nakajima} R.,  2012, \mn@doi
  [\jcap] {10.1088/1475-7516/2012/05/041}, \href
  {http://adsabs.harvard.edu/abs/2012JCAP...05..041B} {5, 041}

\bibitem[\protect\citeauthoryear{{Bos}, {van de Weygaert}, {Dolag}  \&
  {Pettorino}}{{Bos} et~al.}{2012}]{bos2012}
{Bos} E.~G.~P.,  {van de Weygaert} R.,  {Dolag} K.,   {Pettorino} V.,  2012,
  \mn@doi [\mnras] {10.1111/j.1365-2966.2012.21478.x}, \href
  {http://adsabs.harvard.edu/abs/2012MNRAS.426..440B} {426, 440}

\bibitem[\protect\citeauthoryear{{Bruzual} \& {Charlot}}{{Bruzual} \&
  {Charlot}}{2003}]{bruzual2003}
{Bruzual} G.,  {Charlot} S.,  2003, \mn@doi [\mnras]
  {10.1046/j.1365-8711.2003.06897.x}, \href
  {http://adsabs.harvard.edu/abs/2003MNRAS.344.1000B} {344, 1000}

\bibitem[\protect\citeauthoryear{{Cai}, {Padilla}  \& {Li}}{{Cai}
  et~al.}{2015}]{cai2015}
{Cai} Y.-C.,  {Padilla} N.,   {Li} B.,  2015, \mn@doi [\mnras]
  {10.1093/mnras/stv777}, \href
  {http://adsabs.harvard.edu/abs/2015MNRAS.451.1036C} {451, 1036}

\bibitem[\protect\citeauthoryear{{Capaccioli} \& {Schipani}}{{Capaccioli} \&
  {Schipani}}{2011}]{capaccioli2011}
{Capaccioli} M.,  {Schipani} P.,  2011, The Messenger, \href
  {http://adsabs.harvard.edu/abs/2011Msngr.146....2C} {146, 2}

\bibitem[\protect\citeauthoryear{{Carretero}, {Castander}, {Gazta{\~n}aga},
  {Crocce}  \& {Fosalba}}{{Carretero} et~al.}{2015}]{carretero2015}
{Carretero} J.,  {Castander} F.~J.,  {Gazta{\~n}aga} E.,  {Crocce} M.,
  {Fosalba} P.,  2015, \mn@doi [\mnras] {10.1093/mnras/stu2402}, \href
  {http://adsabs.harvard.edu/abs/2015MNRAS.447..646C} {447, 646}

\bibitem[\protect\citeauthoryear{Carretero et~al.}{Carretero
  et~al.}{2017}]{carretero2017}
Carretero J.,  et~al., 2017, in {Proceedings, 2017 European Physical Society
  Conference on High Energy Physics (EPS-HEP 2017): Venice, Italy, July 5-12,
  2017}. p.~488, \mn@doi{10.22323/1.314.0488}

\bibitem[\protect\citeauthoryear{{Cautun}, {Cai}  \& {Frenk}}{{Cautun}
  et~al.}{2016}]{cautun2016}
{Cautun} M.,  {Cai} Y.-C.,   {Frenk} C.~S.,  2016, \mn@doi [\mnras]
  {10.1093/mnras/stw154}, \href
  {http://adsabs.harvard.edu/abs/2016MNRAS.457.2540C} {457, 2540}

\bibitem[\protect\citeauthoryear{{Cautun}, {Paillas}, {Cai}, {Bose}, {Armijo},
  {Li}  \& {Padilla}}{{Cautun} et~al.}{2017}]{cautun2017}
{Cautun} M.,  {Paillas} E.,  {Cai} Y.-C.,  {Bose} S.,  {Armijo} J.,  {Li} B.,
  {Padilla} N.,  2017, preprint, \href
  {http://adsabs.harvard.edu/abs/2017arXiv171001730C} {} (\mn@eprint {arXiv}
  {1710.01730})

\bibitem[\protect\citeauthoryear{{Clampitt} \& {Jain}}{{Clampitt} \&
  {Jain}}{2015}]{clampitt2015}
{Clampitt} J.,  {Jain} B.,  2015, \mn@doi [\mnras] {10.1093/mnras/stv2215},
  \href {http://adsabs.harvard.edu/abs/2015MNRAS.454.3357C} {454, 3357}

\bibitem[\protect\citeauthoryear{{Clampitt}, {Cai}  \& {Li}}{{Clampitt}
  et~al.}{2013}]{clampitt2013}
{Clampitt} J.,  {Cai} Y.-C.,   {Li} B.,  2013, \mn@doi [\mnras]
  {10.1093/mnras/stt219}, \href
  {http://adsabs.harvard.edu/abs/2013MNRAS.431..749C} {431, 749}

\bibitem[\protect\citeauthoryear{{Clifton}, {Ferreira}, {Padilla}  \&
  {Skordis}}{{Clifton} et~al.}{2012}]{clifton2012}
{Clifton} T.,  {Ferreira} P.~G.,  {Padilla} A.,   {Skordis} C.,  2012, \mn@doi
  [\physrep] {10.1016/j.physrep.2012.01.001}, \href
  {http://adsabs.harvard.edu/abs/2012PhR...513....1C} {513, 1}

\bibitem[\protect\citeauthoryear{{Colberg} et~al.,}{{Colberg}
  et~al.}{2008}]{colberg2008}
{Colberg} J.~M.,  et~al., 2008, \mn@doi [\mnras]
  {10.1111/j.1365-2966.2008.13307.x}, \href
  {http://adsabs.harvard.edu/abs/2008MNRAS.387..933C} {387, 933}

\bibitem[\protect\citeauthoryear{{Coles} \& {Jones}}{{Coles} \&
  {Jones}}{1991}]{coles1991}
{Coles} P.,  {Jones} B.,  1991, \mn@doi [\mnras] {10.1093/mnras/248.1.1}, \href
  {http://adsabs.harvard.edu/abs/1991MNRAS.248....1C} {248, 1}

\bibitem[\protect\citeauthoryear{{Colless} et~al.,}{{Colless}
  et~al.}{2001}]{colless2001}
{Colless} M.,  et~al., 2001, \mn@doi [\mnras]
  {10.1046/j.1365-8711.2001.04902.x}, \href
  {http://adsabs.harvard.edu/abs/2001MNRAS.328.1039C} {328, 1039}

\bibitem[\protect\citeauthoryear{{Crocce}, {Castander}, {Gazta{\~n}aga},
  {Fosalba}  \& {Carretero}}{{Crocce} et~al.}{2015}]{crocce2015}
{Crocce} M.,  {Castander} F.~J.,  {Gazta{\~n}aga} E.,  {Fosalba} P.,
  {Carretero} J.,  2015, \mn@doi [\mnras] {10.1093/mnras/stv1708}, \href
  {http://adsabs.harvard.edu/abs/2015MNRAS.453.1513C} {453, 1513}

\bibitem[\protect\citeauthoryear{{Dark Energy Science Collaboration}}{{Dark
  Energy Science Collaboration}}{2012}]{lsst2012}
{Dark Energy Science Collaboration} 2012, preprint, \href
  {http://adsabs.harvard.edu/abs/2012arXiv1211.0310L} {} (\mn@eprint {arXiv}
  {1211.0310})

\bibitem[\protect\citeauthoryear{{Demchenko}, {Cai}, {Heymans}  \&
  {Peacock}}{{Demchenko} et~al.}{2016}]{demchenko2016}
{Demchenko} V.,  {Cai} Y.-C.,  {Heymans} C.,   {Peacock} J.~A.,  2016, \mn@doi
  [\mnras] {10.1093/mnras/stw2030}, \href
  {http://adsabs.harvard.edu/abs/2016MNRAS.463..512D} {463, 512}

\bibitem[\protect\citeauthoryear{{Driver} et~al.,}{{Driver}
  et~al.}{2011}]{driver2011}
{Driver} S.~P.,  et~al., 2011, \mn@doi [\mnras]
  {10.1111/j.1365-2966.2010.18188.x}, \href
  {http://adsabs.harvard.edu/abs/2011MNRAS.413..971D} {413, 971}

\bibitem[\protect\citeauthoryear{{Drlica-Wagner} et~al.,}{{Drlica-Wagner}
  et~al.}{2017}]{drlica2017}
{Drlica-Wagner} A.,  et~al., 2017, preprint, \href
  {http://adsabs.harvard.edu/abs/2017arXiv170801531D} {} (\mn@eprint {arXiv}
  {1708.01531})

\bibitem[\protect\citeauthoryear{{Dvornik} et~al.,}{{Dvornik}
  et~al.}{2017}]{dvornik2017}
{Dvornik} A.,  et~al., 2017, \mn@doi [\mnras] {10.1093/mnras/stx705}, \href
  {http://adsabs.harvard.edu/abs/2017MNRAS.468.3251D} {468, 3251}

\bibitem[\protect\citeauthoryear{{Dvornik} et~al.,}{{Dvornik}
  et~al.}{2018}]{dvornik2018}
{Dvornik} A.,  et~al., 2018, preprint, \href
  {http://adsabs.harvard.edu/abs/2018arXiv180200734D} {} (\mn@eprint {arXiv}
  {1802.00734})

\bibitem[\protect\citeauthoryear{{Edge}, {Sutherland}, {Kuijken}, {Driver},
  {McMahon}, {Eales}  \& {Emerson}}{{Edge} et~al.}{2013}]{edge2013}
{Edge} A.,  {Sutherland} W.,  {Kuijken} K.,  {Driver} S.,  {McMahon} R.,
  {Eales} S.,   {Emerson} J.~P.,  2013, The Messenger, \href
  {http://adsabs.harvard.edu/abs/2013Msngr.154...32E} {154, 32}

\bibitem[\protect\citeauthoryear{{Erben} et~al.,}{{Erben}
  et~al.}{2013}]{erben2013}
{Erben} T.,  et~al., 2013, \mn@doi [\mnras] {10.1093/mnras/stt928}, \href
  {http://adsabs.harvard.edu/abs/2013MNRAS.433.2545E} {433, 2545}

\bibitem[\protect\citeauthoryear{{Falck}, {Koyama}, {Zhao}  \&
  {Cautun}}{{Falck} et~al.}{2017}]{falck2017}
{Falck} B.,  {Koyama} K.,  {Zhao} G.-b.,   {Cautun} M.,  2017, preprint, \href
  {http://adsabs.harvard.edu/abs/2017arXiv170408942F} {} (\mn@eprint {arXiv}
  {1704.08942})

\bibitem[\protect\citeauthoryear{{Fenech Conti}, {Herbonnet}, {Hoekstra},
  {Merten}, {Miller}  \& {Viola}}{{Fenech Conti}
  et~al.}{2017}]{fenechconti2017}
{Fenech Conti} I.,  {Herbonnet} R.,  {Hoekstra} H.,  {Merten} J.,  {Miller} L.,
    {Viola} M.,  2017, \mn@doi [\mnras] {10.1093/mnras/stx200}, \href
  {http://adsabs.harvard.edu/abs/2017MNRAS.467.1627F} {467, 1627}

\bibitem[\protect\citeauthoryear{{Flaugher} et~al.,}{{Flaugher}
  et~al.}{2015}]{flaugher2015}
{Flaugher} B.,  et~al., 2015, \mn@doi [\aj] {10.1088/0004-6256/150/5/150},
  \href {http://adsabs.harvard.edu/abs/2015AJ....150..150F} {150, 150}

\bibitem[\protect\citeauthoryear{{Fosalba}, {Gazta{\~n}aga}, {Castander}  \&
  {Manera}}{{Fosalba} et~al.}{2008}]{fosalba2008}
{Fosalba} P.,  {Gazta{\~n}aga} E.,  {Castander} F.~J.,   {Manera} M.,  2008,
  \mn@doi [\mnras] {10.1111/j.1365-2966.2008.13910.x}, \href
  {http://adsabs.harvard.edu/abs/2008MNRAS.391..435F} {391, 435}

\bibitem[\protect\citeauthoryear{{Fosalba}, {Gazta{\~n}aga}, {Castander}  \&
  {Crocce}}{{Fosalba} et~al.}{2015a}]{fosalba2015a}
{Fosalba} P.,  {Gazta{\~n}aga} E.,  {Castander} F.~J.,   {Crocce} M.,  2015a,
  \mn@doi [\mnras] {10.1093/mnras/stu2464}, \href
  {http://adsabs.harvard.edu/abs/2015MNRAS.447.1319F} {447, 1319}

\bibitem[\protect\citeauthoryear{{Fosalba}, {Crocce}, {Gazta{\~n}aga}  \&
  {Castander}}{{Fosalba} et~al.}{2015b}]{fosalba2015b}
{Fosalba} P.,  {Crocce} M.,  {Gazta{\~n}aga} E.,   {Castander} F.~J.,  2015b,
  \mn@doi [\mnras] {10.1093/mnras/stv138}, \href
  {http://adsabs.harvard.edu/abs/2015MNRAS.448.2987F} {448, 2987}

\bibitem[\protect\citeauthoryear{{Friedrich} et~al.,}{{Friedrich}
  et~al.}{2017}]{friedrich2017}
{Friedrich} O.,  et~al., 2017, preprint, \href
  {http://adsabs.harvard.edu/abs/2017arXiv171005162F} {} (\mn@eprint {arXiv}
  {1710.05162})

\bibitem[\protect\citeauthoryear{{Gruen} et~al.,}{{Gruen}
  et~al.}{2016}]{gruen2016}
{Gruen} D.,  et~al., 2016, \mn@doi [\mnras] {10.1093/mnras/stv2506}, \href
  {http://adsabs.harvard.edu/abs/2016MNRAS.455.3367G} {455, 3367}

\bibitem[\protect\citeauthoryear{{Gruen} et~al.,}{{Gruen}
  et~al.}{2017}]{gruen2017}
{Gruen} D.,  et~al., 2017, preprint, \href
  {http://adsabs.harvard.edu/abs/2017arXiv171005045G} {} (\mn@eprint {arXiv}
  {1710.05045})

\bibitem[\protect\citeauthoryear{{Hamaus}, {Sutter}  \& {Wandelt}}{{Hamaus}
  et~al.}{2014}]{hamaus2014}
{Hamaus} N.,  {Sutter} P.~M.,   {Wandelt} B.~D.,  2014, \mn@doi [Physical
  Review Letters] {10.1103/PhysRevLett.112.251302}, \href
  {http://adsabs.harvard.edu/abs/2014PhRvL.112y1302H} {112, 251302}

\bibitem[\protect\citeauthoryear{{Harnois-D{\'e}raps} \& {van
  Waerbeke}}{{Harnois-D{\'e}raps} \& {van Waerbeke}}{2015}]{harnois2015}
{Harnois-D{\'e}raps} J.,  {van Waerbeke} L.,  2015, \mn@doi [\mnras]
  {10.1093/mnras/stv794}, \href
  {http://adsabs.harvard.edu/abs/2015MNRAS.450.2857H} {450, 2857}

\bibitem[\protect\citeauthoryear{{Harnois-Deraps} et~al.,}{{Harnois-Deraps}
  et~al.}{2018}]{harnois2018}
{Harnois-Deraps} J.,  et~al., 2018, preprint, \href
  {http://adsabs.harvard.edu/abs/2018arXiv180504511H} {} (\mn@eprint {arXiv}
  {1805.04511})

\bibitem[\protect\citeauthoryear{{Heitmann}, {Lawrence}, {Kwan}, {Habib}  \&
  {Higdon}}{{Heitmann} et~al.}{2014}]{heitmann2014}
{Heitmann} K.,  {Lawrence} E.,  {Kwan} J.,  {Habib} S.,   {Higdon} D.,  2014,
  \mn@doi [\apj] {10.1088/0004-637X/780/1/111}, \href
  {http://adsabs.harvard.edu/abs/2014ApJ...780..111H} {780, 111}

\bibitem[\protect\citeauthoryear{{Heymans}, {White}, {Heavens}, {Vale}  \& {van
  Waerbeke}}{{Heymans} et~al.}{2006}]{heymans2006}
{Heymans} C.,  {White} M.,  {Heavens} A.,  {Vale} C.,   {van Waerbeke} L.,
  2006, \mn@doi [\mnras] {10.1111/j.1365-2966.2006.10705.x}, \href
  {http://adsabs.harvard.edu/abs/2006MNRAS.371..750H} {371, 750}

\bibitem[\protect\citeauthoryear{{Higuchi} \& {Shirasaki}}{{Higuchi} \&
  {Shirasaki}}{2016}]{higuchi2016}
{Higuchi} Y.,  {Shirasaki} M.,  2016, \mn@doi [\mnras] {10.1093/mnras/stw814},
  \href {http://adsabs.harvard.edu/abs/2016MNRAS.459.2762H} {459, 2762}

\bibitem[\protect\citeauthoryear{{Hildebrandt} et~al.,}{{Hildebrandt}
  et~al.}{2012}]{hildebrandt2012}
{Hildebrandt} H.,  et~al., 2012, \mn@doi [\mnras]
  {10.1111/j.1365-2966.2012.20468.x}, \href
  {http://adsabs.harvard.edu/abs/2012MNRAS.421.2355H} {421, 2355}

\bibitem[\protect\citeauthoryear{{Hildebrandt} et~al.,}{{Hildebrandt}
  et~al.}{2017}]{hildebrandt2017}
{Hildebrandt} H.,  et~al., 2017, \mn@doi [\mnras] {10.1093/mnras/stw2805},
  \href {http://adsabs.harvard.edu/abs/2017MNRAS.465.1454H} {465, 1454}

\bibitem[\protect\citeauthoryear{{Hoffmann}, {Bel}, {Gazta{\~n}aga}, {Crocce},
  {Fosalba}  \& {Castander}}{{Hoffmann} et~al.}{2015}]{hoffmann2015}
{Hoffmann} K.,  {Bel} J.,  {Gazta{\~n}aga} E.,  {Crocce} M.,  {Fosalba} P.,
  {Castander} F.~J.,  2015, \mn@doi [\mnras] {10.1093/mnras/stu2492}, \href
  {http://adsabs.harvard.edu/abs/2015MNRAS.447.1724H} {447, 1724}

\bibitem[\protect\citeauthoryear{Hunter et~al.}{Hunter
  et~al.}{2007}]{hunter2007matplotlib}
Hunter J.~D.,  et~al., 2007, Computing in science and engineering, 9, 90

\bibitem[\protect\citeauthoryear{{Ilbert} et~al.,}{{Ilbert}
  et~al.}{2009}]{ilbert2009}
{Ilbert} O.,  et~al., 2009, \mn@doi [\apj] {10.1088/0004-637X/690/2/1236},
  \href {http://adsabs.harvard.edu/abs/2009ApJ...690.1236I} {690, 1236}

\bibitem[\protect\citeauthoryear{{Jain} \& {Khoury}}{{Jain} \&
  {Khoury}}{2010}]{jain2010}
{Jain} B.,  {Khoury} J.,  2010, \mn@doi [Annals of Physics]
  {10.1016/j.aop.2010.04.002}, \href
  {http://adsabs.harvard.edu/abs/2010AnPhy.325.1479J} {325, 1479}

\bibitem[\protect\citeauthoryear{{\SortNoop{Jong}}de~Jong
  et~al.,}{{\SortNoop{Jong}}de~Jong et~al.}{2015}]{dejong2015}
{\SortNoop{Jong}}de~Jong J.~T.~A.,  et~al., 2015, \mn@doi [\aap]
  {10.1051/0004-6361/201526601}, \href
  {http://adsabs.harvard.edu/abs/2015A%26A...582A..62D} {582, A62}

\bibitem[\protect\citeauthoryear{{\SortNoop{Jong}}de~Jong
  et~al.,}{{\SortNoop{Jong}}de~Jong et~al.}{2017}]{dejong2017}
{\SortNoop{Jong}}de~Jong J.~T.~A.,  et~al., 2017, \mn@doi [\aap]
  {10.1051/0004-6361/201730747}, \href
  {http://adsabs.harvard.edu/abs/2017A%26A...604A.134D} {604, A134}

\bibitem[\protect\citeauthoryear{{Kacprzak} et~al.,}{{Kacprzak}
  et~al.}{2016}]{kacprzak2016}
{Kacprzak} T.,  et~al., 2016, \mn@doi [\mnras] {10.1093/mnras/stw2070}, \href
  {http://adsabs.harvard.edu/abs/2016MNRAS.463.3653K} {463, 3653}

\bibitem[\protect\citeauthoryear{{Krause}, {Chang}, {Dor{\'e}}  \&
  {Umetsu}}{{Krause} et~al.}{2013}]{krause2013}
{Krause} E.,  {Chang} T.-C.,  {Dor{\'e}} O.,   {Umetsu} K.,  2013, \mn@doi
  [\apjl] {10.1088/2041-8205/762/2/L20}, \href
  {http://adsabs.harvard.edu/abs/2013ApJ...762L..20K} {762, L20}

\bibitem[\protect\citeauthoryear{{Kuijken}}{{Kuijken}}{2008}]{kuijken2008}
{Kuijken} K.,  2008, \mn@doi [\aap] {10.1051/0004-6361:20066601}, \href
  {http://adsabs.harvard.edu/abs/2008A%26A...482.1053K} {482, 1053}

\bibitem[\protect\citeauthoryear{{Kuijken}}{{Kuijken}}{2011}]{kuijken2011}
{Kuijken} K.,  2011, The Messenger, \href
  {http://adsabs.harvard.edu/abs/2011Msngr.146....8K} {146, 8}

\bibitem[\protect\citeauthoryear{{Kuijken} et~al.,}{{Kuijken}
  et~al.}{2015}]{kuijken2015}
{Kuijken} K.,  et~al., 2015, \mn@doi [\mnras] {10.1093/mnras/stv2140}, \href
  {http://adsabs.harvard.edu/abs/2015MNRAS.454.3500K} {454, 3500}

\bibitem[\protect\citeauthoryear{{Lam}, {Clampitt}, {Cai}  \& {Li}}{{Lam}
  et~al.}{2015}]{lam2015}
{Lam} T.~Y.,  {Clampitt} J.,  {Cai} Y.-C.,   {Li} B.,  2015, \mn@doi [\mnras]
  {10.1093/mnras/stv797}, \href
  {http://adsabs.harvard.edu/abs/2015MNRAS.450.3319L} {450, 3319}

\bibitem[\protect\citeauthoryear{{Laureijs} et~al.,}{{Laureijs}
  et~al.}{2011}]{laureijs2011}
{Laureijs} R.,  et~al., 2011, preprint, \href
  {http://adsabs.harvard.edu/abs/2011arXiv1110.3193L} {} (\mn@eprint {arXiv}
  {1110.3193})

\bibitem[\protect\citeauthoryear{{Lavaux} \& {Wandelt}}{{Lavaux} \&
  {Wandelt}}{2010}]{lavaux2010}
{Lavaux} G.,  {Wandelt} B.~D.,  2010, \mn@doi [\mnras]
  {10.1111/j.1365-2966.2010.16197.x}, \href
  {http://adsabs.harvard.edu/abs/2010MNRAS.403.1392L} {403, 1392}

\bibitem[\protect\citeauthoryear{{Li}, {Zhao}  \& {Koyama}}{{Li}
  et~al.}{2012}]{li2012}
{Li} B.,  {Zhao} G.-B.,   {Koyama} K.,  2012, \mn@doi [\mnras]
  {10.1111/j.1365-2966.2012.20573.x}, \href
  {http://adsabs.harvard.edu/abs/2012MNRAS.421.3481L} {421, 3481}

\bibitem[\protect\citeauthoryear{{Liske} et~al.,}{{Liske}
  et~al.}{2015}]{liske2015}
{Liske} J.,  et~al., 2015, \mn@doi [\mnras] {10.1093/mnras/stv1436}, \href
  {http://adsabs.harvard.edu/abs/2015MNRAS.452.2087L} {452, 2087}

\bibitem[\protect\citeauthoryear{{Liu}, {Petri}, {Haiman}, {Hui}, {Kratochvil}
  \& {May}}{{Liu} et~al.}{2015}]{liu2015}
{Liu} J.,  {Petri} A.,  {Haiman} Z.,  {Hui} L.,  {Kratochvil} J.~M.,   {May}
  M.,  2015, \mn@doi [\prd] {10.1103/PhysRevD.91.063507}, \href
  {http://adsabs.harvard.edu/abs/2015PhRvD..91f3507L} {91, 063507}

\bibitem[\protect\citeauthoryear{{Martinet} et~al.,}{{Martinet}
  et~al.}{2018}]{martinet2018}
{Martinet} N.,  et~al., 2018, \mn@doi [\mnras] {10.1093/mnras/stx2793}, \href
  {http://adsabs.harvard.edu/abs/2018MNRAS.474..712M} {474, 712}

\bibitem[\protect\citeauthoryear{{McFarland}, {Verdoes-Kleijn}, {Sikkema},
  {Helmich}, {Boxhoorn}  \& {Valentijn}}{{McFarland}
  et~al.}{2013}]{mcfarland2013}
{McFarland} J.~P.,  {Verdoes-Kleijn} G.,  {Sikkema} G.,  {Helmich} E.~M.,
  {Boxhoorn} D.~R.,   {Valentijn} E.~A.,  2013, \mn@doi [Experimental
  Astronomy] {10.1007/s10686-011-9266-x}, \href
  {http://adsabs.harvard.edu/abs/2013ExA....35...45M} {35, 45}

\bibitem[\protect\citeauthoryear{{Melchior}, {Sutter}, {Sheldon}, {Krause}  \&
  {Wandelt}}{{Melchior} et~al.}{2014}]{melchior2014}
{Melchior} P.,  {Sutter} P.~M.,  {Sheldon} E.~S.,  {Krause} E.,   {Wandelt}
  B.~D.,  2014, \mn@doi [\mnras] {10.1093/mnras/stu456}, \href
  {http://adsabs.harvard.edu/abs/2014MNRAS.440.2922M} {440, 2922}

\bibitem[\protect\citeauthoryear{{Miller}, {Kitching}, {Heymans}, {Heavens}  \&
  {van Waerbeke}}{{Miller} et~al.}{2007}]{miller2007}
{Miller} L.,  {Kitching} T.~D.,  {Heymans} C.,  {Heavens} A.~F.,   {van
  Waerbeke} L.,  2007, \mn@doi [\mnras] {10.1111/j.1365-2966.2007.12363.x},
  \href {http://adsabs.harvard.edu/abs/2007MNRAS.382..315M} {382, 315}

\bibitem[\protect\citeauthoryear{{Miller} et~al.,}{{Miller}
  et~al.}{2013}]{miller2013}
{Miller} L.,  et~al., 2013, \mn@doi [\mnras] {10.1093/mnras/sts454}, \href
  {http://adsabs.harvard.edu/abs/2013MNRAS.429.2858M} {429, 2858}

\bibitem[\protect\citeauthoryear{{Nadathur}, {Hotchkiss}, {Diego}, {Iliev},
  {Gottl{\"o}ber}, {Watson}  \& {Yepes}}{{Nadathur}
  et~al.}{2015}]{nadathur2015}
{Nadathur} S.,  {Hotchkiss} S.,  {Diego} J.~M.,  {Iliev} I.~T.,
  {Gottl{\"o}ber} S.,  {Watson} W.~A.,   {Yepes} G.,  2015, \mn@doi [\mnras]
  {10.1093/mnras/stv513}, \href
  {http://adsabs.harvard.edu/abs/2015MNRAS.449.3997N} {449, 3997}

\bibitem[\protect\citeauthoryear{{Rozo} et~al.,}{{Rozo}
  et~al.}{2016}]{rozo2016}
{Rozo} E.,  et~al., 2016, \mn@doi [\mnras] {10.1093/mnras/stw1281}, \href
  {http://adsabs.harvard.edu/abs/2016MNRAS.461.1431R} {461, 1431}

\bibitem[\protect\citeauthoryear{{Sadeh}, {Abdalla}  \& {Lahav}}{{Sadeh}
  et~al.}{2016}]{sadeh2016}
{Sadeh} I.,  {Abdalla} F.~B.,   {Lahav} O.,  2016, \mn@doi [\pasp]
  {10.1088/1538-3873/128/968/104502}, \href
  {http://adsabs.harvard.edu/abs/2016PASP..128j4502S} {128, 104502}

\bibitem[\protect\citeauthoryear{{S{\'a}nchez} et~al.,}{{S{\'a}nchez}
  et~al.}{2017}]{sanchez2017}
{S{\'a}nchez} C.,  et~al., 2017, \mn@doi [\mnras] {10.1093/mnras/stw2745},
  \href {http://adsabs.harvard.edu/abs/2017MNRAS.465..746S} {465, 746}

\bibitem[\protect\citeauthoryear{{Schaye} et~al.,}{{Schaye}
  et~al.}{2015}]{schaye2015}
{Schaye} J.,  et~al., 2015, \mn@doi [\mnras] {10.1093/mnras/stu2058}, \href
  {http://adsabs.harvard.edu/abs/2015MNRAS.446..521S} {446, 521}

\bibitem[\protect\citeauthoryear{Schneider, Kochanek  \& Wambsganss}{Schneider
  et~al.}{2006}]{schneider2006}
Schneider P.,  Kochanek C.~S.,   Wambsganss J.,  2006, {Gravitational Lensing:
  Strong, Weak and Micro}.
Saas-Fee Advanced Courses, Swiss Society for Astrophysics and Astronomy,
  Springer, Berlin, Heidelberg, \url {http://cds.cern.ch/record/1339023}

\bibitem[\protect\citeauthoryear{{Shan} et~al.,}{{Shan}
  et~al.}{2018}]{shan2018}
{Shan} H.,  et~al., 2018, \mn@doi [\mnras] {10.1093/mnras/stx2837}, \href
  {http://adsabs.harvard.edu/abs/2018MNRAS.474.1116S} {474, 1116}

\bibitem[\protect\citeauthoryear{{Singh}, {Mandelbaum}, {Seljak}, {Slosar}  \&
  {Vazquez Gonzalez}}{{Singh} et~al.}{2017}]{singh2017}
{Singh} S.,  {Mandelbaum} R.,  {Seljak} U.,  {Slosar} A.,   {Vazquez Gonzalez}
  J.,  2017, \mn@doi [\mnras] {10.1093/mnras/stx1828}, \href
  {http://adsabs.harvard.edu/abs/2017MNRAS.471.3827S} {471, 3827}

\bibitem[\protect\citeauthoryear{{Smith}, {Cole}, {Baugh}, {Zheng}, {Angulo},
  {Norberg}  \& {Zehavi}}{{Smith} et~al.}{2017}]{smith2017}
{Smith} A.,  {Cole} S.,  {Baugh} C.,  {Zheng} Z.,  {Angulo} R.,  {Norberg} P.,
   {Zehavi} I.,  2017, preprint, \href
  {http://adsabs.harvard.edu/abs/2017arXiv170106581S} {} (\mn@eprint {arXiv}
  {1701.06581})

\bibitem[\protect\citeauthoryear{{Taylor} et~al.,}{{Taylor}
  et~al.}{2011}]{taylor2011}
{Taylor} E.~N.,  et~al., 2011, \mn@doi [\mnras]
  {10.1111/j.1365-2966.2011.19536.x}, \href
  {http://adsabs.harvard.edu/abs/2011MNRAS.418.1587T} {418, 1587}

\bibitem[\protect\citeauthoryear{{Tinker}, {Robertson}, {Kravtsov}, {Klypin},
  {Warren}, {Yepes}  \& {Gottl{\"o}ber}}{{Tinker} et~al.}{2010}]{tinker2010}
{Tinker} J.~L.,  {Robertson} B.~E.,  {Kravtsov} A.~V.,  {Klypin} A.,  {Warren}
  M.~S.,  {Yepes} G.,   {Gottl{\"o}ber} S.,  2010, \mn@doi [\apj]
  {10.1088/0004-637X/724/2/878}, \href
  {http://adsabs.harvard.edu/abs/2010ApJ...724..878T} {724, 878}

\bibitem[\protect\citeauthoryear{{\SortNoop{Uitert}}van~Uitert \&
  {Schneider}}{{\SortNoop{Uitert}}van~Uitert \&
  {Schneider}}{2016}]{uitert2016b}
{\SortNoop{Uitert}}van~Uitert E.,  {Schneider} P.,  2016, \mn@doi [\aap]
  {10.1051/0004-6361/201628846}, \href
  {http://adsabs.harvard.edu/abs/2016A%26A...595A..93V} {595, A93}

\bibitem[\protect\citeauthoryear{{Viola} et~al.,}{{Viola}
  et~al.}{2015}]{viola2015}
{Viola} M.,  et~al., 2015, \mn@doi [\mnras] {10.1093/mnras/stv1447}, \href
  {http://adsabs.harvard.edu/abs/2015MNRAS.452.3529V} {452, 3529}

\bibitem[\protect\citeauthoryear{{Vogelsberger} et~al.,}{{Vogelsberger}
  et~al.}{2014}]{vogelsberger2014}
{Vogelsberger} M.,  et~al., 2014, \mn@doi [\nat] {10.1038/nature13316}, \href
  {http://adsabs.harvard.edu/abs/2014Natur.509..177V} {509, 177}

\bibitem[\protect\citeauthoryear{{Zehavi} et~al.,}{{Zehavi}
  et~al.}{2011}]{zehavi2011}
{Zehavi} I.,  et~al., 2011, \mn@doi [\apj] {10.1088/0004-637X/736/1/59}, \href
  {http://adsabs.harvard.edu/abs/2011ApJ...736...59Z} {736, 59}

\bibitem[\protect\citeauthoryear{{Zivick}, {Sutter}, {Wandelt}, {Li}  \&
  {Lam}}{{Zivick} et~al.}{2015}]{zivick2015}
{Zivick} P.,  {Sutter} P.~M.,  {Wandelt} B.~D.,  {Li} B.,   {Lam} T.~Y.,  2015,
  \mn@doi [\mnras] {10.1093/mnras/stv1209}, \href
  {http://adsabs.harvard.edu/abs/2015MNRAS.451.4215Z} {451, 4215}

\makeatother
\end{thebibliography}

%%%%%%%%%%%%%%%%%%%%%%%%%%%%%%%%%%%%%%%%%%%%%%%%%%

%%%%%%%%%%%%%%%%%%%%%%%%%%%%%%%%%%%%%%%%%%%%%%%%%%

% Don't change these lines
\bsp	% typesetting comment
\label{lastpage}
\end{document}